\documentclass[aps,prx,twocolumn,reprint,superscriptaddress]{revtex4-2}
\usepackage[pdftex, colorlinks=true, linkcolor=blue, citecolor=blue, urlcolor=blue]{hyperref}

\usepackage{graphicx}
\usepackage{dcolumn}
\usepackage{bm}
\usepackage{amssymb} 

\usepackage{ragged2e}
\usepackage{amsmath} 
\usepackage{amssymb} 
\usepackage{mathrsfs}
\usepackage{physics}
\usepackage{xcolor}
\usepackage{soul} 
\usepackage[dvipsnames]{xcolor}
\setstcolor{Maroon}
\newcommand{\sumint}{%
  \mathop{%
    \ooalign{%
      $\displaystyle\int$\cr
      \hfil$\displaystyle\sum$\hfil\cr
    }%
  }\displaylimits
}

\usepackage[labelfont=normalfont,textfont=normalfont,justification=justified,singlelinecheck=false]{caption}

\begin{document}

\preprint{APS/123-QED}

\title{Fluctuational Quantum Electrodynamics of Dispersive Time-Varying Media}

\author{Jaime E. Sustaeta-Osuna}\thanks{Corresponding author: jaime.echave-sustaeta@uam.es}
\affiliation{Departamento de F\'{i}sica Te\'orica de la Materia Condensada, Universidad Aut\'onoma de Madrid, E-28049 Madrid, Spain}
\affiliation{Condensed Matter Physics Center (IFIMAC), Universidad Aut\'onoma de Madrid, E-28049 Madrid, Spain}
\author{Thomas F. Allard}
\affiliation{Departamento de F\'{i}sica Te\'orica de la Materia Condensada, Universidad Aut\'onoma de Madrid, E-28049 Madrid, Spain}
\affiliation{Condensed Matter Physics Center (IFIMAC), Universidad Aut\'onoma de Madrid, E-28049 Madrid, Spain}
\author{Francisco J. García-Vidal}
\affiliation{Departamento de F\'{i}sica Te\'orica de la Materia Condensada, Universidad Aut\'onoma de Madrid, E-28049 Madrid, Spain}
\affiliation{Condensed Matter Physics Center (IFIMAC), Universidad Aut\'onoma de Madrid, E-28049 Madrid, Spain}
\author{Paloma A. Huidobro}
\affiliation{Departamento de F\'{i}sica Te\'orica de la Materia Condensada, Universidad Aut\'onoma de Madrid, E-28049 Madrid, Spain}
\affiliation{Condensed Matter Physics Center (IFIMAC), Universidad Aut\'onoma de Madrid, E-28049 Madrid, Spain}
\affiliation{Instituto Nicolás Cabrera (INC), Universidad Autónoma de Madrid, E-28049 Madrid, Spain}

\date{\today}

\begin{abstract}
   
In this work, we develop the formalism of fluctuational quantum electrodynamics for frequency-dispersive and dissipative time-varying media. Our framework accounts for the dispersive and lossy nature of the temporal modulation, which we treat in an exact manner, without relying on perturbative methods. Using this theory, we derive a generalized Fermi Golden Rule for time-varying media, which allows us to define the local density of states for these time-dependent systems.  We exemplify our theory with two different periodically driven half-space geometries, considering both a polar insulator and a plasmonic medium. For the polar insulator we work in the slow modulation regime, which reveals a strong enhancement of light-matter interactions at the Floquet sidebands of the surface phonon frequency. On the other hand, for the plasmonic medium we concentrate on the fast modulation regime, in which the coupling to negative frequency replicas results in the appearance of gain and the dynamical Casimir effect. We also analyze the thermal radiation in dispersive and dissipative time-varying media, going well beyond what is allowed by perturbation theory and revealing new features in the thermal spectrum of time-varying systems. Finally, we study the correlation functions of the electromagnetic field, showing how the time modulation enables the creation of entangled polariton pairs exhibiting non-local spatial correlations. 

\end{abstract}


\maketitle
\section{Introduction}
Modulating in time the optical properties of materials lifts the constraints of passivity and energy conservation, leading to the emergence of time-varying media~\cite{galiffi2022photonics,engheta2023four,boltasseva2024photonic,asgari2024theory} and opening exciting new possibilities for both classical and quantum physics.
\par
At the classical level, time-varying media enable novel phenomena not seen in non-modulated materials, such as linear frequency conversion~\cite{shcherbakov2019photon,zhou2020broadband,bohn2021spatiotemporal,apffel2022frequency,ptitcyn2023floquet,globosits2024pseudounitary}, time-reflection and refraction~\cite{moussa2023observation,galiffi2023broadband,lustig2023time,tirole2023double} or synthetic motion effects~\cite{huidobro2019fresnel,bahrami2023electrodynamics,harwood2025space}. Additionally, for the case of periodic temporal modulations, photonic time crystals (PTCs) emerge~\cite{galiffi2022photonics,engheta2023four,boltasseva2024photonic,asgari2024theory}, characterized by exotic band structures with momentum gaps that support modes that grow exponentially in time~\cite{holberg1966parametric,reyes2015observation,wang2023metasurface,khurgin2024photonic,wang2025expanding,lee2026analogs}.
\par
Quantum mechanically, a parametric modulation of material properties has been shown to amplify vacuum fluctuations via the dynamical Casimir effect (DCE)~\cite{nation2012colloquium,lahteenmaki2013dynamical}, whereby pairs of entangled photons are generated from the quantum vacuum through the energy provided by the temporal modulation~\cite{mendoncca2000quantum,bugler2016polariton,sloan2021casimir,horsley2023quantum,ganfornina2024quantum,hassani2024dynamical,sustaeta2025quantum,kim2026classical}. Moreover, several theory works have shown that time-varying media can drastically modify light-matter interactions, such as the emission properties of classical and quantum emitters~\cite{lyubarov2022amplified,dikopoltsev2022light,li2023stationary,bae2025quantum,garg2025inverse,park2025spontaneous,allard2026broadband,sustaeta2026near}, as well as find applications in thermal radiation engineering and heat transfer~\cite{buddhiraju2020photonic,yu2023manipulating,vazquez2023incandescent,yu2025near,ben2026interference,messina2026many,zhu2026enhancing}.
\par
However, all these past works suffer from the same limitation: they lack a complete theory of the electromagnetic field in time-modulated media that properly accounts for the dispersive and dissipative response of the time-varying medium. So far there have been two distinct approaches to deal with dispersion and dissipation in time-varying media: on the one hand, some works have treated the temporal modulation as a perturbation~\cite{sloan2021casimir,vazquez2023incandescent,hassani2024dynamical,ren2026clarification}, with the non-perturbed system being both dispersive and dissipative. This approach, although useful to show the qualitative effects of the time modulation, cannot make quantitative predictions beyond the weak modulation regime, nor meaningful predictions about the strong modulation regime, which is necessarily non-perturbative. On the other hand, there are works that do not rely on perturbative methods, but assume that the temporal modulation is lossless and/or dispersionless~\cite{yu2023manipulating,yu2025near,zhu2026enhancing}. This in turn leads to additional complications, as the effects of the time modulation can be overestimated or underestimated and furthermore, the resulting linear response does not satisfy the Kramers-Kronig relations~\cite{jackson1999classical}. Additionally, there have also been several attempts to develop more microscopic theories for time-varying media~\cite{shirokova2023surface,sloan2024optical,horsley2025macroscopic,mehrpour2026proper,ganfornina2026generalized}, but these works are system-dependent and thus, they do not present a unified modeling of the microscopics of time-modulated materials.
\par
Moreover, state-of-the-art experimental platforms used to implement sub-cycle and order-unity temporal modulations of material properties are all dispersive and dissipative in their operating frequency regimes. This is indeed the case for $\epsilon$-near-zero materials with strong non-linearities in the near-infrared~\cite{caspani116enhanced,alam2016large,vezzoli2018optical,reshef2019nonlinear,bohn2021all,tirole2022saturable,tirole2023double}, plasmonic metamaterials in the THz frequency regime~\cite{guo2025plasmonic}, and transmission line metamaterials in the microwaves~\cite{wang2023metasurface,moussa2023observation,lee2026analogs}. Furthermore, perturbative methods are not guaranteed to work in these setups either. In particular, for the case of $\epsilon$-near-zero media with strong non-linear responses, the relative change in the refractive index induced by the time modulation can reach $100\%$~\cite{reshef2017beyond,reshef2019nonlinear}, thus rendering perturbative approaches ineffective. Therefore, in order to develop a deeper understanding of the electrodynamics of time-modulated materials, it is necessary to account for dispersion and losses in the time modulation, and to be able to treat the latter beyond perturbation theory as well.
\par
In this work, we develop the theoretical formalism of fluctuational electrodynamics in frequency-dispersive and dissipative time-varying media. Crucially, our theory accounts for both dispersion and losses in the time modulation and goes beyond perturbative approaches, enabling us to treat the temporal modulation in an exact manner. Moreover, our framework is quantum mechanical by construction and constitutes the first consistent quantization of the electromagnetic field in time-modulated material bodies. Consequently, our theory is applicable beyond the classical regime, which allows us to find purely quantum mechanical effects in dispersive and dissipative time-varying media.
\par
The outline of the manuscript is as follows: in Sec.~\ref{sec:fqed} we develop the theoretical formalism and present a consistent quantization of the electromagnetic field in dispersive and dissipative time-varying media. In Sec.~\ref{sec:ldos} we derive a Fermi Golden Rule (FGR) for time-varying media, which enables us to define the local density of states (LDOS) for these time-modulated systems. We then apply this framework to a periodically modulated air-medium half-space geometry, considering both a polar insulator and a plasmonic material. In Sec.~\ref{sec:slow} we investigate the slow modulation regime, where coupling to Floquet sidebands dominates, while in Sec.~\ref{sec:fast} we study the fast modulation regime, characterized by the presence of negative frequency replicas and the emergence of gain. In Sec.~\ref{sec:e_corr} we introduce the first-order correlation functions of the electromagnetic field and identify in them signatures of vacuum quantum amplification. Section~\ref{sec:thermal} examines the thermal radiation emitted by a time-modulated polar material beyond perturbative approaches, revealing new features in its thermal spectrum. Finally, in Sec.~\ref{sec:dce} we investigate the dynamical Casimir effect (DCE), demonstrating the existence of non-local correlations between entangled polariton pairs.
\label{sec:intro}
\section{Fluctuational Quantum Electrodynamics}
\label{sec:fqed}
\subsection{Fluctuation-Dissipation Theorem in Dispersive Time-Varying Media}
The starting point of our formalism is Kubo's linear response theory~\cite{kubo1966fluctuation,girvin2019modern}, which establishes a relationship between the response of a system to an external weak perturbation and the properties of the unperturbed system. Crucially, Kubo's theory stands out for its great generality, since it only assumes that the external field probing the system is weak enough so that its effects on the system's dynamics can be treated within first-order perturbation theory~\cite{kubo1966fluctuation,girvin2019modern}. Furthermore, the theory can also be applied to systems that are out of equilibrium~\cite{kubo1966fluctuation,girvin2019modern,dehghani2015out,konopik2019quantum,kumar2020linear,levy2021response}. Despite the aforementioned generality of Kubo's linear response theory, we will particularize it to the case of a macroscopic polarization field $\textbf{P}(\textbf{r},t)$ coupled to an external electric field $\textbf{E}_\text{p}(\textbf{r},t)$ via the interaction Hamiltonian $H_\text{int}(t) = -\int d^3\text{r}\textbf{P}(\textbf{r},t)\cdot\textbf{E}_\text{p}(\textbf{r},t)$. According to Kubo's theory then, the system's electric susceptibility $\boldsymbol{\chi}(\text{r},t;\text{r}',t')$ is given by:

\begin{equation}
    \boldsymbol{\chi}(\text{r},t;\text{r}',t') = -\frac{i}{\hbar}\Theta(t-t')\langle[\textbf{P}(\text{r},t), \textbf{P}(\text{r}',t')]\rangle.
    \label{eq:kubo_macro}
\end{equation}
In Eq.~\eqref{eq:kubo_macro}, the time-evolution of $\textbf{P}(\text{r},t)$ is dictated by the underlying unperturbed Hamiltonian of the system, and the expectation value is calculated using the density matrix of the unperturbed system. Therefore, Eq.~\eqref{eq:kubo_macro} directly relates properties of the unperturbed system, in the form of the polarization-polarization commutator (right-hand side), with its response to a weak probe field (left-hand side).
\par
Taking the Fourier transform of Eq.~\eqref{eq:kubo_macro} leads to:

\begin{equation}
    \begin{split}
       \langle[\textbf{P}(\text{r},\omega), \textbf{P}^\dagger(\text{r}',\omega')]\rangle & = 2\hbar\boldsymbol{\chi}''(\text{r},\omega;\text{r}',\omega'),
    \end{split}
    \label{eq:kubo_macro_freq}
\end{equation}
where we have introduced the anti-Hermitian part of the two-frequency linear susceptibility, $\boldsymbol{\chi}''(\text{r},\omega;\text{r}',\omega')$, which reads:
\begin{equation}
\boldsymbol{\chi}''(\text{r},\omega;\text{r}',\omega') =\frac{1}{i2}\left(\boldsymbol{\chi}(\text{r},\omega;\text{r}',\omega') - \boldsymbol{\chi}^\dagger(\text{r}',\omega';\text{r},\omega)\right).
\label{eq:chi_im}
\end{equation}
The anti-Hermitian part of the electric susceptibility is commonly defined in the case of complex linear responses such as anisotropic or non-local, and generalizes the imaginary part of the same tensor~\cite{rytov1989principles,buhmann2013dispersion1}. Here we write it in its more general form, allowing for two different frequencies. On the other hand, owing to the Hermiticity of the time-domain polarization operator, there is the constraint $\textbf{P}(\textbf{r},-\omega) = \textbf{P}^\dagger(\textbf{r},\omega)$. Therefore, Eq.~\eqref{eq:kubo_macro_freq} establishes that in order for a system to exhibit fluctuations in its polarization field, the system's linear response must be \text{lossy}, in the sense that the anti-Hermitian part of the response function must be non-vanishing. Conversely, it is also possible to understand Eq.~\eqref{eq:kubo_macro_freq} in a different way, whereupon a lossy system must necessarily exhibit fluctuations. The two views are complementary to each other and form what is known as the fluctuation-dissipation theorem~\cite{kubo1966fluctuation,callen1951irreversibility,rytov1989principles,buhmann2013dispersion1,girvin2019modern,novotny2012principles,girvin2019modern}, a connection between the fluctuations of a physical system and the dissipative part of its linear response, and, correspondingly, anti-Hermitian part of its susceptibility.
\par
Hence, Eq.~\eqref{eq:kubo_macro_freq} shows that it is possible to cast a fluctuation-dissipation theorem for time-modulated systems, despite them not being at equilibrium. Additionally, when the system is time-translational invariant, the linear susceptibility satisfies $\boldsymbol{\chi}(\text{r},\omega;\text{r}',\omega') = \boldsymbol{\chi}(\text{r},\text{r}';\omega)2\pi\delta(\omega-\omega')$ and thus, the non-modulated case is recovered. Finally, we mention in here that causality constrains the analytic properties of the Fourier domain electric susceptibility $\boldsymbol{\chi}(\text{r},\omega;\text{r}',\omega')$, similarly to the case of non-modulated media. In particular, it can be proved that $\boldsymbol{\chi}(\text{r},\omega;\text{r}',\omega')$ is analytic in both $\text{Im}(\omega)>0$ and $\text{Im}(\omega')>0$ (see the Supplemental Material~\cite{supmat} for the proof).
\par
So far we have made no assumptions about the underlying properties of the physical systems under consideration; however, from now on we will restrict ourselves to the case of linear media. This is motivated by the observation that time-varying media can be accurately described as linear electromagnetic systems with time-modulated optical parameters~\cite{galiffi2022photonics,engheta2023four,boltasseva2024photonic,asgari2024theory}, as long as they are in a stable phase~\cite{xiong2025observation,kiselev2025symmetry,kyung2026self}. An important feature of linear systems is that the commutator $[\textbf{P}(\text{r},t), \textbf{P}(\text{r}',t')]$ is a c-number (see the SM~\cite{supmat}) and hence, so is its Fourier-domain counterpart. Thus, for the case of linear media, it is always possible to remove the expectation value in Eq.~\eqref{eq:kubo_macro}. 
\par
When describing the electrodynamics of dispersive and dissipative media, bosonic operators that describe joint light-matter excitations are usually introduced. These operators are often called \textit{polaritons}~\cite{hopfield1958theory,huttner1992quantization,matloob1997electromagnetic,schmidt1998quantum,scheel2006quantum,scheel2008macroscopic,philbin2010canonical} and are denoted by $\textbf{f}(\textbf{r},\omega)$ and $\textbf{f}^\dagger(\textbf{r},\omega)$. Furthermore, they exhibit the usual bosonic commutation relations $[\textbf{f}(\textbf{r},\omega), \textbf{f}^\dagger(\textbf{r}',\omega')] = \textbf{I}\delta(\textbf{r}-\textbf{r}')\delta(\omega - \omega')$, where $\textbf{I}$ is the unit dyadic in three dimensions. 
\par
\begin{figure}[t]
 \captionsetup{
        justification=justified,
        singlelinecheck=false
    }
    \centering
    \hspace{0 mm}
    \includegraphics[scale = 0.60]{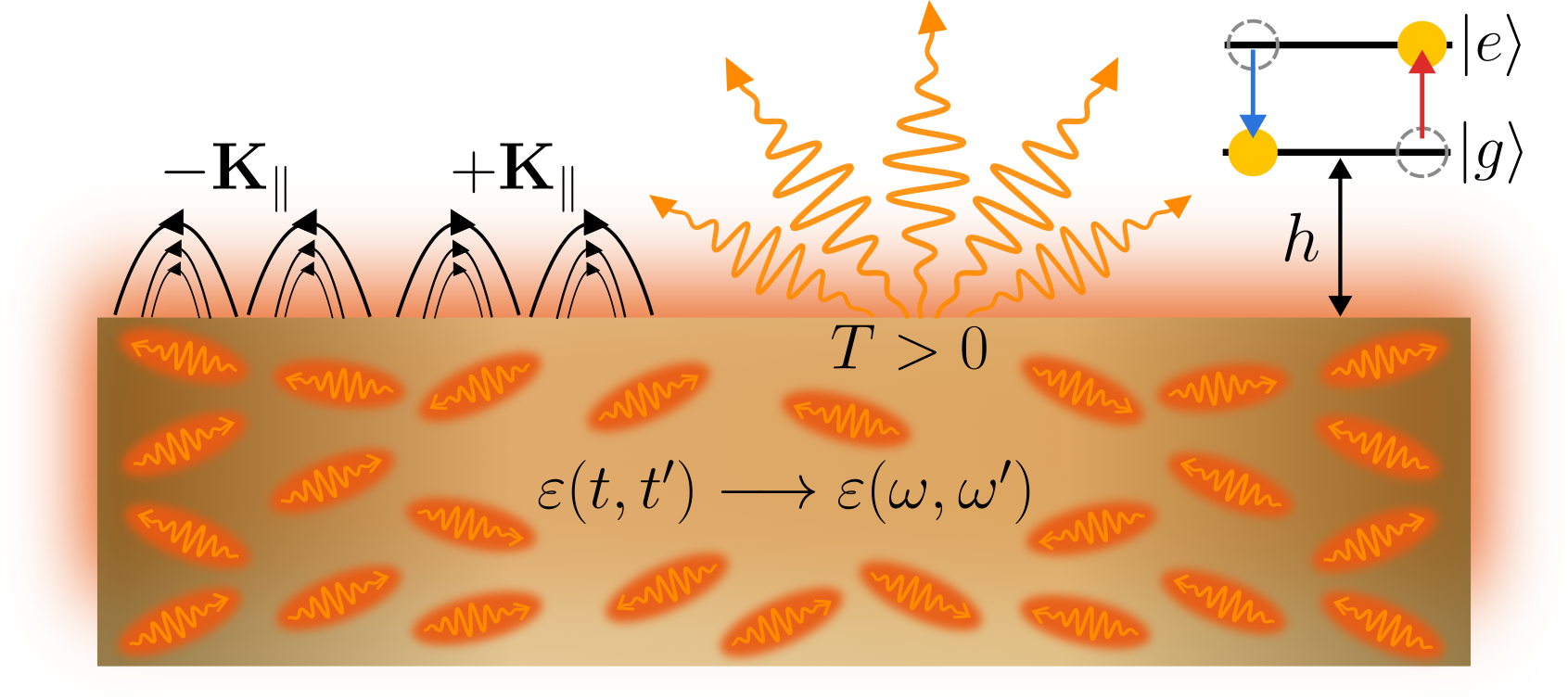} 
    
    \caption{\justifying Diagram of a frequency-dispersive, dissipative and time-modulated half-space. The time modulation breaks time-translational symmetry, leading to a two-frequency permittivity in the Fourier-domain. Owing to losses and by virtue of the fluctuation-dissipation theorem, the medium supports fluctuating currents, which emit thermal radiation for $T>0$. For large modulation frequencies, the time modulation generates entangled pairs of polaritons with opposite lateral momenta $\pm\textbf{K}_\parallel$. Finally, a probe quantum emitter is placed at a height $h$ above the surface; owing to the time modulation, the emitter experiences both loss (spontaneous decay) and gain (spontaneous excitation).}
    \label{fig:fig0} 
\end{figure}
Crucially, it is equally possible to introduce polaritonic operators satisfying bosonic commutation relations for the case of time-varying media (see the SM~\cite{supmat} for more information). First, we note that we can write the polariton commutation relations in a more compact form by introducing negative frequency polaritons, where there is the constraint $\textbf{f}(\textbf{r},-\omega) = \textbf{f}^\dagger(\textbf{r},\omega)$. Then, the commutation relations read $[\textbf{f}(\textbf{r},\omega), \textbf{f}^\dagger(\textbf{r}',\omega')] = \text{sgn}(\omega)\textbf{I}\delta(\textbf{r}-\textbf{r}')\delta(\omega - \omega')$. Next, we introduce the loss kernel of the time-varying medium as $\textbf{K}(\textbf{r}, \omega;\textbf{r}',\omega')$, which connects the microscopic polaritons with the macroscopic polarization:
\begin{equation}
    \begin{split}
        \textbf{P}(\textbf{r},\omega) = \int d^3\text{r}'\int d\omega'\textbf{K}(\textbf{r}, \omega;\textbf{r}';\omega')\cdot\textbf{f}(\textbf{r}',\omega').
    \end{split}
    \label{eq:p_k_f_mod}
\end{equation}
A glance at Eq.~\eqref{eq:p_k_f_mod} shows that there is an additional integral over a frequency variable that is not present in the non-modulated case, for which one has $\textbf{P}(\textbf{r},\omega) = \int d^3\text{r}'\textbf{K}(\textbf{r},\textbf{r}';\omega)\cdot\textbf{f}(\textbf{r}',\omega)$~\cite{scheel2008macroscopic,buhmann2013dispersion1}. Furthermore, recalling that $\textbf{f}(\textbf{r},-\omega) =\textbf{f}^\dagger(\textbf{r},\omega)$, we can write Eq.~\eqref{eq:p_k_f_mod} as:
\begin{equation}
    \begin{split}
        \textbf{P}(\textbf{r},\omega) & =  \int d^3\text{r}'\int_0^\infty d\omega'\textbf{K}(\textbf{r}, \omega;\textbf{r}';\omega')\cdot\textbf{f}(\textbf{r}',\omega')\\&
        + \int d^3\text{r}'\int_0^\infty d\omega'\textbf{K}(\textbf{r}, \omega;\textbf{r}';-\omega')\cdot\textbf{f}^\dagger(\textbf{r}',\omega').
    \end{split}
    \label{eq:p_k_f_mod(2)}
\end{equation}
Therefore, the temporal modulation can couple positive and negative frequencies, which results in the coupling between the positive frequency components of the polarization density, $\textbf{P}(\textbf{r},\omega>0)$ and the polariton creation operators, $\textbf{f}^\dagger(\textbf{r},\omega>0)$. Eq.~\eqref{eq:p_k_f_mod(2)} closely resembles a squeezing transformation~\cite{Loudon2000,barnett2002methods} and we will show below that this is a signature of gain being present in the system. We note that squeezing transformations also appear naturally in idealized, non-dispersive and non-absorbing, models of time varying media~\cite{sustaeta2025quantum,kim2026quantum}.
\par
On the other hand, in order for Eq.~\eqref{eq:p_k_f_mod} to be consistent with the fluctuation-dissipation theorem in Eq.~\eqref{eq:kubo_macro_freq}, the loss kernel must satisfy the following integral formula:
\begin{equation}
    \begin{split}
        \int d^3\text{r}''\int d\omega''\text{sgn}(\omega'')&\textbf{K}(\textbf{r},\omega;\textbf{r}'',\omega'')\cdot\textbf{K}^\dagger(\textbf{r}',\omega';\textbf{r}'',\omega'') \\= 2\hbar\boldsymbol{\chi}''(\text{r},\omega;\text{r}';\omega').
    \end{split}
    \label{eq:loss_kenel_mod}
\end{equation}
In Eq.~\eqref{eq:loss_kenel_mod}, the integral over the frequency variable $\omega''$ involves the sign of the frequency, $\text{sgn}(\omega'')$. The latter functions as a metric for the integral, assigning a negative prefactor to the contribution coming from negative frequencies. Hence, Eq.~\eqref{eq:loss_kenel_mod} shows that to recover the anti-Hermitian part of the electric susceptibility from the loss kernel in frequency-dispersive and dissipative time-varying media, it is necessary to take the difference between positive and negative frequency contributions. 
\par
On the other hand, it must be noted that the frequency $\omega$ appearing in the arguments of the polariton operators is not a Fourier frequency, but a resonant frequency of the continuum of harmonic oscillators that composes the medium~\cite{huttner1992quantization,scheel2006quantum,philbin2010canonical,buhmann2013dispersion1}. For linear and non-modulated media, in which energy is conserved and the frequency of an electromagnetic wave is thus well-defined, Fourier and resonant frequencies coincide~\cite{huttner1992quantization,scheel2008macroscopic,buhmann2013dispersion1}. However, for materials that are nonlinear~\cite{schmidt1998quantum,scheel2006quantum} or time-modulated~\cite{horsley2025macroscopic, park2025spontaneous,allard2026broadband, sustaeta2026near}, this equivalence between Fourier and resonant frequencies no longer holds.
\par
Having discussed in detail Kubo's linear response theory and the extension of the fluctuation-dissipation theorem to time-modulated systems, we now apply them to the study of quantum electrodynamics in dispersive and dissipative time-varying media.
\subsection{Field Quantization in Dispersive Time-Varying Media}
Let us suppose that we have a macroscopic dielectric body with the most general linear permittivity $\boldsymbol{\varepsilon}(\textbf{r},t;\textbf{r}',t')$, which we allow to be frequency-dispersive, spatially non-local, anisotropic and time-varying. Then, Maxwell equations read:
\begin{subequations}
    \begin{equation}
        \curl{\textbf{E}}(\textbf{r},t) = -\partial_t\textbf{B}(\textbf{r},t),
        \label{eq:faraday_law}
    \end{equation}
    \begin{equation}
        \curl{\textbf{B}}(\textbf{r},t) = \mu_0\partial_t\textbf{D}(\textbf{r},t) + \mu_0\partial_t\textbf{P}(\textbf{r},t),
        \label{eq:ampere_law}
    \end{equation}
\end{subequations}
where $\textbf{E}(\textbf{r},t)$ and $\textbf{B}(\textbf{r},t)$ are respectively the electric and magnetic induction fields, and where $\textbf{D}(\textbf{r},t)$ is the displacement field, which connects to the electric field via:
\begin{equation}
    \textbf{D}(\textbf{r},t) = \epsilon_0\int d^3\text{r}'\int dt'\boldsymbol{\varepsilon}(\textbf{r},t;\textbf{r}',t')\cdot\textbf{E}(\textbf{r}',t').
    \label{eq:d_e}
\end{equation}
 In the above, $\epsilon_0$ and $\mu_0$ are the permittivity and permeability of free-space. On the other hand, according to the preceding discussion, the Fourier-domain polarization density $\textbf{P}(\textbf{r},\omega)$ and the anti-Hermitian part of the two-frequency permittivity $\varepsilon(\textbf{r},\omega;\textbf{r}',\omega')$ are connected via the fluctuation-dissipation theorem:
\begin{equation}
    [\textbf{P}(\text{r},\omega), \textbf{P}^\dagger(\text{r}',\omega')]  = 2\hbar\epsilon_0\boldsymbol{\varepsilon}''(\text{r},\omega;\text{r}',\omega').
    \label{eq:kubo_macro_freq(2)}
\end{equation}
Taking the curl in Eq.~\eqref{eq:faraday_law} and using Eq.~\eqref{eq:ampere_law}, we obtain the wave-equation for the electric field:
\begin{equation}
    \begin{split}
        &\curl\curl\textbf{E}(\textbf{r},t) + \\&\frac{1}{c^2}\partial_t^2\int d^3\text{r}'\int dt'\boldsymbol{\varepsilon}(\textbf{r},t;\textbf{r}',t')\cdot\textbf{E}(\textbf{r}',t') = -\mu_0\partial^2_t\textbf{P}(\text{r},t),
    \end{split}
    \label{eq:e_field_time}
\end{equation}
where $c = 1/\sqrt{\mu_0\epsilon_0}$ is the speed of light in free-space. Writing Eq.~\eqref{eq:e_field_time} in the frequency-domain, we have:
\begin{equation}
    \begin{split}
        &\curl\curl\textbf{E}(\textbf{r},\omega) - \\&\frac{\omega^2}{c^2}\int d^3\text{r}'\int\frac{d\omega'}{2\pi}\boldsymbol{\varepsilon}(\textbf{r},\omega;\textbf{r}',\omega')\cdot\textbf{E}(\textbf{r}',\omega')  = \mu_0\omega^2\textbf{P}(\text{r},\omega).
    \end{split}
    \label{eq:e_field_freq}
\end{equation}
In order to solve the wave-equation, we make use of the Green's dyadic of the system, $\textbf{G}(\textbf{r},\omega;\textbf{r}',\omega')$, which solves Eq.~\eqref{eq:e_field_freq} with $\textbf{I}\delta(\textbf{r} - \textbf{r}')2\pi\delta(\omega - \omega')$ as a source-term in the right-hand side:

\begin{equation}
    \begin{split}
    &\curl \curl\mathbf{G}(\mathbf{r},\omega;\mathbf{r}',\omega') - \\& 
     \frac{\omega^2}{c^2}
    \int d^3\mathbf{r}'' \int \frac{d\omega''}{2\pi}
    \boldsymbol{\varepsilon}(\mathbf{r},\omega;\mathbf{r}'',\omega'')
    \cdot \mathbf{G}(\mathbf{r}'',\omega'';\mathbf{r}',\omega')
     \\& = \mathbf{I}\,\delta(\mathbf{r} - \mathbf{r}')\,2\pi\delta(\omega - \omega').
    \end{split}
\label{eq:g_func_freq}
\end{equation}
An important feature of Eqs.~\eqref{eq:e_field_freq} and ~\eqref{eq:g_func_freq} is that they are non-local in the frequency variable, signaling the lack of frequency conservation in time-modulated media. On the other hand, the Green's dyadic satisfies the following integral formula~\cite{supmat}:
\begin{widetext}
\begin{equation}
\begin{aligned}
\iint d^3\text{s }d^3\text{s}' \iint& \frac{d\omega''}{2\pi}\frac{d\omega'''}{2\pi}
\,\mathbf{G}(\mathbf{r},\omega;\mathbf{s},\omega'')
(\omega'')^2
\boldsymbol{\varepsilon}''(\mathbf{s},\omega'';\mathbf{s}',\omega''')
(\omega''')^2
\mathbf{G}^\dagger(\mathbf{r}',\omega';\mathbf{s}',\omega''')
\\&
= \frac{c^2}{i2}\left(
\mathbf{G}(\mathbf{r},\omega;\mathbf{r}',\omega')(\omega')^2
-\omega^2\mathbf{G}^\dagger(\mathbf{r}',\omega';\mathbf{r},\omega)
\right)
\end{aligned}
\label{eq:g_func_integral}
\end{equation}
\end{widetext}
which generalizes the integral formula for the imaginary part of the Green's dyadic in non-modulated media~\cite{scheel2008macroscopic}.
\par
Making use of the Green's dyadic, the Fourier-domain electric field can be written as:
\begin{equation}
    \begin{split}
        &\textbf{E}(\textbf{r},\omega) = \mu_0\int d^3\text{r}'\int\frac{d\omega'}{2\pi}\textbf{G}(\textbf{r},\omega;\textbf{r}',\omega')(\omega')^2\textbf{P}(\textbf{r}',\omega').
        \label{eq:e_field_freq(2)}
    \end{split}
\end{equation}

An important feature of the Green's dyadic representation of the electric field shown in Eq.~\eqref{eq:e_field_freq(2)} is that it describes the total field. Therefore, it contains both the electric field produced by the noise currents associated to the losses of the system [via Eq.~\eqref{eq:kubo_macro_freq(2)}] and the free-space vacuum fluctuations~\cite{huttner1992quantization,hanson2021langevin} (which interact with the time-varying structure and subsequently become scattered by it). 
\par
Next, making use of Eqs.~\eqref{eq:kubo_macro_freq(2)} and~\eqref{eq:e_field_freq(2)} as well as the integral formula presented in Eq.~\eqref{eq:g_func_integral}, we find the commutation relations for the Fourier-domain electric field operator, which we write in component form for clarity:
\begin{equation}
    \begin{split}
         [E_i(\textbf{r},\omega), E_j^\dagger(\textbf{r}',\omega')] & = \frac{\mu_0\hbar}{i}\Big(G_{ij}(\textbf{r},\omega;\textbf{r}',\omega')(\omega')^2 \\&-\omega^2G_{ji}^*(\textbf{r}',\omega';\textbf{r},\omega)\Big)
    \end{split}
\label{eq:e_field_comm}
\end{equation}
In the above, $i\text{, }j=x\text{, }y\text{, }z$ label the Cartesian components of the electric field operator and the Green's dyadic. Crucially, Eq.~\eqref{eq:e_field_comm} shows that, owing to the temporal modulation, the different Fourier components of the electric field operator become correlated via a non-vanishing commutator. On the other hand, it can be readily seen that in the non-modulated case, for which $G_{ij}(\textbf{r},\omega;\textbf{r}',\omega') = G_{ij}(\textbf{r},\textbf{r}';\omega')2\pi\delta(\omega - \omega')$, Eq.~\eqref{eq:e_field_comm} reduces to the usual commutation relations in non-modulated linear media~\cite{scheel2008macroscopic,buhmann2013dispersion1}. Additionally, it is straightforward to see that the right-hand side of Eq.~\eqref{eq:e_field_comm} is the anti-Hermitian part of $2\mu_0\hbar G_{ij}(\textbf{r},\omega;\textbf{r}',\omega')(\omega')^2$. Therefore, it is possible to understand Eq.~\eqref{eq:e_field_comm} as a fluctuation-dissipation theorem for the electric field.
\par
So far it has been shown that the temporal modulation breaks time-translational symmetry, allowing the coupling between different frequency components of the polarization and electric fields and modifying in the process their commutation relations. However, there are properties of the electromagnetic field that must be independent of the details of the material body (whether anisotropic, spatially non-local and/or time-modulated). One of these is the commutator between the electric field $\textbf{E}(\textbf{r},t)$ and the magnetic induction $\textbf{B}(\textbf{r},t)$ at equal times, which must coincide with its free-space value. Importantly, making use of Faraday's law and Eq.~\eqref{eq:e_field_comm}, it can be proved that~\cite{supmat}:
\begin{equation}
    [\textbf{E}(\textbf{r},t),\textbf{B}(\textbf{r}',t)] = -\frac{i\hbar}{\epsilon_0}\curl\delta(\textbf{r} - \textbf{r}')\textbf{I}.
    \label{eq:e_b_field_comm}
\end{equation}
The above is the same commutator obtained from free-space quantum electrodynamics~\cite{maggiore2005modern,buhmann2013dispersion1,scheel2008macroscopic}, and serves as a consistency check of our theory.
\par
The results shown so far are extremely general, since only linearity of the medium has been assumed. However, for practical purposes it is convenient to consider the case of a periodic temporal modulation, also known as Floquet modulation. First, we write the Fourier frequencies as $\omega_n = \tilde{\omega} + n\Omega$, where $-\Omega/2<\tilde{\omega}\leq\Omega/2$ belongs to the first Floquet Brillouin zone, $\Omega$ is the modulation frequency and $n$ labels the $n-$th Floquet band. Then, periodicity of the response function implies that $\varepsilon(t,t') = \varepsilon(t + 2\pi/\Omega,t'+2\pi/\Omega)$, which upon Fourier transforming leads to $\varepsilon(\tilde{\omega} + n\Omega,\tilde{\omega}' + n'\Omega) = \varepsilon_{n,n'}(\tilde{\omega}')2\pi\delta(\tilde{\omega} - \tilde{\omega}')$. Thus, for a periodic temporal modulation, the coupling between different frequencies is limited to different Floquet sidebands of the same frequency. For the Floquet case, Eq.~\eqref{eq:e_field_comm} reduces to:
\begin{equation}
    \begin{split}
         & [E_i(\textbf{r},\tilde{\omega} + n\Omega), E_j^\dagger(\textbf{r}',\tilde{\omega}' + n'\Omega)] = \\& \frac{\mu_0\hbar}{i}\Big(G_{ij;n,n'}(\textbf{r},\textbf{r}',\tilde{\omega}')\omega_{n'}^2 - \omega_{n}^2G_{ji;n',n}^*(\textbf{r}',\textbf{r},\tilde{\omega}')\Big)\\&2\pi\delta(\tilde{\omega} - \tilde{\omega}'),
    \end{split}
\label{eq:e_field_comm(floquet)}
\end{equation}
where $G_{ij;n,n'}(\textbf{r},\textbf{r}',\tilde{\omega})$ are the components of a Cartesian-Floquet tensor. 
\par
Additionally, it is possible to write the electric field in terms of $\textbf{f}(\textbf{r},\omega)$ and $\textbf{f}^\dagger(\textbf{r},\omega)$. To do so, let us first introduce the \textit{dressed} Green's dyadic $\boldsymbol{\mathcal{G}}(\textbf{r},\omega;\textbf{r}',\omega')$, which is defined as:
\begin{equation}
    \begin{split}
        \boldsymbol{\mathcal{G}}(\textbf{r},\omega;\textbf{r}',\omega') = &\mu_0\int d^3\text{r}''\int\frac{d\omega''}{2\pi}\textbf{G}(\textbf{r},\omega;\textbf{r}'',\omega'')(\omega'')^2\cdot\\&\textbf{K}(\textbf{r}'',\omega'';\textbf{r}',\omega'),
    \end{split}
    \label{eq:green_func_dressed}
\end{equation}
which depends on the loss-kernel of the system and where the frequency integral runs from $-\infty$ to $\infty$. Importantly, when the modulation is periodic, the dressed Green's dyadic satisfies $\boldsymbol{\mathcal{G}}(\textbf{r},\tilde{\omega} + n\Omega;\textbf{r}',\tilde{\omega}' + n'\Omega) = \boldsymbol{\mathcal{G}}_{n,n'}(\textbf{r},\textbf{r}';\tilde{\omega}')\delta(\tilde{\omega} - \tilde{\omega}')$, similar to the case of the regular Green's dyadic $\textbf{G}(\textbf{r},\omega;\textbf{r}',\omega')$\footnote{Except for a factor of $2\pi$.}. Hence, the Fourier-domain electric field operator can be written as:
\begin{equation}
    \begin{split}
        \textbf{E}(\textbf{r},\omega) = &\int d^3\text{r}'\int_0^\infty d\omega'\Big(\boldsymbol{\mathcal{G}}(\textbf{r},\omega;\textbf{r}',\omega')\cdot\textbf{f}(\textbf{r}',\omega') \\&+\boldsymbol{\mathcal{G}}(\textbf{r},\omega;\textbf{r}',-\omega')\cdot\textbf{f}^\dagger(\textbf{r}',\omega')\Big),
    \end{split}
    \label{eq:e_field_pols}
\end{equation}
where, again, the coupling to negative frequencies results in a squeezing-like transformation~\cite{Loudon2000,barnett2002methods} that involves both the polariton annihilation operators and the creation ones.
\par
Having presented our formalism, in the next sections we turn our attention to the problem of defining a local density of states in time-varying media and the study of the electric field correlation functions, illustrating our theory with results for two cases of time-varying media.
\section{Local Density of States}
\label{sec:ldos}
A fundamental quantity that describes light-matter interactions is the local density of photonic states (LDOS). In non-modulated media, the LDOS determines the decay rate of a probe quantum emitter (QE)~\cite{purcell1946spontaneous,novotny2012principles,scheel2008macroscopic,buhmann2013dispersion1}, as well as the total electromagnetic power radiated by an oscillating classical dipole~\cite{jackson1999classical,novotny2012principles}. In this section, we derive the LDOS for time-modulated media. To do so, let us consider a probe quantum emitter interacting with a frequency dispersive and dissipative time-varying medium, as shown in Fig.~\ref{fig:fig0}. We model the quantum emitter as a two-level system with transition frequency $\omega_\text{d}$ and transition dipole moment $\textbf{d}$. First, let us note that because we are dealing with a macroscopic material that has a dispersive and dissipative response, the emitter will interact with polaritons and not photons~\cite{scheel2008macroscopic,buhmann2013dispersion1}. On the other hand, since the temporal modulation encompasses gain, the emitter will be able to both absorb and emit polaritons. The interaction Hamiltonian in the interaction picture (where operators are free-evolving) is:
\begin{equation}
    \begin{split}
        H^{(I)}_\text{int}(t) = -(\textbf{d}\sigma^-e^{-i\omega_\text{d}t} + \textbf{d}^*\sigma^+e^{i\omega_\text{d}t})\cdot\textbf{E}(\textbf{r}_\text{d},t),
    \end{split}
    \label{eq:int_hamiltonian}
\end{equation}
where $\sigma^\pm$ are the raising and lowering operators of the quantum emitter and $\textbf{r}_\text{d}$ is its position in space, and where $\textbf{E}(\textbf{r}_\text{d},t) = \int_0^\infty (d\omega/2\pi)e^{-i\omega t}\textbf{E}(\textbf{r}_\text{d},\omega)+\text{H.c.}$ is the electric field operator in the interaction picture.
\par
Next, we calculate the decay and excitation rates of the two-level system and derive a Fermi Golden Rule (FGR) for time-varying media. To do so, we apply time-dependent perturbation theory to first order in the interaction Hamiltonian Eq.~\eqref{eq:int_hamiltonian}, using the explicit form of the electric field in terms of the polariton operators, as shown in Eq.~\eqref{eq:e_field_pols}. Then, we calculate the transition probability for processes of the type $\ket{e,\text{vac}}\longrightarrow\ket{g,1_i(\textbf{r},\omega)}$, corresponding to spontaneous decay, and $\ket{g,\text{vac}}\longrightarrow\ket{e,1_i(\textbf{r},\omega)}$, corresponding to spontaneous excitation. Summing over all processes of each type and assuming the temporal modulation to be periodic, the decay and excitation rates $\Gamma_\text{SD}$ and $\Gamma_\text{SE}$ are given by~\cite{supmat}:
\begin{equation}
    \begin{split}
       \Gamma_\text{SD/SE} =& \frac{1}{\hbar^2}\textbf{d}^*\cdot\int d^3r\sum_{l}\Theta\left(\pm(\tilde{\omega}_\text{d} + l\Omega)\right)\\&\boldsymbol{\mathcal{G}}_{n_\text{d},l}(\textbf{r}_\text{d},\textbf{r};\tilde{\omega}_\text{d})\boldsymbol{\mathcal{G}}^\dagger_{n_\text{d},l}(\textbf{r}_\text{d},\textbf{r};\tilde{\omega}_\text{d})\cdot\textbf{d},
    \end{split}
    \label{eq:gamma_sd_se}
\end{equation}
where $\omega_\text{d} = \tilde{\omega}_\text{d} + n_\text{d}\Omega$. Critically, the step function $\Theta\left(\pm(\tilde{\omega}_\text{d} + l\Omega)\right)$ constrains the summation to positive frequency bands for the case of the spontaneous decay, and to negative frequency bands for the case of the spontaneous excitation. On the other hand, it can be seen from their definitions that both rates are real and positive-valued.
\par
If we now take the difference between the decay and excitation rates and make use of Eqs.~\eqref{eq:g_func_integral} and~\eqref{eq:green_func_dressed}, we find that the total rate $\Gamma_\text{tot} \equiv \Gamma_\text{SD} - \Gamma_\text{SE}$ is given by (see SM~\cite{supmat}):
\begin{equation}
    \begin{split}
        \Gamma_\text{tot} = \frac{2\mu_0\omega^2_\text{d}}{\hbar}\textbf{d}^*\cdot\textbf{G}''_{n_\text{d},n_\text{d}}(\textbf{r}_\text{d},\textbf{r}_\text{d};\tilde{\omega}_\text{d})\cdot\textbf{d}.
    \end{split}
    \label{eq:gamma_total}
\end{equation}

Therefore, the total rate is found to be proportional to the anti-Hermitian part of the Green's dyadic, similar to the case of non-modulated linear media~\cite{scheel2008macroscopic,novotny2012principles,franke2021fermi}. Furthermore, the FGR enables us to define the projected local density of states (LDOS) in time-varying media~\cite{park2025spontaneous}:
\begin{equation}
    \rho_\text{\textbf{n}}(\textbf{r},\omega) = \frac{2\omega}{c^2\pi}\textbf{n}^*\cdot\textbf{G}_{l,l}''(\textbf{r},\textbf{r};\tilde{\omega})\cdot\textbf{n},
    \label{eq:ldos}
\end{equation}
where $\omega = \tilde{\omega} + l\Omega$ and $\textbf{n}$ is a vector of unit modulus (i.e., $\textbf{n}^*\cdot\textbf{n}=1$). Moreover, Eq.~\eqref{eq:ldos} coincides with the formula for the LDOS in non-modulated linear media~\cite{novotny2012principles,scheel2008macroscopic,buhmann2013dispersion1}.
\par
Importantly, in time-varying media it is possible for the projected LDOS to take negative values, since the temporal modulation lifts the constraints of passivity and energy conservation. A negative projected LDOS yields a negative total rate $\Gamma_\text{tot}<0$ [compare Eqs.~\eqref{eq:gamma_total} and~\eqref{eq:ldos}], which occurs when $\Gamma_\text{SE}>\Gamma_\text{SD}$. Since $\Gamma_\text{SE} = 0$ in the non-modulated scenario, for which the coupling to negative frequency modes is forbidden, the negativity of the total rate signals the presence of gain in the system.
\par
Although the results presented in this section have been derived from a purely quantum mechanical calculation, they can also be obtained from a classical radiated power calculation. To do so, let us now consider a classical harmonic point-dipole with polarization density $\textbf{P}(\textbf{r},t) = \text{Re}\big\{\textbf{d}e^{-i\omega_\text{d}t}\delta(\textbf{r} - \textbf{r}_\text{d})\big\}$ and associated current density $\textbf{J}(\textbf{r},t) = \partial_t\textbf{P}(\textbf{r},t)$. Assuming a Floquet modulation, the field produced by such a dipole when embedded in the time-modulated electromagnetic environment is $\textbf{E}(\textbf{r},t) = \mu_0\omega_\text{d}^2\text{Re}\big\{\sum_le^{-i(\tilde{\omega}_\text{d}+l\Omega)t}\textbf{G}_{l,n_\text{d}}(\textbf{r},\textbf{r}_\text{d};\tilde{\omega}_\text{d})\cdot\textbf{d}\big\}$. On the other hand, the power radiated by the dipole is given by $P(t) = \int d^3\text{r}\textbf{J}(\textbf{r},t)\cdot\textbf{E}(\textbf{r},t)$~\cite{jackson1999classical}, where $\textbf{E}(\textbf{r},t)$ is the dipole's own field. Then, if we define $t_0$ as an arbitrary initial time at which the dipole starts radiating, the total emitted power can be calculated as:
\begin{equation}
    \begin{split}
        P_\text{tot} &= \underset{t\longrightarrow\infty}{\lim}\frac{1}{t-t_0}\int_{t_0}^t dt'P(t') \\&= \mu_0\omega^3_\text{d}\textbf{d}^*\cdot\textbf{G}''_{n_\text{d},n_\text{d}}(\textbf{r}_\text{d},\textbf{r}_\text{d};\tilde{\omega}_\text{d})\cdot\textbf{d}.                               
        \label{eq:power}
    \end{split}
\end{equation}
A glance at Eqs.~\eqref{eq:gamma_total} and~\eqref{eq:power} shows that the quantum mechanical total rate and the classical emitted power are connected through $P_\text{tot} = (\hbar\omega_\text{d}/2)\Gamma_\text{tot}$, which shows the equivalence between the quantum FGR and the classical emitted power calculation. Importantly, to the best of our knowledge, in this work we have provided the first explicit proof of this quantum to classical correspondence in time-varying media. Moreover, even though we have proved this equivalence for the case of a periodic temporal modulation, it holds true for any profile of the temporal modulation~\cite{supmat}. However, it must be noted that the total rate and total emitted power and subsequently, the LDOS, are not well-defined for a general time modulation (see the SM~\cite{supmat} for more information).
\par
Next, we exemplify our theory, applying it to a frequency-dispersive and lossy half-space periodically modulated in time (see Fig.~\ref{fig:fig0}). We shall discuss both the slow modulation regime, where the interaction is limited to a few Floquet sidebands, and the fast modulation regime, where the coupling to negative frequency replicas is dominant and for which quantum mechanical effects become prominent. Furthermore, in order to better illustrate the differences between our formalism and those presented in previous works, we shall consider both dispersive and non-dispersive temporal modulations. Finally, we will study two different types of materials: a polar insulator for the slow modulation regime and a plasmonic medium for the fast modulation one. These choices are motivated by state-of-the-art experimental platforms to realize time-varying media in polar~\cite{basini2026terahertz,nazaryan2026terahertz} and plasmonic materials~\cite{vezzoli2018optical,zhou2020broadband,bohn2021all,bohn2021spatiotemporal,tirole2022saturable,tirole2023double,guo2025plasmonic}, as well as to facilitate comparison with previous theoretical approaches~\cite{yu2023manipulating,vazquez2023incandescent,yu2025near,ren2026clarification,zhu2026enhancing}. Finally, we note that while we particularize our results for a time-varying half-space, we find qualitatively similar results for finite slabs.
\subsection{Time-modulated polar insulator: the Floquet sidebands regime}
\label{sec:slow}
\begin{figure*}
 \captionsetup{
        justification=justified,
        singlelinecheck=false
    }
    \centering
    \hspace{0 mm}
    \includegraphics[scale = 0.50]{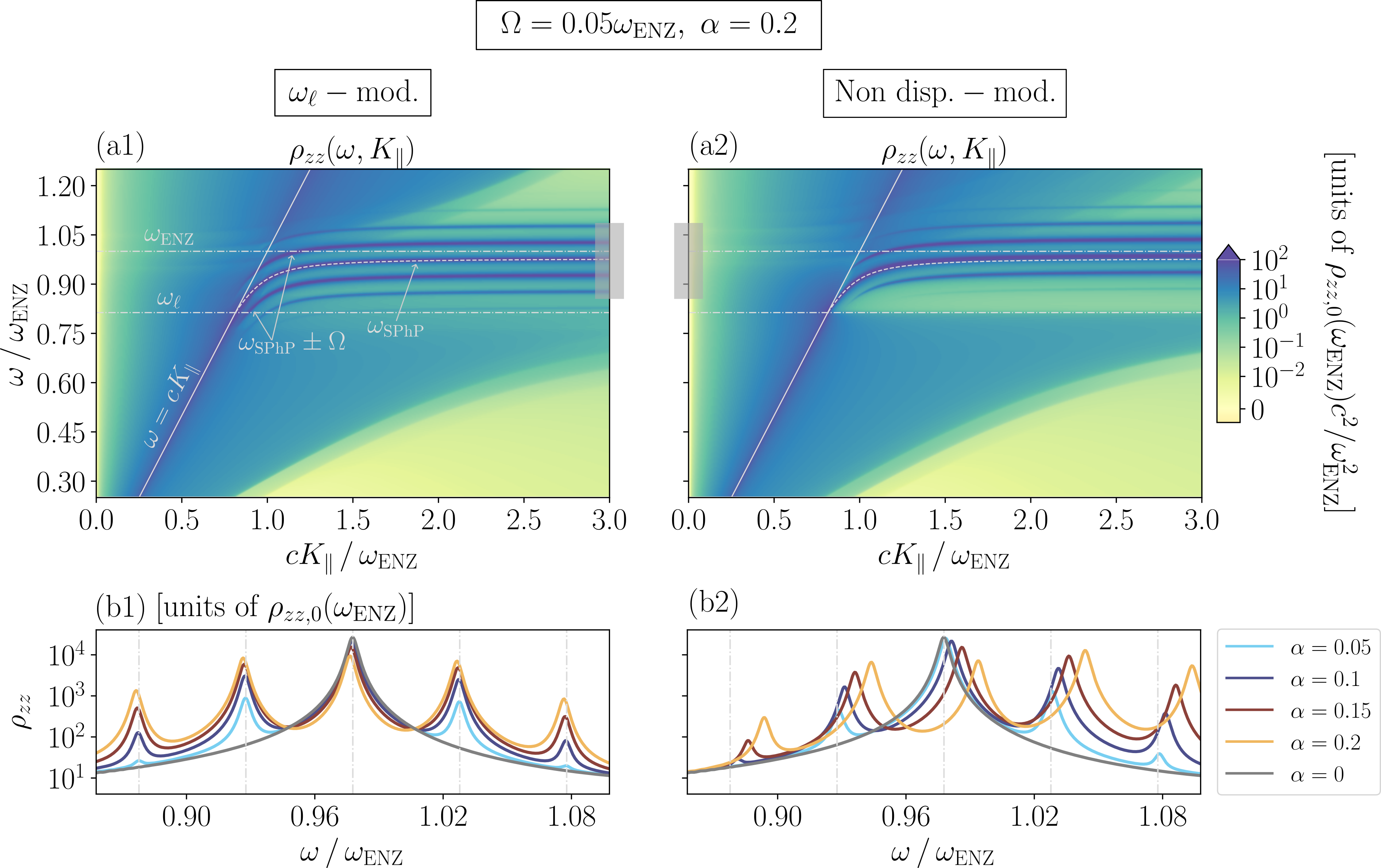} 
    
    \caption{\justifying LDOS for slowly driven polar insulator half-space. (a1)-(a2) Momentum-resolved $z$-projected LDOS for a time-modulated SiC half-space; units of $\rho_{zz,0}(\omega_\text{ENZ})c^2/\omega_\text{ENZ}^2$, where $\rho_{zz,0}(\omega) = \omega^2 / (3\pi^2c^3)$ is the free-space $z$-projected LDOS at a frequency $\omega$. Vertical axis is the frequency (normalized to $\omega_\text{ENZ}$) and horizontal axis is the lateral momentum (normalized to $\omega_\text{ENZ}/c$). In (a1) it is the Lorentz frequency $\omega_ \ell$ which is modulated in time, leading to a dispersive temporal modulation, whereas in (a2) the temporal modulation is non-dispersive. Furthermore, in (a1) the modulation strength is set to $\alpha=0.2$, while in (a2) it is $\alpha_\text{nd}=17.3\alpha=3.46$. (b1) Total $z$-projected LDOS [units of $\rho_{zz,0}(\omega_\text{ENZ})$] vs frequency for the case of a $\omega_\ell$-modulation and for values of $\alpha=0,\text{ }0.05\text{, }0.1\text{, }0.15\text{, }0.2$. In (b2) we consider the non-dispersive case, with varying modulation strength $\alpha_\text{nd} = 17.3\alpha$. In all panels the modulation frequency is set to $\Omega=0.05\omega_\text{ENZ}$.}
    \label{fig:fig1} 
\end{figure*}
Let us consider a polar insulator half-space ($z\leq0$), where $z$ is perpendicular to the surface and $z>0$ is air. The polar insulator is modulated in time through its Lorentz frequency, which we assume to vary periodically in time as:
\begin{equation}
    \omega^2_\ell(t) = \omega^2_\ell\big[1 + \alpha\cos(\Omega t)\big].
    \label{eq:om0_tmod}
\end{equation}
In Eq.~\eqref{eq:om0_tmod}, $\omega_\ell$ is the unmodulated Lorentz frequency of the polar material, $\Omega$ is the modulation frequency and $\alpha$, the modulation strength. Then, the phonons responsible for the optical response of the medium obey the following parametric damped harmonic oscillator equation:
\begin{equation}
    \partial_t^2\textbf{P}^\text{(ind)}(t) + \gamma\partial_t\textbf{P}^\text{(ind)}(t)+\omega_\ell^2(t)\textbf{P}^\text{(ind)}(t) = \varepsilon_0\omega_\text{p}^2\textbf{E}(t).
    \label{eq:lorentz_tmod}
\end{equation}

In the above, $\textbf{P}^\text{(ind)}(t)$ is the polarization density induced by the electric field $\textbf{E}(t)$, which must not be confused with the noise polarization arising from the fluctuation-dissipation theorem~\eqref{eq:kubo_macro_freq(2)}. Indeed, if we denote the latter by $\textbf{P}^\text{(noise)}$ for better clarity, the total polarization density reads $\textbf{P}^\text{(tot)}=\textbf{P}^\text{(ind)} + \textbf{P}^\text{(noise)}$. On the other hand, $\gamma$ is the damping rate of the phonons and $\omega_\text{p}$ is the so-called oscillator strength.
\par
The temporal periodicity of the Lorentz frequency enables us to make Floquet ansatzs for the induced polarization density and the electric field: $P^\text{(ind)}_j(t)=\fontsize{7}{8}{\sumint}(d\omega_n/2\pi)e^{-i\omega_nt}P^\text{(ind)}_{j,n}(\tilde{\omega})$ and $E_j(t)=\fontsize{7}{8}{\sumint}(d\omega_n/2\pi)e^{-i\omega_nt}E_{j,n}(\tilde{\omega})$, where $j=x\text{, }y\text{, }z$ and $n\in\mathbb{Z}$. Furthermore, we have used the shorthand notation $\footnotesize{\sumint} d\omega_n \equiv \int^{\Omega/2}_{-\Omega/2}d\tilde{\omega}\sum_n$. Making use of these Floquet ansatzs, we can solve Eq.~\eqref{eq:lorentz_tmod} and express the induced polarization in terms of the electric field as $P^\text{(ind)}_{j,n}(\tilde{\omega})=\varepsilon_0\sum_{n'}\chi_{n,n'}(\tilde{\omega})E_{j,n'}(\tilde{\omega})$, where $\chi_{n,n'}(\tilde{\omega})$ is the Floquet susceptibility matrix. For the harmonic oscillator equation with time-modulated Lorentz frequency, the Floquet susceptibility matrix reads:
\begin{equation}
    \chi_{n,n'}(\tilde{\omega}) =  \omega_\text{p}^2\big[D^{-1}(\tilde{\omega})\big]_{n,n'},
    \label{eq:chi_lor_1}
\end{equation}
where $D^{-1}(\tilde{\omega})$ stands for the inverse of the Floquet matrix $D(\tilde{\omega})$, which has elements:
\begin{equation}
    \begin{split}
        D_{n,n'}(\tilde{\omega}) &= \left(-\omega_n^2-i\gamma\omega_n+\omega^2_\ell\right)\delta_{n,n'}
        \\&+ \frac{\alpha}{2}\omega^2_\ell\left(\delta_{n+1,n'} + \delta_{n-1,n'}\right).
    \end{split}
    \label{eq:chi_lor_2}
\end{equation}

In Eq.~\eqref{eq:chi_lor_2}, the first line contains the non-modulated part of the linear response, whereas the second line contains the effect of the temporal modulation and describes the coupling to different Floquet bands. Having introduced the Floquet susceptibility matrix, the Floquet permittivity matrix reads:
\begin{equation}
    \begin{split}
        \varepsilon_{n,n'}(\tilde{\omega})= \varepsilon_\infty\delta_{n,n'} +  \omega_\text{p}^2\big[D^{-1}(\tilde{\omega})\big]_{n,n'}.
        \label{eq:eps_lor}
    \end{split}
\end{equation}
In Eq.~\eqref{eq:eps_lor} a background permittivity $\varepsilon_\infty>1$ has been included to comply with the Kramers-Kronig relations~\cite{jackson1999classical}. Importantly, the anti-Hermitian part of the Floquet permittivity, $\varepsilon''_{n,n'}(\tilde{\omega})=1/(i2)\big(\varepsilon_{n,n'}(\tilde{\omega})-\varepsilon_{n',n}^*(\tilde{\omega})\big)$, appears on the right-hand side of Eq.~\eqref{eq:kubo_macro_freq(2)} and thus, defines the commutator between the Fourier components of the noise polarization.
\par
On the other hand, as discussed in the introduction, a widespread approximation when describing the electrodynamics of dispersive time-varying media is to assume the temporal modulation is non-dispersive. This amounts to working with the following Floquet permittivity matrix:
\begin{equation}
    \begin{split}
        \varepsilon_{n,n'}(\tilde{\omega}) = \varepsilon_\text{bg}(\omega_n)\delta_{n,n'} + \frac{\alpha_\text{nd}}{2}\left(\delta_{n+1,n'} + \delta_{n-1,n'}\right).
    \end{split}
    \label{eq:eps_nondisp}
\end{equation}
In Eq.~\eqref{eq:eps_nondisp}, $\varepsilon_\text{bg}(\omega_n)=\varepsilon_\infty-\omega_\text{p}^2/\left(\omega_n^2 +i\gamma\omega_n-\omega_\ell^2\right)$ is the non-modulated background permittivity of the material, and $\alpha_\text{nd}$ is the strength of the non-dispersive temporal modulation. Importantly, a dispersionless temporal modulation leads to physical inconsistencies: on the one hand, the permittivity in Eq.~\eqref{eq:eps_nondisp} does not comply with Kramers-Kronig relations and furthermore, because the temporal modulation does not modify the anti-Hermitian part of the electric permittivity, the commutator in Eq.~\eqref{eq:kubo_macro_freq(2)} is the same as in the equilibrium case~\cite{yu2023manipulating,vazquez2023incandescent,yu2025near,ren2026clarification,zhu2026enhancing}. This approach is thus fundamentally different from the dispersive temporal modulation of Eq.~\eqref{eq:eps_lor}, which respects causality and does not assume the FDT to be the equilibrium one.  
\par
Importantly, it must be emphasized that the meaning of $\alpha$ (dispersive case) and $\alpha_\text{nd}$ (non-dispersive case) is not the same. To show this, let us write the total permittivity as $\varepsilon = \varepsilon_\text{bg}+\varepsilon_\text{mod}$, where $\varepsilon_\text{mod}$ is its time-modulated part. Then, the actual strength of the temporal modulation is determined by the ratio $\abs{\varepsilon_\text{mod} / \varepsilon_\text{bg}}$ at a given frequency $\omega$. Therefore, taking $ \alpha_\text{nd} = \alpha$ can lead to erroneous conclusions when comparing dispersive and non-dispersive temporal modulations. In particular, for the case of a time-varying Lorentz frequency with $\alpha\ll1$, we find that in order to have approximately equal ratios $\abs{\varepsilon_\text{mod} / \varepsilon_\text{bg}}$ for the dispersive ($\omega_\ell$ modulated) and non dispersive cases, one must have $\alpha_\text{nd}\simeq(\omega_\ell/\omega_\text{p})^2\abs{\chi_\text{bg}(\omega)}^2\alpha$~\cite{supmat}, where $\chi_\text{bg}(\omega)=\varepsilon_\text{bg}(\omega)-\varepsilon_\infty$ is the non-modulated electric susceptibility. Then, the rescaling introduced by the factor $(\omega_\ell/\omega_\text{p})^2\abs{\chi_\text{bg}(\omega)}^2$ guarantees that the relative strength of the dispersive and non-dispersive modulations is approximately the same for some reference frequency $\omega$.
\par
In Fig.~\ref{fig:fig1} we study the LDOS of a periodically time-modulated polar insulator half-space according to our new theory. For comparison, we also show the LDOS calculated in a non-dispersive modulation approximation. In order to compare with previous works in the literature~\cite{sloan2021casimir,vazquez2023incandescent,ren2026clarification,zhu2026enhancing} as well as with relevant experimental platforms, we assume the polar insulator to be silicon carbide (SiC). Typical values for the optical response parameters of SiC are $\varepsilon_\infty=6.7,\text{ }\omega_\text{TO}=0.1\text{ eV},\text{ }\omega_\text{LO}=0.12\text{ eV and }\gamma=6\cdot10^{-4}\text{ eV}$~\cite{vazquez2023incandescent}, where we have introduced the transverse and longitudinal optical phonon frequencies, $\omega_\text{TO}$ and $\omega_\text{LO}$. These two frequencies are connected to the parameters of the damped harmonic oscillator equation via $\omega_\text{TO}=\omega_\ell$ and $\omega_\text{LO}=\sqrt{\omega_\ell^2 + \omega_\text{p}^2 / \varepsilon_\infty}$. 
\par
Fig.~\ref{fig:fig1}(a1) shows the momentum-resolved $z$-projected LDOS $\rho_{zz}(\omega,K_\parallel)$ for the case of a time-varying Lorentz frequency~\eqref{eq:chi_lor_2}-\eqref{eq:eps_lor}, where the total $z$-projected LDOS reads:
\begin{equation}
    \rho_{zz}(\omega) = \frac{1}{2\pi}\int_0^\infty dK_\parallel K_\parallel\rho_{zz}(\omega,K_\parallel).
    \label{eq:ldos_zz}
\end{equation}
The modulation parameters are $\Omega=0.05\omega_\text{ENZ}$ and $\alpha=0.2$, where $\omega_\text{ENZ} = \sqrt{\omega_\ell^2+\omega_\text{p}^2/\varepsilon_\infty}$ is the $\epsilon$-near-zero frequency of the material, which coincides with the LO frequency (i.e., $\omega_\text{ENZ}=\omega_\text{LO}$) and for which $\text{Re}[\varepsilon_\text{bg}(\omega_\text{ENZ})]\sim0$. Furthermore, we take the probe emitter's height above the surface to be $h\longrightarrow0^+$, infinitely close to the interface on the vacuum side. The vertical axis of the figure is the frequency, normalized to $\omega_\text{ENZ}$, whereas the horizontal axis is the lateral momentum $K_\parallel$, normalized to $\omega_\text{ENZ}/c$. In the figure we highlight the lightcone (solid grey line) as well as the Lorentz and $\epsilon$-near-zero frequencies of the polar material (horizontal dashed-dotted grey lines).
\par
The spectral region between $\omega_\ell$ and $\omega_\text{ENZ}$ is the Reststrahlen band of the polar insulator, and is characterized by $\text{Re}(\varepsilon_\text{bg})<0$. This negativity of the real part of the permittivity allows the existence of a resonant mode, the surface phonon polariton (SPhP), $\omega_\text{SPhP}$ (highlighted with a dashed grey line). For large $K_\parallel$, the SPhP converges to a flat (i.e., $K_\parallel$-independent) mode, the surface phonon (SPh), whose frequency is $\omega_\text{SPh}=\sqrt{\omega_\ell^2+\omega_\text{p}^2/(1 + \varepsilon_\infty)}$. Crucially, the surface phonon enhances the $z$-projected LDOS $\rho_{zz}(\omega)$ in the proximity of the interface, which in turn enhances the decay rate of a probe emitter in what is known as the quenching of the spontaneous emission~\cite{drexhage1970influence,ford1984electromagnetic,barnes1998fluorescence,anger2006enhancement}. Finally, for frequencies below (i.e., $\omega<\omega_\ell$) and above (i.e., $\omega>\omega_\text{ENZ}$) the Reststrahlen band  there are two continua of guided modes~\cite{jackson1999classical}, another type of surface waves present at the interface between air and material bodies in the frequency regimes for which $\text{Re}(\varepsilon_\text{bg})>0$.
\par
For the time-modulated case, a glance at the Reststrahlen band shows the existence of multiple Floquet sidebands of the SPhP, $\omega_\text{SPhP}+l\Omega$. Thus, the temporal modulation creates multiple replicas of the resonant mode of the structure~\cite{yu2023manipulating,vazquez2023incandescent,ren2026clarification,zhu2026enhancing}, enabling near-field enhancement of the LDOS at the surface phonon sidebands $\omega_\text{SPh}+l\Omega$~\cite{vazquez2023incandescent,ren2026clarification,yu2023manipulating}. This is best seen in Fig.\ref{fig:fig1}(b1), in which we plot $\rho_{zz}(\omega)$ for a frequency interval centered at $\omega_\text{SPh}$ [grey shaded areas in Fig.\ref{fig:fig1}(a1)-(a2)] and for different values of the modulation strength $\alpha=0\text{, }0.05\text{, }0.1\text{, }0.15\text{, }0.2$. Additionally, the probe emitter's height above the surface is set to $h=0.01\lambda_\text{ENZ}$, in the near-field of the structure. It is seen that, as $\alpha$ increases, the central peak at $\omega_\text{SPh}$ gets progressively depleted, which is accompanied by the emergence of sidepeaks at $\omega_\text{SPh}\pm\Omega$ and $\omega_\text{SPh}\pm2\Omega$. 
Furthermore, as $\alpha$ increases the sidepeaks become slightly redshifted. Additionally, it can be seen that the negative index sidebands (i.e., $l<0$) have a larger LDOS than the positive index ones (i.e., $l > 0$). This can be explained by the fact that for a dispersive temporal modulation $\varepsilon_\text{mod}\sim1 / \omega_n^2$, and thus, the coupling to positive index sidebands is comparatively weaker.
\par
In Figs.~\ref{fig:fig1}(a2)-(b2) we consider the case of a non-dispersive temporal modulation, as described by Eq.~\eqref{eq:eps_nondisp}. The modulation frequency remains $\Omega=0.05\omega_\text{ENZ}$, but in order to make a meaningful comparison between dispersive and non-dispersive temporal modulations, we take $\alpha_\text{nd}=(\omega_\ell/\omega_\text{p})^2\abs{\chi_\text{bg}(\omega)}^2\alpha$, as discussed earlier. Taking as reference $\omega_\text{SPh}$, $\alpha_\text{nd}=(\omega_\ell/\omega_\text{p})^2\abs{\chi_\text{bg}(\omega_\text{SPh})}^2\alpha\simeq17.3\alpha$, which for $\alpha=0.2$ gives $\alpha_\text{nd}\simeq3.46$. A glance at the Reststrahlen band of Fig.~\ref{fig:fig1}(a2) shows the presence of multiple Floquet sidebands of the SPhP, as in Fig.~\ref{fig:fig1}(a1). However, these replicas are blueshifted with respect to the dispersive case. This is better displayed in Fig.~\ref{fig:fig1}(b2), which shows how, as the central peak diminishes and the sidebands emerge for increasing $\alpha_\text{nd}$, the peaks slide towards higher frequencies. Additionally, it is seen that the positive index sidebands have a larger LDOS than the negative index ones, opposite to the case of a dispersive temporal modulation. Critically, while for a dispersive temporal modulation one has $\varepsilon_\text{mod}\sim1 / \omega_n^2$, for the non-dispersive case $\varepsilon_\text{mod}$ is frequency-independent. This explains why the non-dispersive modulation overestimates the coupling to Floquet sidebands of higher frequency.
\par
Therefore, our results show that neglecting the dispersive nature of the temporal modulation is not only physically inconsistent, but also results in erroneous predictions about the emergence and evolution of the Floquet sidebands. Moreover, our dispersive theory does not need the use of any artificial cutoffs~\cite{vertiz2025dispersion}.
\subsection{Time modulated plasmonic media: negative frequency replicas}
\label{sec:fast}
\begin{figure*}
    \centering
    \hspace{0 mm}
    \includegraphics[scale = 0.50]{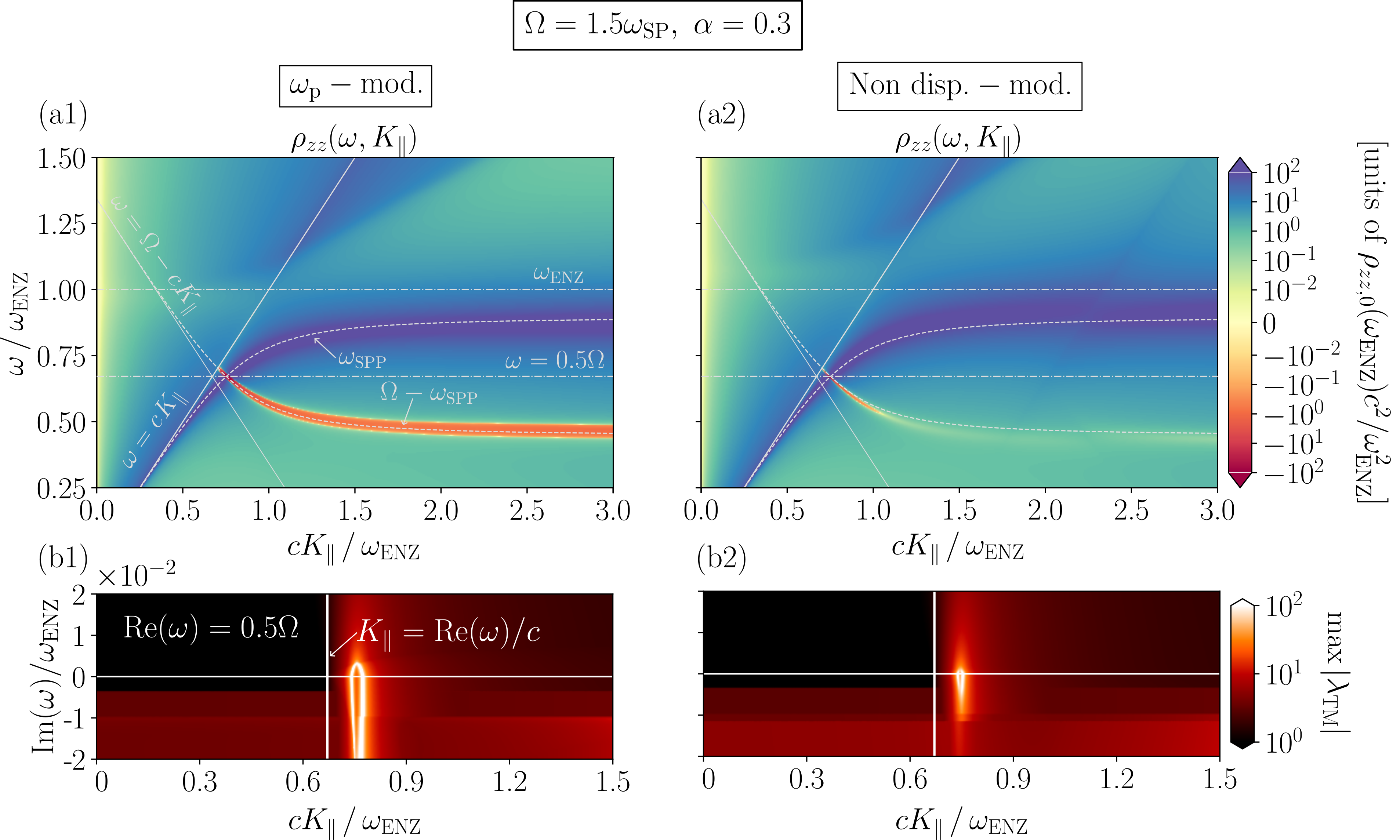} 
    \caption{\justifying LDOS for fast driven plasmonic half-space. (a1)-(a2) Momentum-resolved $z$-projected LDOS for a time-modulated ITO half-space vs frequency and lateral momentum. In (a1) it is the plasma frequency $\omega_\text{p}$ which is modulated in time, whereas in (a2) the modulation is non-dispersive, similar to Fig.~\ref{fig:fig1}(a2). On the other hand, in (a1) the modulation strength is $\alpha=0.3$, while in (a2) it is $\alpha_\text{nd}=1.5$. Blue regions in the plot have positive LDOS and are associated to loss. Conversely, red regions correspond to a negative LDOS and are thus a signature of gain. (b1)-(b2) Maximum eigenvalue in modulus of the scattered Green's dyadic for TM-polarized waves vs imaginary part of the frequency and lateral momentum. The real part of the frequency is set to $\text{Re}(\omega)=0.5\Omega$. In all panels the modulation frequency is $\Omega=1.5\omega_\text{SP}\simeq1.34\omega_\text{ENZ}$.}
    \label{fig:fig2} 
\end{figure*}

We now consider a plasma half-space ($z\leq0$), with air occupying the $z>0$ region. The plasmonic medium is periodically modulated in time through its plasma frequency as:
\begin{equation}
    \omega_\text{p}^2(t) = \omega_\text{p}^2\big[1 + \alpha\cos(\Omega t)\big],
    \label{eq:ompl_mod}
\end{equation}
where $\omega_\text{p}$ is the unmodulated plasma frequency. Then, the conduction band electrons obey the following parametric-Drude model equation with a time-modulated plasma frequency:
\begin{equation}
    \begin{split}
        \partial_t^2\textbf{P}^\text{(ind)}(t) + \gamma\partial_t\textbf{P}^\text{(ind)}(t) = \varepsilon_0\omega_\text{p}^2(t)\textbf{E}(t).
    \end{split}
\end{equation}
Proceeding as we did in the previous section, we make Floquet ansatzs for the induced polarization and the electric field and solve the parametric Drude-model equation. For the case of a time-modulated plasma, the Floquet susceptibility matrix reads:
\begin{equation}
    \begin{split}
        \chi_{n,n'}(\tilde{\omega})& = -\frac{\omega_\text{p}^2}{\omega_n^2+i\gamma\omega_n}\delta_{n,n'}\\&
        -\frac{\alpha}{2}\frac{\omega_\text{p}^2}{\omega_n^2+i\gamma\omega_n}\left(\delta_{n+1,n'} + \delta_{n-1,n'}\right).
    \end{split}
    \label{eq:chi_plasma}
\end{equation}

The first line in Eq.~\eqref{eq:chi_plasma} contains the non-modulated part of the electric susceptibility, while the second line contains its time-modulated part, which is seen to depend on the frequency. Then, the Floquet permittivity matrix of the medium is simply $\varepsilon_{n,n'}(\tilde{\omega})=\varepsilon_\infty\delta_{n,n'}+\chi_{n,n'}(\tilde{\omega})$.
\par
We also consider a non-dispersive temporal modulation in the form of Eq.~\eqref{eq:eps_nondisp}, where the modulation strength is $\alpha_\text{nd}$ and the non-modulated background permittivity is now $\varepsilon_\text{bg}(\omega_n) = \varepsilon_\infty - \omega_\text{p}^2 / \left(\omega_n^2 + i\gamma\omega_n\right)$. As it occurred with the time-varying polar material, in order to compare dispersive and non-dispersive temporal modulations in a plasmonic medium, the ratio $\abs{\varepsilon_\text{mod}/\varepsilon_\text{bg}}$ must be the same for dispersive and non-dispersive $\varepsilon_\text{mod}$. Then, one must have $\alpha_\text{nd}\simeq\abs{\chi_\text{bg}(\omega)}\alpha$~\cite{supmat}, which guarantees that $\abs{\varepsilon_\text{mod} / \varepsilon_\text{bg}}$ is approximately the same for dispersive and non-dispersive $\varepsilon_\text{mod}$ for some reference frequency $\omega$. Note that the scaling factor is different for modulations of the Lorentz and the plasma frequencies.
\par
As it is well known, most features of the spectrum of a polar material find their analogues in a plasmonic one: the $\epsilon$-near-zero frequency becomes $\omega_\text{ENZ} = \omega_\text{p}/\sqrt{\varepsilon_\infty}$, where once again $\text{Re}[\varepsilon_\text{bg}(\omega_\text{ENZ})]\sim0$ and $\text{Re}(\varepsilon_\text{bg})<0$ for $\omega<\omega_\text{ENZ}$. Additionally, the plasmonic half-space supports a resonant mode below its $\epsilon$-near-zero frequency, the surface plasmon polariton (SPP), $\omega_\text{SPP}$, which for large $K_\parallel$ approaches the surface plasmon (SP), with frequency $\omega_\text{SP}=\omega_\text{p}/\sqrt{1 + \varepsilon_\infty}$. Coupling to the SP results in the near-field enhancement of the LDOS and the subsequent quenching of the spontaneous emission~\cite{drexhage1970influence,anger2006enhancement}. Finally, for $\omega>\omega_\text{ENZ}$, $\text{Re}(\varepsilon_\text{bg})>0$ and the plasmonic half-space supports a continuum of guided modes.
\par
Next, we particularize the time-modulated plasmonic medium to the case of indium-tin-oxide (ITO), a plasmonic ENZ material used to implement ultrafast and ultrastrong temporal modulations of the electric permittivity~\cite{vezzoli2018optical,zhou2020broadband,tirole2022saturable,tirole2023double,galiffi2024optical,harwood2026intraband}, whose optical response parameters are $\varepsilon_\infty=4$, $\omega_\text{p}=2\text{ eV}$ and $\gamma=0.08\text{ eV}$~\cite{galiffi2024optical}.
\par
In Fig.~\ref{fig:fig2}(a1) we plot $\rho_{zz}(\omega,K_\parallel)$ for an ITO half-space with time-modulated plasma frequency, where the modulation parameters are set to $\Omega=1.5\omega_\text{SP}\simeq1.34\omega_\text{ENZ}$ (much larger than in Sec.~\ref{sec:slow}, for which $\Omega=0.05\omega_\text{ENZ}$) and $\alpha=0.3$. A fundamental difference between the slow and fast modulation regimes is that the latter enables the coupling to negative frequency replicas of the resonant modes of the system~ \cite{park2025spontaneous,allard2026broadband,sustaeta2026near}. Crucially, these negative frequency modes introduce gain into the system, allowing the LDOS to take negative values.
\par
In Fig.~\ref{fig:fig2}, vertical and horizontal axis are the frequency and the lateral momentum, as in Figs.~\ref{fig:fig1}(a1),(a2). We highlight the lightcone, the SPP and their negative frequency replicas, $\Omega-cK_\parallel$ and $\Omega-\omega_\text{SPP}$, as well as $\omega_\text{ENZ}$ and the parametric resonance condition $\Omega/2$. Blue regions in the plot correspond to $\rho_{zz}(\omega,K_\parallel)>0$, which contribute with a positive value to the total $z$-projected LDOS $\rho_{zz}(\omega)$ and hence, to the total rate $\Gamma_\text{tot}$~\eqref{eq:gamma_total} and total emitted power $P_\text{tot}$~\eqref{eq:power}. Conversely, red regions have $\rho_{zz}(\omega,K_\parallel)<0$ and thus yield a negative contribution to $\rho_{zz}(\omega)$, $\Gamma_\text{tot}$ and $P_\text{tot}$. Therefore, we can identify blue regions with loss and red regions with gain~\cite{park2025spontaneous,allard2026broadband,sustaeta2026near}.
\par
A glance at Fig.~\ref{fig:fig2}(a1) shows that $\rho_{zz}(\omega,K_\parallel)$ takes negative values along $\Omega-\omega_\text{SPP}$, which appears red-colored in the plot. Thus, according to the preceding discussion, it is seen that the negative frequency replica of the SPP introduces gain into the system~\cite{park2025spontaneous,allard2026broadband,sustaeta2026near}. Furthermore, coupling to the negative frequency replica of the surface plasmon, $\Omega-\omega_\text{SP}$, results in $\rho_{zz}(\omega)$ being negative in the vicinity of the air-plasma interface, leading to near-field spontaneous excitation and energy absorption~\cite{sustaeta2026near}.
\par
On the other hand, in Fig.~\ref{fig:fig2}(a2) we plot $\rho_{zz}(\omega,K_\parallel)$ for the case of a non-dispersive temporal modulation, with $\Omega=1.5\omega_\text{SP}$ and $\alpha_\text{nd}=\abs{\chi_\text{bg}(\omega_\text{SP})}\alpha\simeq1.5$. First, we see that $\Omega-\omega_\text{SPP}$ does not lead to any prominent features in the LDOS, nor that does it carry a large negative value of $\rho_{zz}(\omega,K_\parallel)$ either, in sharp contrast with the dispersive case. Additionally, $\omega_\text{SPP}$ is seen to deviate from the non-modulated curve (highlighted with a dashed grey line), moving to higher frequencies. Similarly, the continuum of guided modes is pushed to higher frequencies as well. Thus, dispersive and non-dispersive descriptions of the time modulation can lead to very different physics in the fast modulation regime. 
\par
It must be emphasized that fast temporal modulations can push the system to an unstable regime if the modulation strength is strong enough. This occurs due to the crossing between a resonant mode of the system and its negative frequency replica, such as $\omega_\text{SPP}$ and $\Omega-\omega_\text{SPP}$, at the parametric resonance condition $\omega=\Omega/2$~\cite{allard2026broadband,sustaeta2026near}. For large enough $\alpha$, this interaction between resonant modes and their negative frequency replicas results in the opening of a $K_\parallel$-gap, which is accompanied by one or more poles of the Green's dyadic crossing from $\text{Im}(\omega_\text{pole})<0$ to $\text{Im}(\omega_\text{pole})>0$~\cite{kim2025complex,lee2026analogs,allard2026broadband,sustaeta2026near,globosits2025exceptional}. Hence, in the unstable regime, the parametrically-driven system supports resonant modes that grow exponentially in time. However, this exponential growth would eventually saturate due to material non-linearities~\cite{kiselev2025symmetry,lee2026analogs,kyung2026self}. Subsequently, if the system is unstable, the linear theory is valid only at times much shorter than the growth rate of the system, $t\ll1 / \text{max}[\text{Im}(\omega_\text{pole})]$, beyond which it is necessary to include material non-linearities. 
\par
Let us now analyze in detail the stability of the system. First, let us note that the total Green's dyadic can be decomposed into its free-space $(\text{i.e., }\textbf{G}^{(0)}_\sigma)$ and scattered $(\text{i.e., }\textbf{G}^\text{(scatt.)}_\sigma)$ parts for each polarization $\sigma=\text{TE, TM}$~\cite{supmat}, where the total Green's dyadic then reads $\textbf{G}_\sigma = \textbf{G}^{(0)}_\sigma + \textbf{G}^\text{(scatt.)}_\sigma$~\cite{rotter2017light}. Since $\textbf{G}^\text{(scatt.)}_\sigma$ contains all the effects of the material body and the SPPs are TM-polarized~\cite{otto1968excitation,barnes2003surface,greffet2012introduction}, we can study in full depth the stability of the system by analyzing the eigenvalues of $\textbf{G}^\text{(scatt.)}_\text{TM}$.
\par
In Figs.~\ref{fig:fig2}(b1),(b2) we plot the maximum eigenvalue (in modulus) of the TM-polarized scattered Green's dyadic, $\textbf{G}^\text{(scatt.)}_\text{TM}$, versus the imaginary part of the frequency, $\text{Im}(\omega)$, and the lateral momentum. The real part of the frequency is set to $\text{Re}(\omega) = \Omega/2$, at the crossing point between $\omega_\text{SPP}$ and $\Omega-\omega_\text{SPP}$. The vertical white line marks the lightcone, $K_\parallel = \text{Re}(\omega) / c$, to whose right we can see the poles of the system, which appear as bright yellow curves in the plot. Critically, for both a dispersive and non-dispersive $\varepsilon_\text{mod}$, there are poles at $\text{Im}(\omega)>0$, showing that the system is indeed unstable for the modulation parameters considered. Additionally, the unstable regime will also have an imprint on the quantum phenomena supported by the system; in particular, it will be shown to modify the amplification of vacuum fluctuations in the DCE.
\section{Electric Field Correlation Functions}
\label{sec:e_corr}
In order to provide a deeper characterization of the electromagnetic field in dispersive time-varying media, we discuss its first order correlation. These correlation functions reveal intrinsic features of the electromagnetic field, and are important as well in the study of light-matter interactions~\cite{vogel2006quantum,gerry2023introductory}. First, there are correlation functions associated to polariton-conserving processes. Making use of Eq.~\eqref{eq:e_field_pols}, these read:
\begin{subequations}
    \begin{equation}
        \begin{split}
            &\langle E_i(\textbf{r},\omega)E^\dagger_j(\textbf{r}',\omega')\rangle   = \int d^3\textbf{r}''\int d\omega''\sum_k\\&
            \mathcal{G}_{ik}(\textbf{r},\omega;\textbf{r}'',\omega'')\left(n_\text{th}(\abs{\omega''})+\Theta(\omega'')\right)\mathcal{G}^*_{jk}(\textbf{r}',\omega';\textbf{r}'',\omega'')
        \end{split}
        \label{eq:corr_gamma}
    \end{equation}
    and
    \begin{equation}
        \begin{split}
            &\langle E^\dagger_i(\textbf{r},\omega)E_j(\textbf{r}',\omega')\rangle   = \int d^3\textbf{r}''\int d\omega''\sum_k\\&
            \mathcal{G}^*_{ik}(\textbf{r},\omega;\textbf{r}'',\omega'')\left(n_\text{th}(\abs{\omega''})+\Theta(-\omega'')\right)\mathcal{G}_{jk}(\textbf{r}',\omega';\textbf{r}'',\omega''),
        \end{split}
        \label{eq:corr_g1}
    \end{equation}
    \label{eq:corr_single}
\end{subequations}
where $\omega,\text{ }\omega'>0$. Eq.~\eqref{eq:corr_gamma} describes the creation of a $j$-polarized polariton of frequency $\omega'$ at the point $\textbf{r}'$ and the subsequent annihilation of an $i$-polarized polariton of frequency $\omega$ at $\textbf{r}$. Conversely, Eq.~\eqref{eq:corr_g1} describes the opposite process, whereby a $j$-polarized polariton of frequency $\omega'$ is annihilated at $\textbf{r}'$ and an $i$-polarized polariton of frequency $\omega$ is later created at the point $\textbf{r}$. Moreover, in the above equations the expectation value equals $\langle...\rangle = \Tr{\rho_B...}$, where $\rho_B$ is the density matrix of the electromagnetic environment prior to the start of the time modulation. In particular, we have assumed $\rho_B$ to be a thermal state, where $n_\text{th}(\omega) = 1 / (e^{\beta\hbar\omega} -1)$ is the number of thermal photons at a frequency $\omega>0$ and $\beta=(k_BT)^{-1}$ is the inverse of the thermal energy at a temperature $T$.
\par
Additionally, there are two features of Eqs.~\eqref{eq:corr_single} which must be highlighted: first, there is an absolute value in the frequency variable of the polariton number, $n_\text{th}(\abs{\omega''})$, which stems from the constraint $\textbf{f}(\textbf{r},-\omega) = \textbf{f}^\dagger(\textbf{r},\omega)$ and furthermore, guarantees that negative frequency modes have positive energies. Secondly, there are step functions $\Theta(\pm\omega'')$, which account for vacuum fluctuations. In particular, for Eq.~\eqref{eq:corr_gamma} one has $\Theta(\omega'')$, since the expectation value $\langle EE^\dagger\rangle$ already incorporates vacuum fluctuations in the absence of the temporal modulation~\cite{vogel2006quantum,gerry2023introductory}. However, for Eq.~\eqref{eq:corr_g1} we have $\Theta(-\omega'')$, which describes the amplification of the quantum vacuum through the coupling to negative frequencies~\cite{mendoncca2000quantum,sloan2021casimir,horsley2023quantum,hassani2024dynamical,bae2025quantum,sustaeta2025quantum,kim2026classical}, and is rooted in the squeezing term in Eq.~\eqref{eq:e_field_pols}. Importantly, for the case of an initial vacuum state $\rho_B = \ket{\text{vac}}\bra{\text{vac}}$, $n_\text{th}(\omega) = 0$. Consequently, in that case, Eq.~\eqref{eq:corr_g1} being non-vanishing is a signature of the dynamical Casimir effect, in which virtual photons are converted into real ones through the energy provided by the time modulation~\cite{nation2012colloquium,lahteenmaki2013dynamical,mendoncca2000quantum,ganfornina2024quantum,sustaeta2025quantum}.  
\par
On the other hand, there are also correlation functions that represent processes that do not conserve the number of polaritons:
\begin{subequations}
    \begin{equation}
        \begin{split}
            &\langle E^\dagger_i(\textbf{r},\omega)E^\dagger_j(\textbf{r}',\omega')\rangle   = \int d^3\textbf{r}''\int d\omega''\sum_k\\&
            \mathcal{G}_{ik}(\textbf{r},-\omega;\textbf{r}'',\omega'')\left(n_\text{th}(\abs{\omega''})+\Theta(\omega'')\right)\mathcal{G}^*_{jk}(\textbf{r}',\omega';\textbf{r}'',\omega'')
        \end{split}
        \label{eq:corr_gamma(dce)}
    \end{equation}
    and
    \begin{equation}
        \begin{split}
            &\langle E_i(\textbf{r},\omega)E_j(\textbf{r}',\omega')\rangle   = \int d^3\textbf{r}''\int d\omega''\sum_k\\&
            \mathcal{G}^*_{ik}(\textbf{r},-\omega;\textbf{r}'',\omega'')\left(n_\text{th}(\abs{\omega''})+\Theta(-\omega'')\right)\mathcal{G}_{jk}(\textbf{r}',\omega';\textbf{r}'',\omega''),
        \end{split}
        \label{eq:corr_g1(dce)}
    \end{equation}
    \label{eq:corr_two}
\end{subequations}
where once again, it is $\omega,\text{ }\omega'>0$. Eq.~\eqref{eq:corr_gamma(dce)} describes the creation of a $j$-polarized polariton with frequency $\omega'$ at $\textbf{r}'$, followed by the creation of a second polariton at the point $\textbf{r}$ with polarization along the $i$-axis and frequency $\omega$. Conversely, Eq.~\eqref{eq:corr_g1(dce)} describes the annihilation of two polaritons, one at the point $\textbf{r}'$ with frequency $\omega'$ and polarization $j$ and a second one at $\textbf{r}$ with frequency $\omega$ and polarization $i$. Moreover, a glance at Eqs.~\eqref{eq:corr_two} shows that two-polariton processes originate from the coupling between positive and negative frequencies. Critically, this coupling between frequencies of opposite sign was also responsible for the amplification of vacuum fluctuations discussed in Eq.~\eqref{eq:corr_g1}. However, unlike the case of Eq.~\eqref{eq:corr_g1}, which is non-zero for thermal states, Eqs.~\eqref{eq:corr_two} can only be non-vanishing if positive and negative frequencies interact, even if the system is in a thermal state. Thus, two-polariton process constitute an unmistakable signature of the DCE.
\par
Finally, it will be convenient to particularize Eqs.~\eqref{eq:corr_single} and~\eqref{eq:corr_two} for a periodic temporal modulation. In this case, for photon-conserving processes we can write:
\begin{subequations}
    \begin{equation}
        \begin{split}
            &\langle E_i(\textbf{r},\tilde{\omega}+n\Omega)E^\dagger_j(\textbf{r}',\tilde{\omega}'+n'\Omega)\rangle \\&\equiv \Gamma_{ij;nn'}(\textbf{r},\textbf{r}';\tilde{\omega}')2\pi\delta(\tilde{\omega} - \tilde{\omega}')
        \end{split}
        \label{eq:corr_gamma_floquet(0)}
    \end{equation}
    and
    \begin{equation}
        \begin{split}
            &\langle E^\dagger_i(\textbf{r},\tilde{\omega}+n\Omega)E_j(\textbf{r}',\tilde{\omega}'+n'\Omega)\rangle \\&\equiv \text{G}^{(1)}_{ij;nn'}(\textbf{r},\textbf{r}';\tilde{\omega}')2\pi\delta(\tilde{\omega} - \tilde{\omega}'),
        \end{split}
        \label{eq:corr_g1_floquet(0)}
    \end{equation}
    \label{eq:corr_floquet(0)}
\end{subequations}
where the following quantities have been introduced:
\begin{subequations}
    \begin{equation}
        \begin{split}
            &\Gamma_{ij;nn'}(\textbf{r},\textbf{r}';\tilde{\omega}') = \int d^3\text{r}''\sum_{l}\sum_k\\&\mathcal{G}_{ik;nl}(\textbf{r},\textbf{r}'';\tilde{\omega}')\left(n_\text{th}(\abs{\omega_l})+\Theta(\omega_l)\right)\mathcal{G}^*_{jk;n'l}(\textbf{r}',\textbf{r}'';\tilde{\omega}'),
        \end{split}
        \label{eq:corr_gamma_floquet}
    \end{equation}
    and
    \begin{equation}
        \begin{split}
            &\text{G}^{(1)}_{ij;nn'}(\textbf{r},\textbf{r}';\tilde{\omega}') = \int d^3\text{r}''\sum_{l}\sum_k\\&\mathcal{G}^*_{ik;nl}(\textbf{r},\textbf{r}'';\tilde{\omega}')\left(n_\text{th}(\abs{\omega_l})+\Theta(-\omega_l)\right)\mathcal{G}_{jk;n'l}(\textbf{r}',\textbf{r}'';\tilde{\omega}'),
        \end{split}
        \label{eq:corr_g1_floquet}
    \end{equation}
    \label{eq:corr_floquet}
\end{subequations}
In Eqs.~\eqref{eq:corr_floquet}, $\omega_l = \tilde{\omega}' + l\Omega$. Importantly, the Dirac delta $\delta(\tilde{\omega}-\tilde{\omega}')$, which arises from the periodicity of the temporal modulation, constrains polariton-conserving processes to the case $\omega-\omega' = l\Omega$, where $\omega,\text{ }\omega'>0$ and $l\in\mathbb{Z}$. Thus, these processes are limited to polaritons whose frequencies differ only by an integer multiple of the modulation frequency and hence, they do not necessitate a large modulation frequency (with the exception of $\text{G}^{(1)}>0$ for a vacuum state).
\par
Similarly, for two-polariton processes we can write:
\begin{subequations}
    \begin{equation}
        \begin{split}
            &\langle E^\dagger_i(\textbf{r},\tilde{\omega}+n\Omega)E^\dagger_j(\textbf{r}',\tilde{\omega}'+n'\Omega)\rangle \\&\equiv \Gamma_{ij;-n,n'}(\textbf{r},\textbf{r}';\tilde{\omega}')2\pi\delta(\tilde{\omega} + \tilde{\omega}')
        \end{split}
        \label{eq:corr_gamma_floquet(dce)(0)}
    \end{equation}
    and
    \begin{equation}
        \begin{split}
            &\langle E_i(\textbf{r},\tilde{\omega}+n\Omega)E_j(\textbf{r}',\tilde{\omega}'+n'\Omega)\rangle \\&\equiv \text{G}^{(1)}_{ij;-n,n'}(\textbf{r},\textbf{r}';\tilde{\omega}')2\pi\delta(\tilde{\omega} + \tilde{\omega}'),
        \end{split}
        \label{eq:corr_g1_floquet(dce)(0)}
    \end{equation}
    \label{eq:corr_dce_floquet}
\end{subequations}
where $\Gamma_{ij;-n,n'}(\textbf{r},\textbf{r}';\tilde{\omega}')$ and $\text{G}^{(1)}_{ij;-n,n'}(\textbf{r},\textbf{r}';\tilde{\omega}')$ are defined as in Eqs.~\eqref{eq:corr_floquet}, but with the change $n\longrightarrow-n$ in the Floquet band index. Crucially, the Dirac delta $\delta(\tilde{\omega} + \tilde{\omega}')$ restricts polariton non-conserving interactions to the case $\omega + \omega' = l\Omega$, where once again it is $\omega,\text{ }\omega'>0$ and $l\in\mathbb{Z}$. Therefore, two-polariton processes necessitate a large modulation frequency, which constitutes a standard feature of the DCE, for which the driving frequency must be comparable to the frequencies of the excitations of the system~\cite{nation2012colloquium,lahteenmaki2013dynamical}. As a last comment, for the case of periodic temporal modulations, $\rho_B$ must be understood as the state of the electromagnetic environment at $t\longrightarrow-\infty$, prior to the start of the time modulation.

\begin{figure*}
    \centering
    \hspace{0 mm}
    \includegraphics[scale = 0.50]{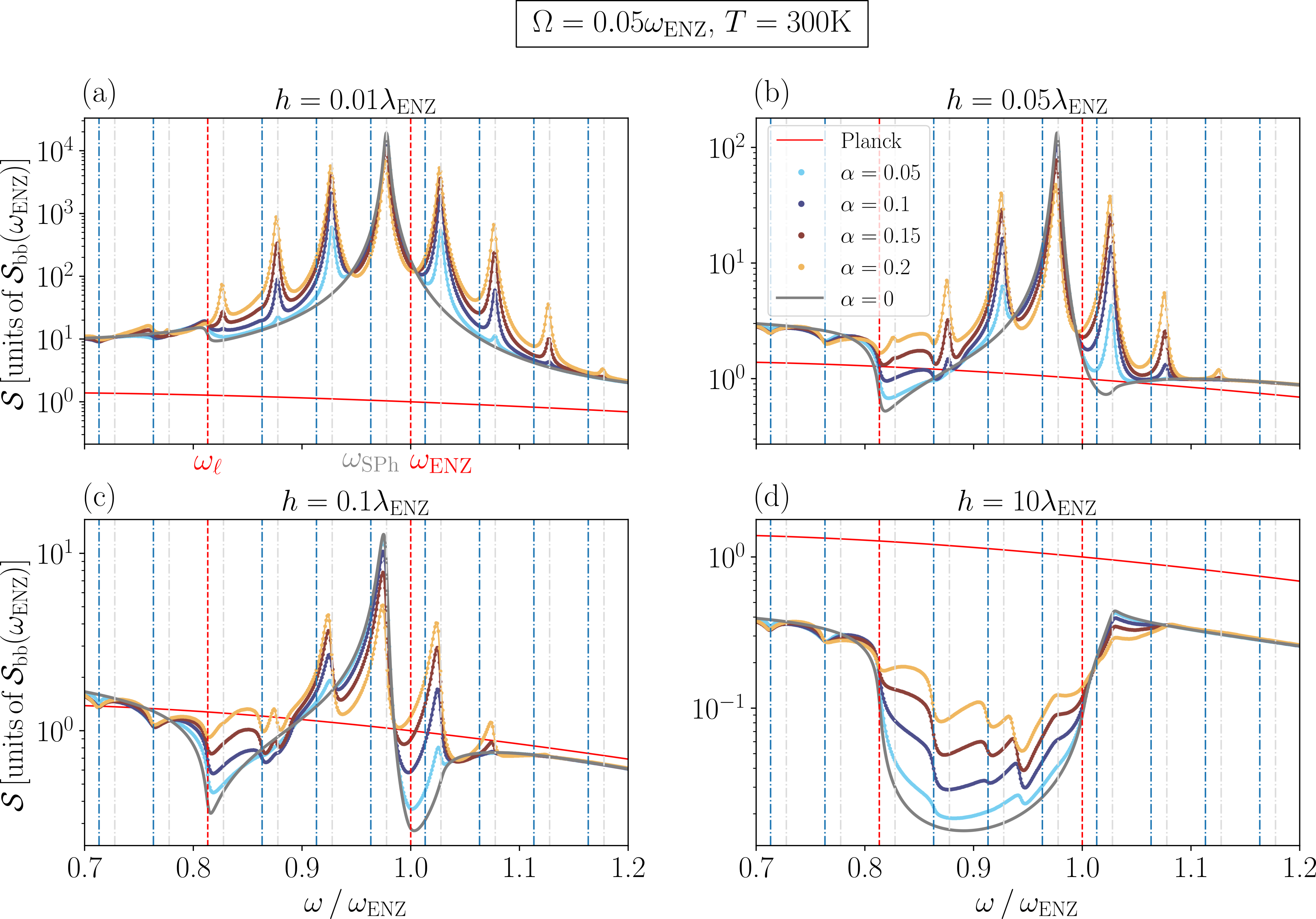} 
    \caption{\justifying Thermal radiation in a time-modulated polar insulator. The panels show the spectral density $\mathcal{S}$ [normalized to the black-body spectral density evaluated at $\omega_\text{ENZ}$, $\mathcal{S}_\text{bb}(\omega_\text{ENZ})$] vs frequency for a SiC half-space with a periodically time-varying Lorentz frequency. We consider values different values of $\alpha$ and compare with Planck's law (red line) as well. On the other hand, we study different values of the height above the surface, $h =0.01\lambda_\text{ENZ}\text{(a), }0.05\lambda_\text{ENZ}\text{(b), }0.1\lambda_\text{ENZ}\text{(c), }10\lambda_\text{ENZ}\text{(d)}$. Finally, the modulation frequency is set to $\Omega=0.05\omega_\text{ENZ}$ and the polar insulator is initially (prior to the start of the temporal modulation) in a thermal state with temperature $T = 300$K.}
    \label{fig:fig2.5} 
\end{figure*}
\subsection{Thermal Radiation in Dispersive Time-Varying Media}
\label{sec:thermal}
Once the field correlation functions have been introduced, we now study thermal radiation by time-varying media. We consider a periodically time-modulated SiC half-space, as in Sec.~\ref{sec:slow}, with the Lorentz frequency varying in time as in Eq.~\eqref{eq:om0_tmod}. Moreover, in this section we only study the case of a dispersive temporal modulation (the non-dispersive case can be found in the SM~\cite{supmat}). On the other hand, we set the modulation frequency to $\Omega=0.05\omega_\text{ENZ}$, the same as in Fig.~\ref{fig:fig1}, and consider the polar insulator to be initially in a thermal state at $T = 300$K.
\par
The thermal radiation emitted by thermal sources is characterized by the spectral density of the electromagnetic field~\cite{vazquez2023incandescent,ren2026clarification,zhu2026enhancing}:
\begin{equation}
    \mathcal{S}(\textbf{r},\omega) \equiv\sum_{i=x,y,z}\text{G}^{(1)}_{ii}(\textbf{r},\textbf{r};\omega),
    \label{eq:spectral_density}
\end{equation}
where we have set $n = n'$ in Eq.~\eqref{eq:corr_g1_floquet} and subsequently dropped the Floquet indices for the sake of simplicity. A particularly interesting baseline to compare with is the black-body spectrum or Planckian distribution, $\mathcal{S}_\text{bb}(\omega) =\hbar\omega^3/(\pi c^3\epsilon_0)n_\text{th}(\omega)$. Then, the black-body spectrum enables us to classify the spectral density of different thermal sources as sub-Planckian (if $\mathcal{S}<\mathcal{S}_\text{bb}$), Planckian (if $\mathcal{S}\simeq\mathcal{S}_\text{bb}$) and super-Planckian (if $\mathcal{S}>\mathcal{S}_\text{bb}$).
\par
In Fig.~\ref{fig:fig2.5} we plot the spectral density as a function of the frequency, and for different values of the modulation strength and the height above the surface. The vertical grey dashed-dotted lines are $\omega_\text{SPh}$ and its Floquet sidebands $\omega_\text{SPh} +l\Omega$, while the vertical red dashed lines are $\omega_\ell$ and $\omega_\text{ENZ}$, which limit the Reststrahlen band. Furthermore, the vertical blue dashed-dotted lines are the replicas of the Lorentz frequency, $\omega_\ell+l\Omega$. In Fig.~\ref{fig:fig2.5}(a) we consider the case of $ h = 0.01\lambda_\text{ENZ}$, in the near-field of the structure. The grey curve corresponds to the $\alpha=0$ case, which is seen to peak at $\omega_\text{SPh}$. As is known, thermal emission is super-Planckian in the near-field of a structure (compare with $\mathcal{S}_\text{bb}$, the red curve). When the medium is time-modulated (see light blue, dark blue, brown and yellow dots for $\alpha=0.05,0.1,0.15,0.2$), Floquet sidebands of the SPh emerge at $\omega_\text{SPh}+l\Omega$, both at higher and lower frequencies. With increasing modulation strength, the height of the central peak at $\omega_\text{SPh}$ diminishes, while that of the Floquet sidebands increases, similar to the behavior of the LDOS reported in Sec.~\ref{sec:slow}. For the parameters considered, the number of visible sidebands ranges from four for $\alpha=0.05$ (with the $l = \pm2$ ones being small but nonetheless non-vanishing) to eight for $\alpha=0.2$.
\par
As the distance to the surface becomes larger, we transition to the intermediate-field of the structure, as shown in Figs.~\ref{fig:fig2.5}(b),(c), for which $h = 0.05\lambda_\text{ENZ},\text{ }0.1\lambda_\text{ENZ}$, respectively. In the intermediate-field, the coupling to $\omega_\text{SPh}$ and its sidebands weakens; this in turn leads to an overall reduction of the spectral density, which becomes sub-Planckian for the $\alpha=0$ case for some frequencies, most notably in the vicinity of $\omega_\ell$ and $\omega_\text{ENZ}$. However, for large enough $\alpha$, the temporal modulation amplifies the spectral density at the surface phonon sidebands, making the thermal emission of the time-modulated polar material super-Planckian at these frequencies. Furthermore, it is seen that as $\alpha$ increases, $\mathcal{S}$ slightly diminishes at the negative sidebands of the Lorentz frequency (i.e., $\omega_\ell+l\Omega$ for $l<0$). On the other hand, and in sharp contrast with what was previously reported in the literature~\cite{vazquez2023incandescent,ren2026clarification}, there is no enhancement of the spectral density at the sidebands of $\omega_\text{ENZ}$.   
\par
In the far-field of the structure, for $h=10\lambda_\text{ENZ}$ [Fig.~\ref{fig:fig2.5}(d)], the coupling to the SPh and its sidebands has completely died out. Furthermore, even though the temporal modulation enhances the thermal emission in the Reststrahlen band, the spectral density is sub-Planckian at all frequencies for the values of $\alpha$ considered. On the other hand, as was the case for the intermediate-field, we do not see any enhancement of the spectral density at the sidebands of $\omega_\text{ENZ}$, nor do we report any super-Planckian features in the far-field thermal spectrum of the time-modulated material. 
\par
\begin{figure*}[t]
    \centering
    \hspace{0 mm}
    \includegraphics[scale = 0.50]{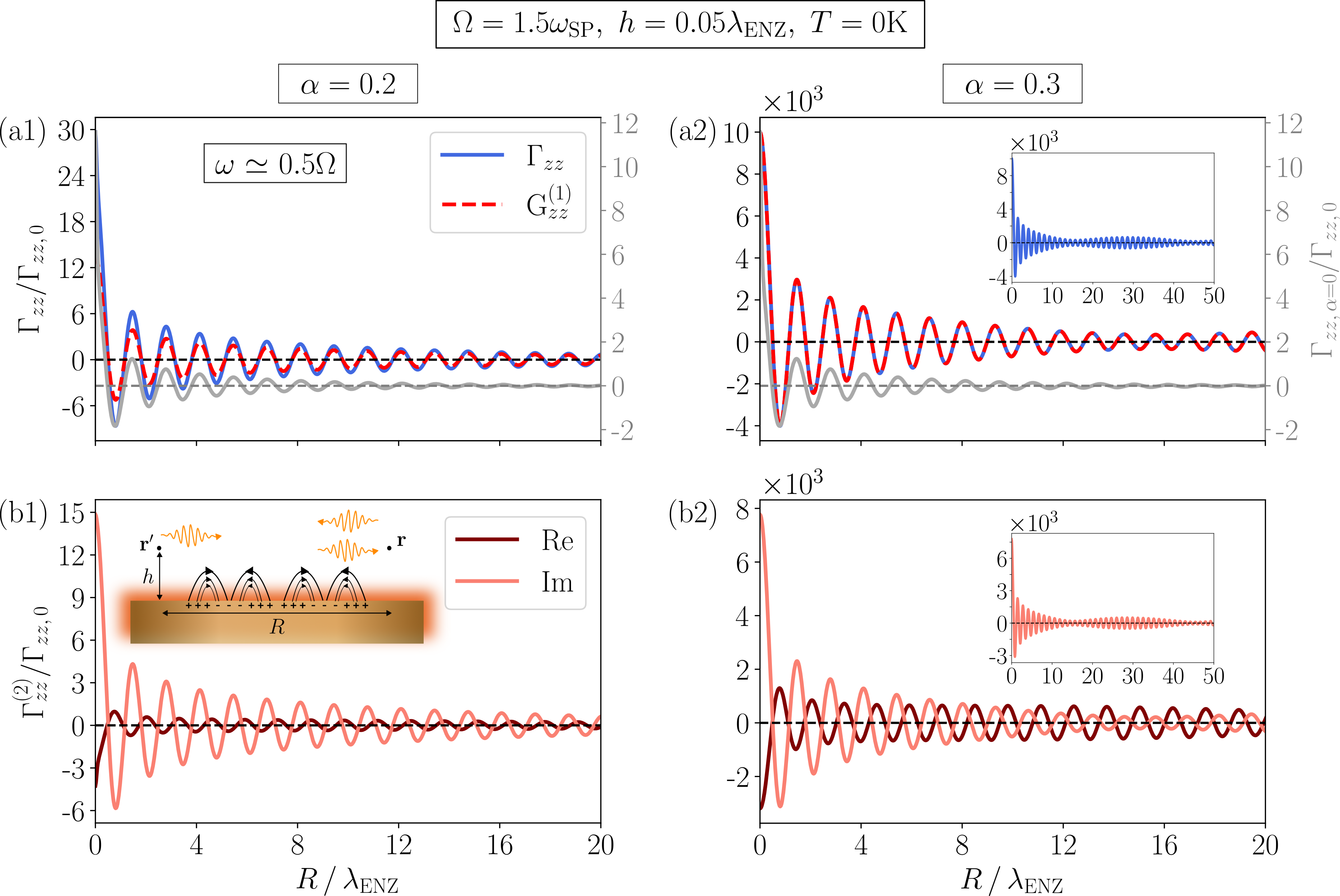} 
    \caption{\justifying Dynamical Casimir effect in a time-modulated plasmonic medium. The panels show the first order correlation functions of the electric field generated by an ITO half-space with periodically time-varying plasma frequency. (a1)-(a2) Polariton-conserving correlation functions (normalized to $\Gamma_{zz,0}$) vs the lateral separation between points $R$ (normalized to $\lambda_\text{ENZ}$). The solid blue line and the red dashed line are $\Gamma_{zz}$ and $\text{G}^{(1)}_{zz}$, respectively, while the grey line is $\Gamma_{zz,\alpha=0}$. (b1)-(b2) Real (maroon) and imaginary (pink) parts of $\Gamma_{zz}^{(2)}$, the two-polariton absorption correlation function. In (a1)-(b1) the modulation strength is $\alpha=0.2$, for which the system is stable, whereas in (a2)-(b2) it is $\alpha=0.3$, for which the system is already unstable. On the other hand, the insets in (a2)-(b2) show spatially long-lived oscillations in the correlation functions in the unstable regime of the system. The modulation frequency is $\Omega=1.5\omega_\text{SP}$ in all figures, while the points' height above the surface is $h = 0.05\lambda_\text{ENZ}$. Finally, the electromagnetic field is assumed to be initially in its vacuum state, with $T = 0\text{K}$.}
    \label{fig:fig3} 
\end{figure*}
It must be emphasized that our approach is fundamentally different from those previous works: on the one hand, they apply time-dependent perturbation theory to obtain the time-evolution of the polariton operators. Then, they substitute $\textbf{f}(\textbf{r},\omega)\longrightarrow\textbf{f}(\textbf{r},\omega;t)$ in the formula for the electric field in non-modulated linear media. Critically, this approach does not take into account that, in time-varying media, the electric field wave equation is also modified by the temporal modulation and subsequently, so is the Green's dyadic. 
\par
On the other hand, our approach is based on using the FDT for time-varying media, which connects the fluctuating currents of the system with the dissipative part of its linear response. In particular, the FDT shows that consistency between fluctuations and dissipation demands that both the noise currents and the electric permittivity be modified by the temporal modulation. Furthermore, our results for the thermal spectrum are consistent with those of the LDOS presented in Sec.~\ref{sec:slow}, in which it is seen that the time-modulation produces sidebands of the resonant modes of the system.

\subsection{Dynamical Casimir Effect}
\label{sec:dce}
We now study the DCE in a time-periodic plasmonic half-space. First, let us note that for a planar system, the $(\omega,\textbf{K}_\parallel)-$domain electric field operator can be written as:
\begin{equation}
\begin{split}
    &\textbf{E}(z;\omega,\textbf{K}_\parallel) = \\&   \int_0^\infty d\omega'\int dz'\Big(\boldsymbol{\mathcal{G}}(z,\omega;z',\omega';\textbf{K}_\parallel)\cdot\textbf{f}(z';\omega',\textbf{K}_\parallel)\\& +\boldsymbol{\mathcal{G}}(z,\omega;z',-\omega';\textbf{K}_\parallel)\cdot\textbf{f}^\dagger(z';\omega',-\textbf{K}_\parallel)\Big),
\end{split}
\label{eq:e_field_bogo}
\end{equation}
where the frequency domain electric field operator reads $\textbf{E}(\textbf{r},\omega) = \textbf{E}(\textbf{R},z,\omega) =\int d^2K_\parallel/(2\pi)^2e^{i\textbf{K}_\parallel\cdot\textbf{R}}\textbf{E}(z;\omega,\textbf{K}_\parallel)$, and where we have introduced the momentum-resolved dressed Green's dyadic $\mathcal{G}(z,\omega;z',\omega';\textbf{K}_\parallel)$ and polaritons $\textbf{f}(z;\omega,\textbf{K}_\parallel)$. Crucially, Eq.~\eqref{eq:e_field_bogo} shows explicitly that the temporal modulation and the subsequent coupling to negative frequencies results in the production of entangled pairs of polaritons with opposite lateral momenta $\pm\textbf{K}_\parallel$~\cite{mendoncca2000quantum,ganfornina2024quantum,sustaeta2025quantum}. 
\par
Returning to the calculation of the correlation functions, we take the electromagnetic field to be initially in its vacuum state, $\rho_B = \ket{\text{vac}}\bra{\text{vac}}$, which allows us to isolate the signatures of vacuum quantum amplification. Furthermore, we set $i = j = z$ in Eqs.~\eqref{eq:corr_single},~\eqref{eq:corr_two} in order to maximize the coupling to the SPP, and fix the height above the surface of the two points to $h = 0.05\lambda_\text{ENZ}$. Additionally, we only consider in here the case of a dispersive temporal modulation, where the plasma frequency of the system is periodically modulated in time, as in Sec.~\ref{sec:fast}.
\par
On the other hand, we work with a modulation frequency of $\Omega=1.5\omega_\text{SP}$, as in Fig.~\ref{fig:fig2}. Crucially, $\Omega/2=0.75\omega_\text{SP}$ crosses the SPP dispersion relation, as can be seen from Fig.~\ref{fig:fig2}(a1). An important feature of SPPs (and surface waves in general) is that they dramatically enhance the spatial coherence of the electromagnetic field~\cite{yu2023manipulating,carminati1999near,shchegrov2000near}, leading to non-vanishing correlations between points separated by distances much larger than in free-space. Thus, in order to better understand the role of the temporal modulation in modifying the spatial coherence of the two-point correlation functions, we set $\omega,\text{ }\omega' \simeq \Omega/2=0.75\omega_\text{SP}$.
\par
In Fig.~\ref{fig:fig3}(a1) we plot the polariton-conserving electric field correlation functions, $\Gamma_{zz}$ and $\text{G}^{(1)}_{zz}$, as a function of the in-plane separation length between the two points, $R$, which we normalize to $\lambda_\text{ENZ} = 2\pi c/\omega_\text{ENZ}$. We set the modulation strength to $\alpha=0.2$, for which the system is stable, with all the poles of its Green's dyadic lying at $\text{Im}(\omega)<0$. The grey and blue lines are $\Gamma_{zz}$ for $\alpha=0$ and $\alpha=0.2$, respectively, while the red-dashed line represents $\text{G}^{(1)}_{zz}$. Additionally, we normalize all the correlation functions to $\Gamma_{zz,0}=\hbar\omega^3/(3\pi c^3\epsilon_0)$, the free-space value at $R = 0$. On the one hand, a glance at the figure shows that $\Gamma_{zz}$ is approximately two times larger than $\Gamma_{zz,\alpha=0}$ and subsequently, that it is non-zero at larger distances than in the non-modulated case. On the other hand, it is seen that $\text{G}^{(1)}_{zz}$ is non-zero, with values smaller than those of $\Gamma_{zz}$ but comparable to those of $\Gamma_{zz,\alpha=0}$. Crucially, as discussed earlier, a non-vanishing $\text{G}^{(1)}_{zz}$ for $\rho_B = \ket{\text{vac}}\bra{\text{vac}}$ is a signature of DCE, whereby virtual photons are transformed into real ones through the energy provided by the temporal modulation. Therefore, it is seen that the time modulation amplifies the quantum vacuum and extends as well the coherence length of single-polariton processes.
\par
Next, we study the polariton-non-conserving correlation functions. First, let us write $\langle E_z^\dagger E_z^\dagger\rangle = \Gamma_{zz}^\text{(2)}2\pi\delta$ and $\langle E_z E_z\rangle = \text{G}_{zz}^\text{(2)}2\pi\delta$, where we have omitted the position and frequency labels in the electric field operators and Dirac deltas for the sake of simplicity. Crucially, from Eqs.~\eqref{eq:corr_dce_floquet} it can be readily seen that for $\omega,\text{ }\omega'=\Omega/2$, $\Gamma_{zz}^{(2)} = \big(\text{G}_{zz}^\text{(2)}\big)^*$. Consequently, it is sufficient to study either of them, and so we focus on $\Gamma_{zz}^\text{(2)}$. In Fig.~\ref{fig:fig3}(b1) we plot the real and imaginary parts of $\Gamma_{zz}^\text{(2)}$ as a function of $R$. First, it is seen that $\Gamma_{zz}^\text{(2)}$ is mostly dominated by its imaginary part, whose maxima are approximately four times larger than those of the real part. Furthermore, real and imaginary parts are approximately $\pi$-dephased, with the maxima of the real part coinciding with the minima of the imaginary part. Additionally, $\text{Im}\big(\Gamma_{zz}^\text{(2)}\big)$ is seen to be larger than $\Gamma_{zz,\alpha=0}$ and to survive at larger distances as well. Thus, the combined action of frequency dispersion, spatial confinement and the temporal modulation enables long-range two-polariton processes, giving rise to a spatially non-local form of the dynamical Casimir effect.
\par
Let us now consider the case of $\alpha=0.3$, for which the system is known to be unstable, as discussed earlier. Figs.~\ref{fig:fig3}(a2),(b2) show the single polariton and two polariton correlation functions. First, it is seen that all correlation functions have maxima and minima $\sim10^{3}$ times larger than those of $\Gamma_{zz,\alpha=0}$. Additionally, the correlation functions do not exhibit the same decay as in the $\alpha=0$ and $\alpha=0.2$ cases, characterized by a Bessel function~\cite{carminati1999near,shchegrov2000near}. This is best seen in the insets, which show how the correlation functions exhibit oscillations characterized by two different wavelengths, with the subsequent modulation of the amplitude. This can be understood by looking at Fig.~\ref{fig:fig2}(b1), where it is seen how the opening of the $K_\parallel$-gap results in two different poles existing at $\text{Im}(\omega)=0$. Crucially, each pole has a different lateral momentum and thus, a different wavelength, which explains the oscillations in the envelope. Hence, these modulated oscillations are a feature of the unstable regime of the system~\cite{sustaeta2026near}. On the other hand, it is worth mentioning that in the unstable regime, the Fourier transform of the electromagnetic field does not strictly exist and thus, neither do the frequency-domain correlation functions in Eqs.~\eqref{eq:corr_floquet},~\eqref{eq:corr_dce_floquet}. This is so because when the system is unstable, the linear theory is only valid for times $t\ll1/\text{max}[\text{Im}(\omega_\text{pole})]$, beyond which material non-linearities must be included to describe saturation (see Sec.~\ref{sec:fast}). 
\par
Finally, we do not find any signatures of vacuum quantum amplification for the case of a small modulation frequency (i.e., $\Omega\ll\omega_\text{ENZ}$)~\cite{supmat}, in contrast to what was reported in Ref.~\cite{vazquez2023incandescent}. Indeed, in order to amplify vacuum fluctuations, it is necessary for positive frequency modes to couple to negative frequency ones, as can be seen from Eq.~\eqref{eq:e_field_bogo} of our theory; this in turn demands a fast modulation frequency, in accordance with the general features of the DCE~\cite{nation2012colloquium,lahteenmaki2013dynamical}.
\section*{Conclusions}
In this work, we have presented the theoretical formalism of fluctuational quantum electrodynamics of frequency-dispersive and dissipative time-varying media. In contrast to previous works in the literature, our framework accounts for dispersion and losses in the temporal modulation, which we treat in an exact manner, without relying on perturbative methods. Furthermore, we connect the fluctuations of the system with its dissipation through the fluctuation-dissipation theorem extended to time-varying systems; this way, our theory produces the first consistent quantization of the electromagnetic field in time-modulated material bodies. Additionally, we show that gain and loss must both be accounted for to define the local density of states in time-varying media, and provide a mathematical proof for the equivalence between the quantum mechanical Fermi Golden Rule and the classical power radiated by a harmonic point dipole. We also demonstrate that neglecting the dispersive and dissipative nature of the temporal modulation leads to erroneous predictions both in the slow and fast modulation regimes. Moreover, we analyze the thermal spectrum of a time-modulated material body beyond perturbation theory, showcasing new features in the amplification of thermal radiation in time-varying media. Finally, we study the dynamical Casimir effect, revealing how the temporal modulation amplifies vacuum fluctuations and generates entangled pairs of polaritons exhibiting spatially long-lived correlations. 
\par
Thus, our work sets the foundations of the theory of quantum electrodynamics of time-varying media, properly accounting for dispersion, dissipation and the time modulation. Furthermore, it paves the way towards exciting new research directions in the field, such as quantum optics in time-varying media, thermal radiation engineering and quantum light generation, among others.

\begin{acknowledgments}
\textit{Acknowledgments---} We acknowledge funding from the European Union through the ERC grant TIMELIGHT under GA101115792 \textcolor{black}{and from the Spanish Ministry for Science, Innovation, and Universities – Agencia Estatal de Investagación (AEI) through Grants No. PID2022-141036NA-I00, PID2021-125894NB-I00, No. CEX2018-000805-M (through the María de Maeztu program for Units of Excellence in Research and Developments) and Grant No. RYC2021-031568-I (Ramón y Cajal program)}. JESO also acknowledges support from the CAM Consejería de Educación, Ciencia y Universidades, Viceconsejería de Universidades, Investigación y Ciencia, Dirección General de Investigación e Innovación Tecnológica (CAM FPI grant Ref. A281).

\end{acknowledgments}

\bibliography{main}
\end{document}


\maketitle
\thispagestyle{empty} 
\setstcolor{Maroon}

\vspace*{\fill} 

\vspace*{\fill} 

\tableofcontents

\newpage
\pagenumbering{arabic} 
\setcounter{page}{1} 
\renewcommand{\thepage}{S\arabic{page}} 
\section{Fluctuation-Dissipation Theorem in Dispersive Time-Varying Media}

Our starting point is Kubo's formula for the linear response of a macroscopic dielectric system~\cite{kubo1966fluctuation}:
\begin{equation}
    \boldsymbol{\chi}(\text{r},t;\text{r}',t') = -\frac{i}{\hbar}\Theta(t-t')\langle[\textbf{P}(\text{r},t), \textbf{P}(\text{r}',t')]\rangle.
    \label{eq:sm_kubo_macro}
\end{equation}
Taking the Fourier transform in both temporal variables $t$ and $t'$, we have:
\begin{equation*}
    \begin{split}
       &\boldsymbol{\chi}(\textbf{r},\omega;\textbf{r}',\omega') = \iint dt\text{ }dt'e^{i\omega t}e^{-i\omega' t'}\boldsymbol{\chi}(\textbf{r},t;\textbf{r}',t') \\&= \iint dt\text{ }dt'e^{i\omega t}e^{-i\omega' t'}\frac{i}{\hbar}\Theta(t-t')\langle[\textbf{P}(\textbf{r},t),\textbf{P}(\textbf{r},t')]\rangle \\&
       = \iint dt\text{ }dt'e^{i\omega t}e^{-i\omega' t'}\frac{i}{\hbar}\Theta(t-t')\iint\frac{d\omega''}{2\pi}\frac{d\omega'''}{2\pi}e^{-i\omega'' t}e^{i\omega''' t'}\langle[\textbf{P}(\textbf{r},\omega''),\textbf{P}^\dagger(\textbf{r},\omega''')]\rangle.
    \end{split}
\end{equation*}
Now, changing variables to $\tau = t - t'$, we find that:
\begin{equation}
    \begin{split}
        &\boldsymbol{\chi}(\textbf{r},\omega;\textbf{r}',\omega') 
       = \iint\frac{d\omega''}{2\pi}\frac{d\omega'''}{2\pi}\int d\tau e^{i(\omega - \omega'') \tau}\Theta(\tau)\int dt'e^{-i(\omega' - \omega''' - \omega + \omega'')t'}\frac{i}{\hbar}\langle[\textbf{P}(\textbf{r},\omega''),\textbf{P}^\dagger(\textbf{r}',\omega''')]\rangle \\&
       = \iint\frac{d\omega''}{2\pi}\frac{d\omega'''}{2\pi}\left(\pi\delta(\omega - \omega'') + i\mathcal{P}\left(\frac{1}{\omega - \omega''}\right)\right)2\pi\delta(\omega' - \omega''' - \omega + \omega'')\frac{i}{\hbar}\langle[\textbf{P}(\textbf{r},\omega''),\textbf{P}^\dagger(\textbf{r}',\omega''')]\rangle\\&
       = \frac{i}{2\hbar}\langle[\textbf{P}(\textbf{r},\omega),\textbf{P}^\dagger(\textbf{r}',\omega')]\rangle - \frac{1}{\hbar}\int\frac{d\omega''}{2\pi}\mathcal{P}\left(\frac{1}{\omega - \omega''}\right)\langle[\textbf{P}(\textbf{r},\omega''),\textbf{P}^\dagger(\textbf{r}',\omega' - \omega + \omega'')]\rangle,
    \end{split}
    \label{eq:sm_kubo_macro_v1}
\end{equation}
where $\mathcal{P}(...)$ denotes the Cauchy principal value and where we have used the Sokhotski–Plemelj theorem for the Fourier transform of the step function~\cite{scheel2008macroscopic}. Next, we take the Hermitian conjugate of the above and arrive to:

\begin{equation*}
    \begin{split}
        &\boldsymbol{\chi}^\dagger(\textbf{r},\omega;\textbf{r}',\omega') = -\frac{i}{2\hbar}\langle[\textbf{P}(\textbf{r}',\omega'),\textbf{P}^\dagger(\textbf{r},\omega)]\rangle - \frac{1}{\hbar}\int\frac{d\omega''}{2\pi}\mathcal{P}\left(\frac{1}{\omega - \omega''}\right)\langle[\textbf{P}(\textbf{r}',\omega' - \omega + \omega''),\textbf{P}^\dagger(\textbf{r},\omega'')]\rangle.
    \end{split}
\end{equation*}
Now, exchanging the spatial and frequency arguments in the above equation (i.e., $\textbf{r}\longleftrightarrow\textbf{r}'$ and $\omega\longleftrightarrow\omega'$) and doing the change of variables $\omega'' \longrightarrow \omega'' - \omega + \omega'$ in the principal value's integral, leads to:
\begin{equation}
    \begin{split}
        &\boldsymbol{\chi}^\dagger(\textbf{r}',\omega';\textbf{r},\omega) = -\frac{i}{2\hbar}\langle[\textbf{P}(\textbf{r},\omega),\textbf{P}^\dagger(\textbf{r}',\omega')]\rangle - \frac{1}{\hbar}\int\frac{d\omega''}{2\pi}\mathcal{P}\left(\frac{1}{\omega - \omega''}\right)\langle[\textbf{P}(\textbf{r},\omega''),\textbf{P}^\dagger(\textbf{r}',\omega' - \omega + \omega'')]\rangle.
    \end{split}
    \label{eq:sm_kubo_macro_v2}
\end{equation}
Thus, taking the difference between Eqs.~\eqref{eq:sm_kubo_macro_v1} and~\eqref{eq:sm_kubo_macro_v2}, we derive the frequency-domain fluctuation-dissipation theorem [introduced in Eq.~(2) of the main text]:
\begin{equation}
    \begin{split}
       \langle[\textbf{P}(\text{r},\omega), \textbf{P}^\dagger(\text{r}',\omega')]\rangle & = 2\hbar\boldsymbol{\chi}''(\text{r},\omega;\text{r}',\omega').
    \end{split}
    \label{eq:sm_kubo_macro_freq}
\end{equation}
So far no assumptions have been made about the physical system being probed; however, we will now assume that the system has an underlying bilinear Hamiltonian. First, let us denote the initial canonical variables of the system by $\textbf{Q}(\textbf{r},\omega_r;t_0)$ and $\boldsymbol{\Pi}(\textbf{r},\omega_r;t_0)$, where $\omega_r$ labels a frequency of the continuum of modes that constitute the macroscopic medium~\cite{scheel2008macroscopic,huttner1992quantization,philbin2010canonical}, and where $t_0$ labels an initial time before the start of the temporal evolution, when the system is assumed to have been at equilibrium. Then, the canonical commutation relations are $[\textbf{Q}(\textbf{r},\omega_r;t_0),\boldsymbol{\Pi}(\textbf{r}',\omega'_r;t_0)] = i\hbar\delta(\textbf{r}-\textbf{r}')\delta(\omega_r - \omega'_r)\textbf{I}$, where $\textbf{I}$ is the unit dyadic in three dimensions. Next, if the system Hamiltonian is bilinear, it must be of the form:
\begin{equation}
    \begin{split}
        &H_\text{sys}(t) = \iint d\text{r}\text{ }d\text{r}'\iint d\omega_r\text{ }d\omega'_r\sum_{i,j}\Big(A_{ij}(\textbf{r},\omega_r;\textbf{r}',\omega'_r;t)\Pi_i(\textbf{r},\omega_r)\Pi_j(\textbf{r}',\omega'_r) \\& + B_{ij}(\textbf{r},\omega_r;\textbf{r}',\omega'_r;t)Q_i(\textbf{r},\omega_r)Q_j(\textbf{r}',\omega'_r) + C_{ij}(\textbf{r},\omega_r;\textbf{r}',\omega'_r;t)\Pi_i(\textbf{r},\omega_r)Q_j(\textbf{r}',\omega'_r) \\&+C^*_{ij}(\textbf{r},\omega_r;\textbf{r}',\omega'_r;t)Q_j(\textbf{r}',\omega'_r)\Pi_i(\textbf{r},\omega_r)\Big),
    \end{split}
    \label{eq:sm_h_sys}
\end{equation}
where $i,j=x,y,z$ and where we have allowed the coupling tensors $A$, $B$ and $C$ to depend on time. Eq.~\eqref{eq:sm_h_sys} then constitutes the most general bilinear Hamiltonian for a frequency-dispersive and dissipative time-varying system. Furthermore, it is Hermitian. This is standard in macroscopic media, since the continuum of modes that compose the material body is responsible for the dissipative dynamics of the electromagnetic field~\cite{scheel2008macroscopic,huttner1992quantization,philbin2010canonical,rytov1989principles}. 
\par
Crucially, owing to the bilinearity of the system Hamiltonian, the solutions to the Heisenberg equations of motion $d_tQ_i(\textbf{r},\omega_r) = 1/(i\hbar)[Q_i(\textbf{r},\omega_r),H_\text{sys}(t)]$ and $d_t\Pi_i(\textbf{r},\omega_r) = 1/(i\hbar)[\Pi_i(\textbf{r},\omega_r),H_\text{sys}(t)]$ will be functions of $Q_i(\textbf{r},\omega_r;t_0)$ and $\Pi_i(\textbf{r},\omega_r;t_0)$. Furthermore, the Hermiticity of the Hamiltonian implies that the equal-time commutation relations between the canonical variables are preserved during the time-evolution. On the other hand, the most general form for the macroscopic polarization field is:
\begin{equation}
    \begin{split}
        P_i(\textbf{r},t)=\int d^3\text{r}'\int d\omega_r\sum_j\left(\mathcal{K}^{(Q)}_{ij}(\textbf{r},\textbf{r}';\omega_r;t)Q_j(\textbf{r}',\omega_r;t) + \mathcal{K}^{(\Pi)}_{ij}(\textbf{r},\textbf{r}';\omega_r;t)\Pi_j(\textbf{r}',\omega_r;t)\right),
    \end{split}
    \label{eq:sm_pol_q_pi}
\end{equation}
where $\mathcal{K}^{(Q)}_{ij}(\textbf{r},\textbf{r}';\omega_r;t)$ and $\mathcal{K}^{(\Pi)}_{ij}(\textbf{r},\textbf{r}';\omega_r;t)$ are kernels that can depend on time; importantly, the time dependence in the kernels does not come from solving the Heisenberg equations of motion, but from a parametric temporal modulation of the system parameters. This implies that, in general, the total time derivative of $P_i(\textbf{r},t)$  will be given by $d_tP_i(\textbf{r},t) = 1/(i\hbar)[P_i(\textbf{r},t),H_\text{sys}(t)] + \partial_tP_i(\textbf{r},t)$, where $\partial_tP_i(\textbf{r},t)$ is the time derivative due to the time dependence of the kernels. Now, writing Eq.~\eqref{eq:sm_pol_q_pi} in terms of the initial canonical variables $Q_i(\textbf{r},\omega_r;t_0)$ and $\Pi_i(\textbf{r},\omega_r;t_0)$, the polarization field reads:
\begin{equation}
    \begin{split}
        P_i(\textbf{r},t)=\int d^3\text{r}'\int d\omega_r\sum_j\left(\widetilde{\mathcal{K}}^{(Q)}_{ij}(\textbf{r},\textbf{r}';\omega_r;t)Q_j(\textbf{r}',\omega_r;t_0) + \widetilde{\mathcal{K}}^{(\Pi)}_{ij}(\textbf{r},\textbf{r}';\omega_r;t)\Pi_j(\textbf{r}',\omega_r;t_0)\right),
    \end{split}
    \label{eq:sm_pol_q_pi(v2)}
\end{equation}
where we have introduce new kernels $\widetilde{\mathcal{K}}^{(Q)}_{ij}(\textbf{r},\textbf{r}';\omega_r;t)$ and $\widetilde{\mathcal{K}}^{(\Pi)}_{ij}(\textbf{r},\textbf{r}';\omega_r;t)$ that capture the full time dependence (i.e., parametric and Hamiltonian) of the system. Now, using Eq.~\eqref{eq:sm_pol_q_pi(v2)}, it is trivial to prove that $[\textbf{P}(\text{r},t), \textbf{P}(\text{r}',t')]$ is indeed a c-number. This in turn enables us to get rid of the expectation value in Eqs.~\eqref{eq:sm_kubo_macro} and~\eqref{eq:sm_kubo_macro_freq}.
\par
Let us now write the initial canonical variables of the system in terms of bosonic operators $\textbf{f}(\textbf{r},\omega_r;t_0)$ and $\textbf{f}^\dagger(\textbf{r},\omega_r;t_0)$ as:
\begin{subequations}
    \begin{equation}
        \textbf{Q}(\textbf{r},\omega_r;t_0) = \sqrt{\frac{\hbar\omega_r\rho}{2}}\left(\textbf{f}(\textbf{r},\omega_r;t_0) + \textbf{f}^\dagger(\textbf{r},\omega_r;t_0)\right)
        \label{eq:sm_q_f}
    \end{equation}
    and
    \begin{equation}
        \boldsymbol{\Pi}(\textbf{r},\omega_r;t_0) = -i\sqrt{\frac{\hbar}{2\omega_r\rho}}\left(\textbf{f}(\textbf{r},\omega_r;t_0) - \textbf{f}^\dagger(\textbf{r},\omega_r;t_0)\right),
        \label{eq:sm_p_f}
    \end{equation}
    \label{eq:sm_qp_f}
\end{subequations}
where $\rho$ is the mass per unit volume of the continuum of modes that make up the material body and $[\textbf{f}(\textbf{r},\omega_r;t_0),\textbf{f}^\dagger(\textbf{r}',\omega'_r;t_0)] = \delta(\textbf{r} - \textbf{r}')\delta(\omega_r - \omega_r')\textbf{I}$. Then, one can further write Eq.~\eqref{eq:sm_pol_q_pi(v2)} as:
\begin{equation}
    \begin{split}
        P_i(\textbf{r},t) = \int d^3\text{r}'\int d\omega_r\sum_jK_{ij}(\textbf{r},\textbf{r}';\omega_r;t)f_j(\textbf{r}',\omega_r;t_0) + \text{H.c.},
    \end{split}
    \label{eq:sm_pol_p_f}
\end{equation}
where $K_{ij}(\textbf{r},\textbf{r}';\omega_r;t) \equiv \sqrt{\hbar\omega_r\rho/2}\widetilde{\mathcal{K}}^{(Q)}_{ij}(\textbf{r},\textbf{r}';\omega_r;t) - i\sqrt{\hbar/(2\omega_r\rho)}\widetilde{\mathcal{K}}^{(\Pi)}_{ij}(\textbf{r},\textbf{r}';\omega_r;t)$ and where H.c. stands for the Hermitian conjugate. Finally, taking the Fourier transform of the above, we arrive to:
\begin{equation}
    \begin{split}
        P_i(\textbf{r},\omega) = \int d^3\text{r}'\int d\omega_r\sum_j\Big(K_{ij}(\textbf{r},\textbf{r}';\omega_r;\omega)f_j(\textbf{r}',\omega_r;t_0) + K^*_{ij}(\textbf{r},\textbf{r}';\omega_r;-\omega)f^\dagger_j(\textbf{r}',\omega_r;t_0)\Big).
    \end{split}
    \label{eq:sm_pol_p_f(v2)}
\end{equation}
In particular, if we change the notation to $f_i(\textbf{r},\omega_r;t_0)\longrightarrow f_i(\textbf{r},\omega)$ and $K_{ij}(\textbf{r},\textbf{r}';\omega_r;\omega)\longrightarrow K_{ij}(\textbf{r},\omega; \textbf{r}',\omega')$, we arrive to:
\begin{equation}
    \begin{split}
        P_i(\textbf{r},\omega) = \int d^3\text{r}'\int_0^\infty d\omega'\sum_j\Big(K_{ij}(\textbf{r},\omega; \textbf{r}',\omega')f_j(\textbf{r}',\omega') + K_{ij}(\textbf{r},\omega; \textbf{r}',-\omega')f^\dagger_j(\textbf{r}',\omega')\Big).
    \end{split}
    \label{eq:sm_pol_p_f(v3)}
\end{equation}
Crucially, Eq.~\eqref{eq:sm_pol_p_f(v3)} is exactly Eq.~(5) of the main text in component form. Thus, we have proved that is it indeed possible to introduce polaritonic operators in time-varying media and furthermore, that these coincide with the initial polaritons of the system, prior to the start of the temporal modulation. Finally, substituting Eq.~\eqref{eq:sm_pol_p_f(v3)} into Eq.~\eqref{eq:sm_kubo_macro_freq}, the loss kernel $K_{ij}(\textbf{r},\omega; \textbf{r}',\omega')$ is found to satisfy the following integral formula:
\begin{equation}
    \begin{split}
        \int d^3\text{r}''\int d\omega''\text{sgn}(\omega'')&\textbf{K}(\textbf{r},\omega;\textbf{r}'',\omega'')\cdot\textbf{K}^\dagger(\textbf{r}',\omega';\textbf{r}'',\omega'') = 2\hbar\boldsymbol{\chi}''(\text{r},\omega;\text{r}';\omega'),
    \end{split}
    \label{eq:sm_loss_kenel_mod}
\end{equation}
which is Eq.~(6) of the main text.
\section{Field Quantization in Dispersive Time-Varying Media}
In this section, we provide more details about the quantization of the electromagnetic field in frequency-dispersive and dissipative time-varying media. In particular, we will prove the integral formula for the Green's dyadic [Eq.~(13) of the main text] and that the commutator between the electric and magnetic field operators equals its free-space value:
\begin{equation}
    [\textbf{E}(\textbf{r}, t),\textbf{B}(\textbf{r}', t)] = -i\frac{\hbar}{\epsilon_0}\boldsymbol{\nabla}\cross\delta(\textbf{r} - \textbf{r}').
\end{equation}
We start from Maxwell curl equations:
\begin{subequations}
    \begin{equation}
        \curl\textbf{E} = -\partial_t\textbf{B},
    \end{equation}
    \begin{equation}
        \curl\textbf{H} = \partial_t\textbf{D} + \textbf{J},
    \end{equation}
\end{subequations}
Where $\textbf{J}(\textbf{r},t) = \partial_t\textbf{P}(\textbf{r},t)$ is the noise current associated to the noise polarization field. On the other hand, the constitutive relations that relate the displacement and magnetic fields to the electric field and the magnetic induction are:
\begin{subequations}
    \begin{equation}
        \textbf{D}(\textbf{r},t) = \epsilon_0\int d^3\text{r}'\int dt'\boldsymbol{\varepsilon}(\textbf{r},t;\textbf{r}',t')\cdot\textbf{E}(\textbf{r}',t'),
    \end{equation}
    \begin{equation}
        \textbf{B}(\textbf{r},t) = \mu_0\textbf{H}(\textbf{r},t).
    \end{equation}
\end{subequations}
Thus, we assume the material to be non-magnetic, with the electric response on the other hand being as general as possible: anisotropic, spatially non-local, dispersive and time-varying. However, the electric permittivity tensor will be causal, naturally; causality is built in our formalism, since our starting point is Kubo's formula~\eqref{eq:sm_kubo_macro}. By taking the curl in Faraday's law and taking the Fourier transform of the resulting equation, we can derive the following wave-equation for the electric field:
\begin{equation}
    \curl\curl\textbf{E}(\textbf{r},\omega) - \frac{\omega^2}{c^2}\int d^3\text{r}'\int\frac{d\omega'}{2\pi}\boldsymbol{\varepsilon}(\textbf{r},\omega;\textbf{r}',\omega')\cdot\textbf{E}(\textbf{r}',\omega') = \mu_0\omega^2\textbf{P}(\textbf{r},\omega),
\end{equation}
with the Green's tensor solving:
\begin{equation}
    \curl\curl\textbf{G}(\textbf{r},\omega;\textbf{r}',\omega') - \frac{\omega^2}{c^2}\int d^3\text{r}''\int\frac{d\omega''}{2\pi}\boldsymbol{\varepsilon}(\textbf{r},\omega;\textbf{r}'',\omega'')\cdot\textbf{G}(\textbf{r}'',\omega'';\textbf{r}',\omega') = 2\pi\textbf{I}\delta(\textbf{r} - \textbf{r}')\delta(\omega - \omega').
\end{equation}
Writing the electric field in terms of the noise polarization, we have:
\begin{equation}
    \textbf{E}(\textbf{r},\omega) = \mu_0\int d^3\text{r}'\int\frac{d\omega'}{2\pi}\textbf{G}(\textbf{r},\omega;\textbf{r}',\omega')(\omega')^2\textbf{P}(\textbf{r}',\omega').
\end{equation}
Importantly, the time-domain permittivity tensor $\boldsymbol{\varepsilon}(\textbf{r},t;\textbf{r}',t')$ is causal, and as such, is analytic in both $\Im(\omega) > 0$ and $\Im(\omega') > 0$. Any two-frequency kernel that is causal must be analytic in both $\Im(\omega) > 0$ and $\Im(\omega') > 0$, as we prove now. In the time-domain,
\begin{equation*}
    \varepsilon(t,t') = \int\frac{d\omega}{2\pi}e^{-i\omega t}\int\frac{d\omega'}{2\pi}e^{i\omega' t'}\varepsilon_{(t,t')}(\omega,\omega'),
\end{equation*}
where the $(t,t')$-subindex is there to remind us that $\varepsilon_{(t,t')}(\omega,\omega')$ is the Fourier transform of $\varepsilon(t,t')$ and not of $\varepsilon(t,t - t')$ or $\varepsilon(t - t',t')$; obviously, all three choices for the way the two-time permittivity depends on $t$ and $t'$ share the same functional form, but their Fourier transforms do not coincide. For $t\leq t'$, the above is zero due to causality. we can write it as
\begin{equation*}
    \varepsilon(t,t') = \int\frac{d\omega}{2\pi}e^{-i\omega(t - t')}\int\frac{d\omega'}{2\pi}e^{i(\omega' - \omega)t'}\varepsilon_{(t,t')}(\omega,\omega').
\end{equation*}
Now, we write $t - t' = \tau$ and change variables $\omega' - \omega \longrightarrow \omega'$, so that the above looks like
\begin{equation*}
    \varepsilon(t,t') = \int\frac{d\omega}{2\pi}e^{-i\omega\tau}\int\frac{d\omega'}{2\pi}e^{i\omega't'}\varepsilon_{(t,t')}(\omega,\omega' + \omega) = \varepsilon(\tau,t').
\end{equation*}
Now, exchanging the order of integration,
\begin{equation*}
    \varepsilon(t,t') = \int\frac{d\omega'}{2\pi}e^{i\omega't'}\int\frac{d\omega}{2\pi}e^{-i\omega\tau}\varepsilon_{(t,t')}(\omega,\omega' + \omega) = \varepsilon(\tau,t') = \int\frac{d\omega'}{2\pi}e^{i\omega't'}\int\frac{d\omega}{2\pi}e^{-i\omega\tau}\varepsilon_{(\tau,t')}(\omega,\omega')
\end{equation*}
If we now move the integral in $\omega$ to the complex plane, for $\tau \leq 0$ we can close the integration contour in the $\text{Im}(\omega)>0$ half plane. Since the above has to vanish due to causality, we prove that $\varepsilon_{(\tau,t')}(\omega,\omega')$ is analytic in $\text{Im}(\omega)>0$, which means $\varepsilon_{(t,t')}(\omega,\omega' + \omega) = \varepsilon_{(\tau,t')}(\omega,\omega')$ is too analytic in $\Im(\omega)>0$ for arbitrary $\omega'$. Now, proceeding in a similar fashion, we can write the two-time permittivity as:
\begin{equation*}
    \varepsilon(t,t') = \int\frac{d\omega}{2\pi}e^{i\omega t}\int\frac{d\omega'}{2\pi}e^{-i\omega'\tau}\varepsilon_{(t,t')}(\omega + \omega',\omega') = \varepsilon(t,\tau) = \int\frac{d\omega}{2\pi}e^{i\omega t}\int\frac{d\omega'}{2\pi}e^{-i\omega'\tau}\varepsilon_{(t,\tau)}(\omega,\omega').
\end{equation*}
Since the above is zero for $\tau \leq 0$, we can close the $\omega'$-integral in the upper complex half-plane and thus prove that $\varepsilon_{(t,\tau)}(\omega,\omega')$ is analytic in $\Im(\omega') > 0$, which makes $\varepsilon_{(t,t')}(\omega + \omega',\omega')$ analytic in $\Im(\omega') > 0$ for every $\omega$. Hence, we have proved that $\varepsilon_{(t,t')}(\omega,\omega')$ is analytic in both $\Im(\omega)>0$ and $\Im(\omega')>0$.
\par
After this brief parenthesis, we return to our problem. The magnetic induction can be written in terms of the electric field using Faraday's law:
\begin{equation}
    \textbf{B}(\textbf{r},\omega) = \frac{1}{i\omega}\curl\textbf{E}(\textbf{r},\omega).
\end{equation}
Now, writing the commutator explicitly,
\begin{equation*}
    \begin{split}
        [\textbf{E}(\textbf{r}, t),\textbf{B}(\textbf{r}', t)] & = \int\frac{d\omega}{2\pi}e^{-i\omega t}\int\frac{d\omega'}{2\pi}e^{i\omega' t}[\textbf{E}(\textbf{r}, \omega),\textbf{B}^\dagger(\textbf{r}', \omega)] \\& = \boldsymbol{\nabla}'\cross i\int\frac{d\omega}{2\pi}\int\frac{d\omega'}{2\pi}e^{-i(\omega - \omega')t}\frac{1}{\omega'}[\textbf{E}(\textbf{r}, \omega),\textbf{E}^\dagger(\textbf{r}', \omega)].
    \end{split}
\end{equation*}
Next, we calculate the commutator between different frequency components of the electric field. To do so, we follow an operator approach that greatly simplifies the algebra:
\begin{equation}
    E_{i}(\textbf{r},\omega) = \bra{\textbf{r},\omega,i}\text{E},
\end{equation}
where $\text{E}$ is the coordinate-free electric field operator. Similarly,
\begin{equation}
    G_{ij}(\textbf{r},\omega;\textbf{r}',\omega') = \bra{\textbf{r},\omega,i}\text{G}\ket{\textbf{r}',\omega',j},
\end{equation}
and
\begin{equation}
    G^*_{ji}(\textbf{r}',\omega';\textbf{r},\omega) = \bra{\textbf{r},\omega,i}\text{G}^\dagger\ket{\textbf{r}',\omega',j},
\end{equation}
where $\dagger$ refers to the full/total Hermitian conjugate (complex conjugation followed by transposition of all indices: polarization, frequency and position ones). We now introduce the following diagonal operator $\text{W}$, where $\bra{\textbf{r},\omega,i}\text{W}\ket{\textbf{r}',\omega',j} = \omega\delta_{ij}\delta(\textbf{r} - \textbf{r}')2\pi\delta(\omega - \omega')$. This enables us to write the electric field operator in a coordinate-free representation as:
\begin{equation}
    \text{E} = \mu_0\text{G}\cdot\text{W}^2\cdot\text{P}.
\end{equation}
Importantly, E and P are at the same time operators and vectors: on the one hand, they are quantum mechanical operators, and thus have a Hilbert space associated with them (namely, the Fock space of the polariton operators $\textbf{f}(\textbf{r},\omega)$ and $\textbf{f}^\dagger(\textbf{r},\omega)$). However, they are also vectors with respect to the differential operator of the electric field wave-equation. Using this coordinate-free representation of the electric field, we can write its commutation relations as $[E_i(\textbf{r}, \omega),E_j^\dagger(\textbf{r}', \omega)] = \bra{\textbf{r}, \omega,i}[\text{E},\text{E}^\dagger]\ket{\textbf{r}', \omega',j}$. Then, one has:
\begin{equation*}
    \begin{split}
        [\text{E},\text{E}^\dagger] & = \mu^2_0\text{G}\text{W}^2[\text{P},\text{P}^\dagger]\text{W}^2\text{G}^\dagger\\
        & = \mu_0^2\text{G}\text{W}^2(2\hbar\epsilon_0\mathcal{E}'')\text{W}^2\text{G}^\dagger,
    \end{split}
\end{equation*}
where we have made use of the fluctuation-dissipation theorem and where $\bra{\textbf{r},\omega,i}\mathcal{E}''\ket{\textbf{r}',\omega',j} = \\\varepsilon''_{ij}(\textbf{r},\omega;\textbf{r}',\omega')$. Now, the electric field Green's dyadic solves, in coordinate-free representation, the following:
\begin{equation}
    (\text{D} - c^{-2}\text{W}^2\mathcal{E})\cdot \text{G} = \text{I},
\end{equation}
where, $\bra{\textbf{r},\omega,i}\text{D}\ket{\textbf{r}',\omega',j} = (\curl\curl)_{ij}\delta(\textbf{r} - \textbf{r}')2\pi\delta(\omega - \omega')$. Going back to the electric field commutator, we now focus our attention on $\text{G}\text{W}^2\mathcal{E}''\text{W}^2\text{G}^\dagger$:
\begin{equation*}
    \begin{split}
        \text{G}\text{W}^2\mathcal{E}''\text{W}^2\text{G}^\dagger & = \frac{1}{i2}\text{G}\text{W}^2(\mathcal{E} - \mathcal{E}^\dagger)\text{W}^2\text{G}^\dagger = \frac{1}{i2}(\text{G}\text{W}^2\mathcal{E}\text{W}^2\text{G}^\dagger - \text{G}\text{W}^2\mathcal{E}^\dagger \text{W}^2\text{G}^\dagger).
    \end{split}
\end{equation*}
Now, since $(\text{D} - c^{-2}\text{W}^2\mathcal{E}) = \text{G}^{-1}$, as long as $(\text{D} - c^{-2}\text{W}^2\mathcal{E})$ is well-defined, the inverse of the Green's tensor exists. Additionally, we can make use of
\begin{equation}
    c^{-2}\text{W}^2\mathcal{E}\text{G} = \text{D}\text{G} - \text{I}
\end{equation}
and
\begin{equation}
    c^{-2}\text{G}^\dagger\mathcal{E}^\dagger \text{W}^2 = \text{G}^\dagger \text{D} - \text{I}.
\end{equation}
Bringing all these ingredients together, we have:
\begin{equation}
    \begin{split}
        \text{G}\text{W}^2\mathcal{E}\text{W}^2\text{G}^\dagger = c^2\text{G}(\text{D} -\text{G}^{-1})\text{W}^2\text{G}^\dagger = c^2\text{G}\text{D}\text{W}^2\text{G}^\dagger - c^2\text{W}^2\text{G}^\dagger,
    \end{split}
\end{equation}
while its Hermitian conjugate gives:
\begin{equation}
    \begin{split}
        \text{G}\text{W}^2\mathcal{E}^\dagger \text{W}^2\text{G}^\dagger = c^2\text{G}\text{W}^2\text{D}\text{G}^\dagger - c^2\text{G}\text{W}^2.
    \end{split}
\end{equation}
Since $\text{D}$ and $\text{W}$ commute, $\text{D}\text{W}^2 = \text{W}\text{D}\text{W} = \text{W}^2\text{D}$ and thus, given that both $\text{D}$ and $\text{W}$ are Hermitian, $(\text{G}\text{W}\text{D}\text{W}\text{G}^\dagger)'' = 0$. Therefore, we finally find that:
\begin{equation}
    \text{G}\text{W}^2\mathcal{E}''\text{W}^2\text{G}^\dagger = c^2(\text{G}\text{W}^2)'' = c^2\frac{1}{i2}(\text{G}\text{W}^2 - \text{W}^2\text{G}^\dagger),
    \label{eq:sm_g_int}
\end{equation}
which is nothing but Eq.~(13) of the main text in a coordinate free-representation. Hence, we find that: 
\begin{equation}
    [\text{E},\text{E}^\dagger] = 2\mu^2_0\epsilon_0c^2\hbar(\text{G}\text{W}^2)'' = 2\hbar\mu_0(\text{G}\text{W}^2)''
\end{equation}
which in coordinate representation reads:
\begin{equation}
    [E_i(\textbf{r}, \omega),E_j^\dagger(\textbf{r}', \omega')] = \frac{\hbar\mu_0}{i}\left(G_{ij}(\textbf{r}, \omega;\textbf{r}', \omega')(\omega')^2 - \omega^2G^*_{ji}(\textbf{r}', \omega';\textbf{r}, \omega)\right),
    \label{eq:sm_e_comm}
\end{equation}
which proves Eq.~(15) of the main text.
\par
Making use of Eq.~\eqref{eq:sm_e_comm}, the equal-time commutator between the electric field and the magnetic induction reads:
\begin{equation}
    \begin{split}
        [\textbf{E}(\textbf{r}, t),\textbf{B}(\textbf{r}', t)] & =   \boldsymbol{\nabla}'\cross i\int\frac{d\omega}{2\pi}\int\frac{d\omega'}{2\pi}e^{-i(\omega - \omega')t}\frac{1}{\omega'}[\textbf{E}(\textbf{r}, \omega),\textbf{E}^\dagger(\textbf{r}', \omega')]\\&
        = \boldsymbol{\nabla}'\cross i\int\frac{d\omega}{2\pi}\int\frac{d\omega'}{2\pi}e^{-i(\omega - \omega')t}\frac{1}{\omega'}\frac{\hbar\mu_0}{i}\left(\textbf{G}(\textbf{r}, \omega;\textbf{r}', \omega')(\omega')^2 - \omega^2\textbf{G}^\dagger(\textbf{r}', \omega';\textbf{r}, \omega)\right).
    \end{split}
    \label{eq:sim_e_b_comm(v2)}
\end{equation}
Let's now have a look at the following:
\begin{equation}
    \begin{split}
         \int\frac{d\omega}{2\pi}\int\frac{d\omega'}{2\pi}e^{-i(\omega - \omega')t}&\frac{1}{\omega'}\hbar\mu_0\textbf{G}(\textbf{r}, \omega;\textbf{r}', \omega')(\omega')^2  = \int\frac{d\omega}{2\pi}\int\frac{d\omega'}{2\pi}e^{-i(\omega - \omega')t}\hbar\mu_0\textbf{G}(\textbf{r}, \omega;\textbf{r}', \omega')\omega'\\&
        = \int\frac{d\omega}{2\pi}\int\frac{d\omega'}{2\pi}e^{i\omega't}\mu_0\hbar\textbf{G}(\textbf{r}, \omega;\textbf{r}', \omega' + \omega)(\omega + \omega'),
    \end{split}
\end{equation}
where between the first and second lines, we have made the change of variables $\omega'-\omega\longrightarrow\omega'$. In the $\omega\longrightarrow\infty$ limit, where material response necessarily becomes transparent to light~\cite{scheel2008macroscopic}, we have the following:
\begin{equation}
    \textbf{G}(\textbf{r}, \omega;\textbf{r}', \omega') \sim -\frac{c^2}{\omega^2}2\pi\textbf{I}\delta(\textbf{r} - \textbf{r}')\delta(\omega - \omega'),
\end{equation}
and so
\begin{equation}
    \textbf{G}(\textbf{r}, \omega;\textbf{r}', \omega' + \omega) \sim -\frac{c^2}{\omega^2}2\pi\textbf{I}\delta(\textbf{r} - \textbf{r}')\delta(\omega').
\end{equation}
Thus, $\textbf{G}(\textbf{r}, \omega;\textbf{r}', \omega' + \omega)(\omega + \omega')\sim-(c^2/\omega)2\pi\textbf{I}\delta(\textbf{r} - \textbf{r}')\delta(\omega')$ has a simple pole when $\omega\longrightarrow\infty$. Crucially, since we are working with a causal prescription of the Green's dyadic, this pole must be understood as having an infinitesimal negative imaginary part $-i\delta$, where $\delta\longrightarrow0^+$~\cite{scheel2008macroscopic}. Then, to be consistent with causality, we do $\omega \longrightarrow \omega + i\delta$ in the denominator and calculate the resulting integral using Cauchy's residue theorem. This way, it is found that:
\begin{equation}
    \int\frac{d\omega}{2\pi}\int\frac{d\omega'}{2\pi}e^{i\omega't}\hbar\mu_0\textbf{G}(\textbf{r}, \omega;\textbf{r}', \omega' + \omega)(\omega + \omega') = -\frac{i}{2}\frac{\hbar}{\epsilon_0}\textbf{I}\delta(\textbf{r} - \textbf{r}').
    \label{eq:sm_int_1}
\end{equation}
Now, let us proceed in a similar manner with the second term in the sum of Eq.~\eqref{eq:sim_e_b_comm(v2)}: first, we change variables to $\omega\longrightarrow-\omega$ and $\omega'\longrightarrow-\omega'$ and use the realness of the time-domain Green's tensor, which implies that $\textbf{G}(\textbf{r}, -\omega;\textbf{r}', -\omega') = \textbf{G}^*(\textbf{r}, \omega;\textbf{r}', \omega')$. Finally, exchanging $\omega$ and $\omega'$ as a last step, we arrive to:
\begin{equation}
    -\int\frac{d\omega}{2\pi}\int\frac{d\omega'}{2\pi}e^{-i(\omega - \omega')t}\frac{1}{\omega'}\hbar\mu_0\omega^2\textbf{G}^\dagger(\textbf{r}', \omega';\textbf{r}, \omega) = -\frac{i}{2}\frac{\hbar}{\epsilon_0}\textbf{I}\delta(\textbf{r} - \textbf{r}').
    \label{eq:sm_int_2}
\end{equation}
Therefore, summing Eqs.~\eqref{eq:sm_int_1} and~\eqref{eq:sm_int_2}, we finally prove that
\begin{equation}
    [\textbf{E}(\textbf{r}, t),\textbf{B}(\textbf{r}', t)] = -i\frac{\hbar}{\epsilon_0}\boldsymbol{\nabla}\cross\delta(\textbf{r} - \textbf{r}').
    \label{eq:sm_e_b_comm}
\end{equation}
\section{Fermi Golden Rule}
We now study the interaction between a probe quantum emitter (QE) and the electromagnetic field. Modeling the emitter as a two-level system, and further assuming the coupling to be of the dipolar type, the interaction Hamiltonian in the interaction picture is:
\begin{equation}
    H_\text{int}(t) = -(\sigma^-e^{-i\omega_\text{d}t}\textbf{d} + \sigma^+e^{i\omega_\text{d}t}\textbf{d}^*)\cdot\textbf{E}(\textbf{r}_\text{d},t).
\end{equation}
The Fourier-domain electric field operator reads:
\begin{equation}
   \textbf{E}(\textbf{r},\omega) = \int d^3\text{r}\int_0^\infty d\omega'\left(\boldsymbol{\mathcal{G}}(\textbf{r},\omega;\textbf{r},\omega')\cdot\textbf{f}(\textbf{r},\omega') + \boldsymbol{\mathcal{G}}(\textbf{r},\omega;\textbf{r},-\omega')\cdot\textbf{f}^\dagger(\textbf{r},\omega')\right),
\end{equation}
where we recall the definition of the \textit{dressed} Green's dyadic:
\begin{equation}
    \begin{split}
        \boldsymbol{\mathcal{G}}(\textbf{r},\omega;\textbf{r}',\omega') = &\mu_0\int d^3\text{r}''\int\frac{d\omega''}{2\pi}\textbf{G}(\textbf{r},\omega;\textbf{r}'',\omega'')(\omega'')^2\cdot\textbf{K}(\textbf{r}'',\omega'';\textbf{r}',\omega'),
    \end{split}
    \label{eq:sm_green_func_dressed}
\end{equation}
and where the total electric field is given by:
\begin{equation}
    \textbf{E}(\textbf{r},t) = \int_0^\infty\frac{d\omega}{2\pi}e^{-i\omega t}\textbf{E}(\textbf{r},\omega) + \text{H.c.}
\end{equation}
We study two different type of processes: the spontaneous decay (SD) of the quantum emitter, which encompasses all processes of the type $\ket{e,\text{vac}}\longrightarrow\ket{g,1_{i}(\textbf{r},\omega)}$, and its spontaneous excitation (SE), represented by all transitions of the type $\ket{g,\text{vac}}\longrightarrow\ket{e,1_{i}(\textbf{r},\omega)}$.
\par
To first-order in time-dependent perturbation theory, the wave-function of the joint QE-electromagnetic field system is given by:
\begin{equation}
    \ket{\psi(t)} = \ket{\psi(t_0)} - \frac{1}{i\hbar}\int_{t_0}^tdt'H_\text{int}(t')\ket{\psi(t_0)} + \mathcal{O}(H_\text{int}^2).
    \label{eq:sm_psi}
\end{equation}
For the case of SD, the initial ket is $\ket{\psi(t_0)} = \ket{e,\text{vac}}$, whereas for the case of SE, it is $\ket{\psi(t_0)} = \ket{g,\text{vac}}$.
\par
Let us first study the case of SD: multiplying Eq.~\eqref{eq:sm_psi} by $\bra{g,1_{i}(\textbf{r},\omega)}$ and performing the time-integrals, we arrive to:
\begin{equation}
    \begin{split}
        \langle g,1_{i}(\textbf{r},\omega)|\psi(t)\rangle & = \frac{1}{\hbar}\sum_j d_j\int_0^\infty\frac{d\omega'}{2\pi}\Big({\mathcal{G}}^*_{ji}(\textbf{r}_\text{d},\omega';\textbf{r},\omega)\frac{\sin\left(\frac{1}{2}(\omega_\text{d} - \omega')(t - t_0)\right)}{\frac{1}{2}(\omega_\text{d} - \omega')}e^{-i\frac{1}{2}(\omega_\text{d} - \omega')(t + t_0)} + \\&
        {\mathcal{G}}_{ji}(\textbf{r}_\text{d},\omega';\textbf{r},-\omega)\frac{\sin\left(\frac{1}{2}(\omega_\text{d} + \omega')(t - t_0)\right)}{\frac{1}{2}(\omega_\text{d} + \omega')}e^{-i\frac{1}{2}(\omega_\text{d} + \omega')(t + t_0)}\Big).
    \end{split}
\end{equation}
If we take the $t\longrightarrow\infty$ limit, the above expression transforms into: 
\begin{equation}
    \begin{split}
        \langle g,1_{i}(\textbf{r},\omega)|\psi(\infty)\rangle & = \frac{1}{\hbar}\sum_j d_j\int_0^\infty\frac{d\omega'}{2\pi}\Big({\mathcal{G}}^*_{ji}(\textbf{r}_\text{d},\omega';\textbf{r},\omega)2\pi\delta(\omega_\text{d} - \omega') + 
        {\mathcal{G}}_{ji}(\textbf{r}_\text{d},\omega;\textbf{r},-\omega)2\pi\delta(\omega_\text{d} + \omega')\Big) \\& =  \frac{1}{\hbar}\sum_jd_j{\mathcal{G}}^*_{ji}(\textbf{r}_\text{d},\omega_\text{d};\textbf{r},\omega).
    \end{split}
\end{equation}
It it seen that the counter-rotating term contributes only to the transient of the scattering process and not to its asymptotic value. This is a standard feature of the weak coupling regime~\cite{scheel2008macroscopic}. Now we calculate $\abs{\langle g,1_{i}(\textbf{r},\omega)|\psi(\infty)\rangle}^2$ and sum over $i=x,y,z$ and integrate over $\textbf{r}$ and $\omega$ to derive the total probability for the spontaneous decay of the quantum emitter:
\begin{equation}
    \text{Pr}_\text{SD} = \int d^3\text{r}\int_0^\infty d\omega\sum_i\abs{\langle g,1_{i}(\textbf{r},\omega)|\psi(\infty)\rangle}^2 = \frac{1}{\hbar^2}\textbf{d}^*\cdot\int d^3\text{r}\int_0^{\infty}d\omega\boldsymbol{\mathcal{G}}(\textbf{r}_\text{d},\omega_\text{d};\textbf{r},\omega)\cdot\boldsymbol{\mathcal{G}}^\dagger(\textbf{r}_\text{d},\omega_\text{d};\textbf{r},\omega)\cdot\textbf{d}.
    \label{eq:sm_prob_sd}
\end{equation}
Crucially, $\text{Pr}_\text{SD}$ denotes a probability and not a probability per unit time, in contrast with the usual decay rate calculation in non-modulated media~\cite{scheel2008macroscopic}. Importantly, for a general temporal modulation, there is no frequency conservation law (or generalized conservation law) that produces a Dirac delta in $\mathcal{G}_{ij}(\textbf{r},\omega;\textbf{r}',\omega')$. However, for the particular case of a periodic temporal modulation, $\mathcal{G}_{ij}(\textbf{r},\tilde{\omega}+n\Omega;\textbf{r}',\tilde{\omega}'+n'\Omega) = \mathcal{G}_{ij;nn'}(\textbf{r},\textbf{r}';\tilde{\omega})\delta(\tilde{\omega} - \tilde{\omega}')$; then, the Dirac delta makes the limit $\underset{t\longrightarrow\infty}{\lim}1 / (t-t_0)\abs{\langle g,1_{i}(\textbf{r},\omega)|\psi(t)\rangle}^2$ well-defined, which enables us to recover a decay rate for the quantum emitter:
\begin{equation}
    \begin{split}
       \Gamma_\text{SD} =& \frac{1}{\hbar^2}\textbf{d}^*\cdot\int d^3r\sum_{l}\Theta\left(\tilde{\omega}_\text{d} + l\Omega\right)\boldsymbol{\mathcal{G}}_{n_\text{d},l}(\textbf{r}_\text{d},\textbf{r};\tilde{\omega}_\text{d})\boldsymbol{\mathcal{G}}^\dagger_{n_\text{d},l}(\textbf{r}_\text{d},\textbf{r};\tilde{\omega}_\text{d})\cdot\textbf{d},
    \end{split}
    \label{eq:sm_gamma_sd}
\end{equation}
where $\omega_\text{d} = \tilde{\omega}_\text{d} + n_\text{d}\Omega$. As a second comment, we also highlight that neither $\text{Pr}_\text{SD}$ nor $\Gamma_\text{SD}$ are proportional to the anti-Hermitian part of the Green's dyadic, in contrast to case of non-modulated media~\cite{scheel2008macroscopic}; this is rooted in the time modulation, which introduces gain into the system and allows the quantum emitter to become spontaneously excited from the ground state.
\par
We now calculate the probability for the spontaneous excitation of the emitter. The initial state of the system is now $\ket{\psi(t_0)} = \ket{g,\text{vac}}$, which we substitute into Eq.~\eqref{eq:sm_psi} and then project into $\ket{e,1_i(\textbf{r},\omega)}$. Then, proceeding as we did for the spontaneous decay calculation, the probability for the spontaneous excitation of the emitter is found to be:
\begin{equation}
    \begin{split}
        \text{Pr}_\text{SE} & = \int d^3\text{r}\int_0^\infty d\omega\sum_\alpha\abs{\langle e,1_{\alpha}(\textbf{r},\omega)|\psi(\infty)\rangle}^2 = \frac{1}{\hbar^2}\textbf{d}^*\cdot\int d^3\text{r}\int_0^\infty d\omega\boldsymbol{\mathcal{G}}(\textbf{r}_\text{d},\omega_\text{d};\textbf{r},-\omega)\cdot\boldsymbol{\mathcal{G}}^\dagger(\textbf{r}_\text{d},\omega_\text{d};\textbf{r},-\omega)\cdot\textbf{d}\\&
        = \frac{1}{\hbar^2}\textbf{d}^*\cdot\int d^3\text{r}\int_{-\infty}^0 d\omega\boldsymbol{\mathcal{G}}(\textbf{r}_\text{d},\omega_\text{d};\textbf{r},\omega)\cdot\boldsymbol{\mathcal{G}}^\dagger(\textbf{r}_\text{d},\omega_\text{d};\textbf{r},\omega)\cdot\textbf{d}.
    \end{split}
    \label{eq:sm_prob_se}
\end{equation}
In Eq.~\eqref{eq:sm_prob_se}, the integral over the frequency variable $\omega$ is restricted to negative frequencies only. Thus, for the emitter to become spontaneously excited from the ground state, positive and negative frequencies must interact. On the other hand, if the temporal modulation is periodic, one can derive a rate for the spontaneous excitation of the emitter:
\begin{equation}
    \begin{split}
       \Gamma_\text{SE} =& \frac{1}{\hbar^2}\textbf{d}^*\cdot\int d^3r\sum_{l}\Theta\left(-(\tilde{\omega}_\text{d} + l\Omega)\right)\boldsymbol{\mathcal{G}}_{n_\text{d},l}(\textbf{r}_\text{d},\textbf{r};\tilde{\omega}_\text{d})\boldsymbol{\mathcal{G}}^\dagger_{n_\text{d},l}(\textbf{r}_\text{d},\textbf{r};\tilde{\omega}_\text{d})\cdot\textbf{d}.
    \end{split}
    \label{eq:sm_gamma_se}
\end{equation}
Next, we take the difference between $\text{Pr}_\text{SD}$ and $\text{Pr}_\text{SE}$ and make use of Eqs.~\eqref{eq:sm_loss_kenel_mod},~\eqref{eq:sm_g_int} and~\eqref{eq:sm_green_func_dressed} to derive the following expression for the total probability:
\begin{equation}
    \begin{split}
        \text{Pr}_\text{tot} & = \text{Pr}_\text{SD} - \text{Pr}_\text{SE} \\&
        = \frac{2\mu_0^2\varepsilon_0}{\hbar}\textbf{d}^*\cdot\iint d^3\text{r }d^3\text{r}'\iint\frac{d\omega}{2\pi}\frac{d\omega'}{2\pi}\textbf{G}(\textbf{r}_\text{d},\omega_\text{d};\textbf{r},\omega)\omega^2\cdot\boldsymbol{\varepsilon}''(\textbf{r},\omega;\textbf{r}',\omega')\cdot(\omega')^2\textbf{G}^\dagger(\textbf{r}_\text{d},\omega_\text{d};\textbf{r}',\omega')\cdot\textbf{d}
        \\& = \frac{2\mu_0\omega_\text{d}^2}{\hbar}\textbf{d}^*\cdot\textbf{G}''(\textbf{r}_\text{d},\omega_\text{d};\textbf{r}_\text{d},\omega_\text{d})\cdot\textbf{d}.
        \label{eq:sm_prob_tot}
    \end{split}
\end{equation}
For the particular case of a periodic temporal modulation, Eq.~\eqref{eq:sm_prob_tot} reduces to Eq.(22) of the main text. Now, let us calculate the electromagnetic energy radiated by a classical harmonic point dipole with polarization density $\textbf{P}(\textbf{r},t) = \text{Re}\{\textbf{d}e^{-i\omega_\text{d}t}\delta(\textbf{r} - \textbf{r}_\text{d})\}$ and associated current density $\textbf{J}(\textbf{r},t) = \partial_t\textbf{P}(\textbf{r},t)$. On the other hand, the electric field generated by the dipole is $\textbf{E}(\textbf{r},t) = \mu_0\omega^2_\text{d}\text{Re}\{\int d\omega/(2\pi)e^{-i\omega t}\textbf{G}(\textbf{r},\omega;\textbf{r}_\text{d},\omega_\text{d})\cdot\textbf{d}\}$. Then, the total electromagnetic energy emitted by the dipole is given by:
\begin{equation}
    W_\text{tot} = \underset{t\longrightarrow\infty}{\lim} \int_{t_0}^t\int d^3\text{r}\textbf{J}\cdot\textbf{E} = \mu_0\omega_\text{d}^3\textbf{d}^*\cdot\textbf{G}''(\textbf{r}_\text{d},\omega_\text{d};\textbf{r}_\text{d},\omega_\text{d})\cdot\textbf{d} = \frac{\hbar\omega_\text{d}}{2}\text{Pr}_\text{tot}.
\end{equation}
Thus, the total energy emitted by the classical dipole is found to be proportional to the total probability for decay-excitation of the quantum emitter; crucially, this equivalence between classical and quantum calculations is general and applies to any temporal modulation. Importantly, as it was the case for the total probability, $W_\text{tot}$ is the total energy emitted by the dipole and not the energy per unit time, since the latter is not well-defined for a general time modulation. However, for the particular case of a periodic temporal modulation, the limit $\underset{t\longrightarrow\infty}{\lim}1 / (t-t_0)\int_{t_0}^t\int d^3\text{r}\textbf{J}\cdot\textbf{E} = P_\text{tot}$ is indeed well-defined and thus, it is possible to define a total emitted power.
\section{Green's dyadic for a planar time-periodic half-space}
In this section, we calculate the Green's dyadic for a planar half-space that is frequency-dispersive, dissipative and periodically modulated in time, where the half-space occupies the $z<0$ region and the $z>0$ region is filled with air. Our approach will be very similar to the one presented in the Supplementary Information of Ref.~\cite{sustaeta2026near}. First, let us make Floquet \textit{Ansätze} for the fields and the currents inside the half-space: $E_j(\textbf{r},t)=\fontsize{7}{8}{\sumint}(d\omega_n/2\pi)e^{-i\omega_nt}E_{j,n}$$(\textbf{r},\tilde{\omega})$, $H_j(\textbf{r},t)=\fontsize{7}{8}{\sumint}(d\omega_n/2\pi)e^{-i\omega_nt}H_{j,n}$$(\textbf{r},\tilde{\omega})$ and so on, where $j=x\text{, }y\text{, }z$, $n\in\mathbb{Z}$ and $\footnotesize{\sumint} d\omega_n \equiv \int^{\Omega/2}_{-\Omega/2}d\tilde{\omega}\sum_n$. Then, the Fourier-domain Maxwell equations read:
\begin{subequations}
    \begin{equation}
        \curl\textbf{E}_n(\textbf{r},\tilde{\omega}) = i\mu_0\omega_n\textbf{H}_n(\textbf{r},\tilde{\omega}),
        \label{eq:sm_faraday}
    \end{equation}
    \begin{equation}
        \curl\textbf{H}_n(\textbf{r},\tilde{\omega}) = -i\epsilon_0\omega_n\sum_{n'}\varepsilon_{n,n'}(\tilde{\omega})\textbf{E}_{n'}(\textbf{r},\tilde{\omega}) + \textbf{J}_n(\textbf{r},\tilde{\omega}).
        \label{eq:sm_ampere}
    \end{equation}
\end{subequations}
For practical purposes, we have to set a bound on the number of Floquet modes included in the field expansions. Therefore, we take $-N_\text{F}\leq n,n'\leq N_\text{F}$, where the total number of Floquet bands is $N_\text{tot} = 2N_\text{F} + 1$. On the other hand, owing to the conservation of the in-plane momentum $\textbf{K}_\parallel$, we can write $E_{j,n}(\textbf{r},\tilde{\omega}) = E_{j,n}(\textbf{R},z,\tilde{\omega})  =  \int d^2K_\parallel / (2\pi)^2e^{i\textbf{K}_\parallel\cdot\textbf{R}}E_{j,n}(z;\textbf{K}_\parallel,\tilde{\omega})$ and similarly for the magnetic field and the current. Furthermore, we can work in the basis expanded by $\{\hat{\textbf{s}},\hat{\textbf{k}},\hat{\textbf{z}}\}$, where $\hat{\textbf{k}} = \textbf{K}_\parallel / K_\parallel$ and $\hat{\textbf{s}} = \hat{\textbf{k}}\cross\hat{\textbf{z}}$; this basis enables us to decouple the two polarizations $\sigma=\text{TE, TM}$, as we will see. Next, dropping the $(z;\textbf{K}_\parallel,\tilde{\omega})$ argument for the sake of simplicity, we can write Maxwell equations for TE-polarized fields:
\begin{subequations}
    \begin{equation}
        \partial_zE_{s,n} = i\mu_0\omega_nH_{k,n},
    \end{equation}
       \begin{equation}
        -iK_\parallel E_{s,n} = i\mu_0\omega_nH_{z,n},
    \end{equation}
       \begin{equation}
        iK_\parallel H_{z,n} - \partial_zH_{k,n} = -i\epsilon_0\sum_{n'}\varepsilon_{n,n'}E_{s,n'} + J_{s,n},
    \end{equation}
    \label{eq:sm_eqs_te}
\end{subequations}
and for TM-polarized ones:
\begin{subequations}
    \begin{equation}
        iK_\parallel E_{z,n} - \partial_zE_{k,n} = i\mu_0\omega_nH_{s,n},
    \end{equation}
    \begin{equation}
        \partial_zH_{s,n} = -i\epsilon_0\sum_{n'}\varepsilon_{n,n'}E_{k,n'} + J_{k,n},
    \end{equation}
    \begin{equation}
        -iK_\parallel H_{s,n} = -i\epsilon_0\sum_{n'}\varepsilon_{n,n'}E_{z,n'} + J_{z,n}.
    \end{equation}
    \label{eq:sm_eqs_tm}
\end{subequations}
From the above set of equations, it can be readily seen that the $z$-components can be eliminated as:
\begin{equation}
    H_{z,n} = -\frac{K_\parallel}{\mu_0\omega_n}E_{s,n},
    \label{eq:sm_h_z}
\end{equation}
\begin{equation}
    E_{z,n} = \frac{i\mu_0\omega_nH_{s,n}+\partial_zE_{k,n}}{iK_\parallel}.
    \label{eq:sm_e_z}
\end{equation}
Next, we expand the tangential components of the electric and magnetic fields in terms of the system eigenvectors:
\begin{subequations}
    \begin{equation}
        E_{s,n} = \sum_p\mathcal{E}_{n,p}e^{ik_{z,p}z}A^\text{(TE)}_{p},
    \end{equation}
        \begin{equation}
        H_{k,n} = \sum_p\mathcal{H}^\text{(TE)}_{n,p}e^{ik_{z,p}z}A^\text{(TE)}_{p}
    \end{equation}
\end{subequations}
for TE-modes and:
\begin{subequations}
    \begin{equation}
        E_{k,n} = \sum_p\mathcal{E}_{n,p}e^{ik_{z,p}z}A^\text{(TM)}_{p},
    \end{equation}
        \begin{equation}
        H_{s,n} = \sum_p\mathcal{H}^\text{(TM)}_{n,p}e^{ik_{z,p}z}A^\text{(TM)}_{p}
    \end{equation}
\end{subequations}
for TM-ones. In the above, $\mathcal{E}_{n,p}$ denotes the $n$-th Floquet component of the $p$-th electric eigenvector and $\mathcal{H}^{(\sigma)}_{n,p}$ denotes the $n$-th Floquet component of the $p$-th magnetic eigenvector for $\sigma=$TE, TM polarization. Additionally, $A^{(\sigma)}_p$ is the amplitude of the $p$-th eigenvector for a polarization $\sigma$, while $k_{z,p}$ is the $p$-th wave-vector. Substituting these formulas in Eqs.~\eqref{eq:sm_eqs_te},~\eqref{eq:sm_eqs_tm} and making use of Eqs.~\eqref{eq:sm_h_z},~\eqref{eq:sm_e_z} to eliminate the $z$-components, we find the following eigenvalue equation for the electric field eigenvectors:
\begin{equation}
    \left(c^{-2}\textbf{W}^2\cdot\boldsymbol{\varepsilon}-K_\parallel^2\textbf{I}\right)\cdot\boldsymbol{{\mathcal{E}}}_p = k^2_{z,p}\boldsymbol{{\mathcal{E}}}_p,
    \label{eq:sm_eig},
\end{equation}
where $W_{n,n'} = \omega_{n}\delta_{n,n'}$, $[\boldsymbol{\varepsilon}]_{n,n'} = \varepsilon_{n,n'}$ and $I_{n,n'} = \delta_{n,n'}$. Crucially, the square in $k^2_{z,p}$ means that both $k_{z,p}$ and $-k_{z,p}$ are eigenvalues; these two form a pair of wave-vectors, with one being forward or right-propagating [i.e., $\text{Im}(k_{z,p})>0$] and the other being backward or left-propagating [i.e., $\text{Im}(k_{z,p})<0$]. We also define the displacement field eigenvectors as $\boldsymbol{\mathcal{D}}_p = \boldsymbol{\varepsilon}\cdot\boldsymbol{\mathcal{E}}_p$. Furthermore, we can write the magnetic field eigenvectors in terms of the electric ones as:
\begin{equation}
    \mathcal{H}_{n,p}^\text{(TE)} = \frac{k_{z,p}}{\mu_0\omega_n}\mathcal{E}_{n,p},
    \label{eq:sm_hTE_m}
\end{equation}
and
\begin{equation}
    \mathcal{H}_{n,p}^\text{(TM)} = -\frac{\epsilon_0\omega_n}{k_{z,p}}\mathcal{D}_{n,p}.
    \label{eq:sm_hTM_m}
\end{equation}
Now, since we aim to solve for the Green's dyadic, we take the source current to be ${J}_{i,n} = -i\omega_n\delta_{nm}\delta_{ij}\delta(z-z')$; this is the current associated to a $j$-polarized harmonic dipole of unit amplitude that oscillates at the frequency $\omega_m$ and is located at $z = z'$. Integrating each set of equations with respect to $z$ along an interval centered at $z'$ as $\int_{z'-\delta z}^{z'+\delta z}dz(...)$ and taking the $\delta z\longrightarrow0^+$ limit, we derive the boundary conditions for the electric and magnetic field components at $z = z'$. For TE-polarization, these read:
\begin{subequations}
    \begin{equation}
        \Delta E_{s,n} = 0,
    \end{equation}
    \begin{equation}
        \Delta H_{k,n} = i\omega_n\delta_{s,j}\delta_{n,m},
    \end{equation}
    \label{eq:sm_jump_te}
\end{subequations}
while for TM-polarization one finds:
\begin{subequations}
    \begin{equation}
        \Delta H_{s,n} = -i\omega_n\delta_{k,j}\delta_{n,m},
    \end{equation}
    \begin{equation}
        \Delta E_{k,n} = -\frac{iK_\parallel}{\epsilon_0}\delta_{z,j}[\boldsymbol{\varepsilon}^{-1}]_{n,m},
    \end{equation}
    \label{eq:sm_jump_tm}
\end{subequations}
where $\boldsymbol{\varepsilon}^{-1}$ denotes the inverse of the Floquet permittivity matrix. Importantly, for the case of TM polarization we ought to consider both $j=k,z$. Additionally, owing to the boundary condition equations Eqs.~\eqref{eq:sm_jump_te},~\eqref{eq:sm_jump_tm}, we must consider separately the $z<z'$ and $z>z'$ cases. Thus, we write the TE-polarized fields as $E_{s,n} = E^{(l)}_{s,n}\Theta(z'-z) + E^{(r)}_{s,n}\Theta(z-z')$ and $H_{k,n} = H^{(l)}_{k,n}\Theta(z'-z) + H^{(r)}_{k,n}\Theta(z-z')$ and similarly for the case of TM polarization. On the other hand, the temporal modulation couples different Floquet harmonics; this in turn means that the Green's dyadic will be a Cartesian-Floquet tensor, with each polarization having their own Floquet matrix with components $[\textbf{g}^{(\sigma)(l,r)}]_{n,n'} = g^{(\sigma)(l/r)}_{n,n'}$.  Then, using Eqs.~\eqref{eq:sm_jump_te},~\eqref{eq:sm_jump_tm}, we find that the Green's dyadic Floquet matrices read:
\begin{equation}
        \textbf{g}^{\text{(TE)}(l/r)} = \textbf{g}^{\text{(TE)}} = \frac{i}{2}\text{diag}\left(1 / \textbf{k}_z\right)\cdot\boldsymbol{\mathcal{E}}^{-1}\cdot\textbf{W}^2\delta_{s,j}
        \label{eq:sm_gTE}
\end{equation}
and
\begin{equation}
        \textbf{g}^{\text{(TM)}(l/r)} = \frac{i}{2}\text{diag}\left( \textbf{k}_z\right)\cdot\boldsymbol{\mathcal{D}}^{-1}\delta_{k,j} \text{  }\mp\frac{i}{2}K_\parallel\boldsymbol{\mathcal{D}}^{-1}\delta_{z,j} = \textbf{g}^{\text{(TM)}}_k + \textbf{g}^{\text{(TM)}(l/r)}_z.
        \label{eq:sm_gTM}
\end{equation}
In the above equations, $[\text{diag}\left(1 / \textbf{k}_z\right)]_{p,p'} = 1 / k_{z,p}\delta_{p,p'}$ and $[\text{diag}\left(\textbf{k}_z\right)]_{p,p'} = k_{z,p}\delta_{p,p'}$, while $[\mathcal{E}]_{n,p} = \mathcal{E}_{n,p}$ and $[\mathcal{D}]_{n,p} = \mathcal{D}_{n,p}$. Importantly, in all the previous matrices, the eigenvalue index $p$ is constrained to $\text{Im}(k_{z,p})>0$~\cite{sustaeta2026near}\footnote{For the electric and displacement field eigenvectors this is not important, as both $\pm k_{z,p}$ lead to the same $\boldsymbol{\mathcal{E}}_p$ and $\boldsymbol{\mathcal{D}}_p$.}. Crucially, Eqs.~\eqref{eq:sm_gTE},~\eqref{eq:sm_gTM} give the electric field amplitudes corresponding to the Green's dyadic, but not the total fields. To obtain these, one must multiply Eqs.~\eqref{eq:sm_gTE},~\eqref{eq:sm_gTM} by the electric field eigenvector matrix $\boldsymbol{\mathcal{E}}$.
\par
So far we have assumed that $z'<0$, corresponding to emission from inside the semi-infinite slab. For $z'>0$, the medium is air and thus, there is no time modulation, with all Floquet matrices consequently becoming diagonal. In this case, one can proceed as for the $z'<0$ case, and calculate the forward-propagating vacuum wave-vectors:
\begin{equation}
    k^{\text{(vac)}}_{z,n} = \text{sgn}(\omega_n)\sqrt{\frac{\omega_n^2}{c^2}-K_\parallel^2}\Theta(\abs{\omega_n}-cK_\parallel) + i\sqrt{K_\parallel^2-\frac{\omega_n^2}{c^2}}\Theta(cK_\parallel-\abs{\omega_n}),
    \label{eq:sm_kz_vac}
\end{equation}
and magnetic field amplitudes for TE and TM-modes:
\begin{equation}
    \mathcal{H}^{\text{(vac)(TE)}}_{n,\pm} = \pm\frac{k^\text{(vac)}_{z,n}}{\mu_0\omega_n},
    \label{eq:sm_hTE_vac}
\end{equation}
\begin{equation}
    \mathcal{H}^{\text{(vac)(TM)}}_{n,\pm} = \mp\frac{\epsilon_0\omega_n}{k^\text{(vac)}_{z,n}}.
     \label{eq:sm_hTM_vac}
\end{equation}
In the above, $\pm$ distinguishes between the magnetic field components of forward and backward propagating modes. On the other hand, $\text{sgn}(\omega_n)$ guarantees that for negative frequencies and $\abs{\omega_n}>cK_\parallel$, Eq.~\eqref{eq:sm_kz_vac} still describes forward-propagating modes, in accordance with causality~\cite{sustaeta2026near}. Furthermore, the free-space magnetic field amplitudes define the vacuum admittances via $\mathcal{H}^{\text{(vac)}(\sigma)}_{n,\pm} = \pm Y^{\text{(vac)}(\sigma)}_n$; these in turn allow us to define the vacuum impedances as $Z^{\text{(vac)}(\sigma)}_n = 1 / Y^{\text{(vac)}(\sigma)}_n$. Finally, to obtain the amplitudes of the free-space Green's dyadic, it is sufficient to substitute Eq.~\eqref{eq:sm_kz_vac} into Eqs.~\eqref{eq:sm_gTE},~\eqref{eq:sm_gTM} and use $\boldsymbol{\mathcal{E}}^\text{(vac)},\text{ }\boldsymbol{\mathcal{D}}^\text{(vac)} = \textbf{I}$.
\par
Now that we have the Green's dyadic Floquet matrices, we calculate the field transmitted from the time-varying medium to the outside. To do so, we first calculate the Floquet reflection and transmission matrices at the air-medium interface. First, let us assume that light is incident from the vacuum side. Then, the continuity equations for the tangential electric and magnetic field components at $z = 0$ read~\cite{sustaeta2026near}:
\begin{subequations}
    \begin{equation}
        A^{\text{(vac)}(\sigma)}_{n,+} + A^{\text{(vac)}(\sigma)}_{n,-} = \sum_{p<0}\mathcal{E}_{n,p}A^{\text{(med)}(\sigma)}_{p},
    \end{equation}
    \begin{equation}
        \mathcal{H}^{\text{(vac)}(\sigma)}_{n,+}A^{\text{(vac)}(\sigma)}_{n,+} + \mathcal{H}^{\text{(vac)}(\sigma)}_{n,-}A^{\text{(vac)}(\sigma)}_{n,-} = \sum_{p<0}\mathcal{H}^{(\sigma)}_{n,p}A^{\text{(med)}(\sigma)}_{p}.
    \end{equation}
\end{subequations}
In the above, $p<0$ constrains the summation over the eigenvalue index $p$ to $\text{Im}(k_{z,p})<0$, since the fields transmitted from the free-space region to the half-space propagate towards $z < 0$. Then, some simple matrix algebra leads to:
\begin{subequations}
    \begin{equation}
        \textbf{A}^{\text{(med)}(\sigma)}_- = 2\left(\boldsymbol{\mathcal{E}}+\textbf{Z}^{\text{(vac)}(\sigma)}\cdot\boldsymbol{\mathcal{H}}^{(\sigma)}_+\right)^{-1}\cdot\textbf{A}^{\text{(vac)}(\sigma)}_-,
    \end{equation}
        \begin{equation}
        \textbf{A}^{\text{(vac)}(\sigma)}_+ = \left(\boldsymbol{\mathcal{E}}-\textbf{Z}^{\text{(vac)}(\sigma)}\cdot\boldsymbol{\mathcal{H}}^{(\sigma)}_+\right)\cdot\left(\boldsymbol{\mathcal{E}}+\textbf{Z}^{\text{(vac)}(\sigma)}\cdot\boldsymbol{\mathcal{H}}^{(\sigma)}_+\right)^{-1}\cdot\textbf{A}^{\text{(vac)}(\sigma)}_-,
    \end{equation}
\end{subequations}
where now the matrix elements of $\boldsymbol{\mathcal{H}}^{(\sigma)}_+$ are constrained to $\text{Im}(k_{z,p})>0$, where we have made use of $\boldsymbol{\mathcal{H}}^{(\sigma)}_- = -\boldsymbol{\mathcal{H}}^{(\sigma)}_+$ [see Eqs.~\eqref{eq:sm_hTE_m},~\eqref{eq:sm_hTM_m}]. Then, the transmission and reflection Floquet matrices when light is incident from the vacuum side are:
\begin{subequations}
    \begin{equation}
        \textbf{t}^{(\sigma)} = 2\left(\boldsymbol{\mathcal{E}}+\textbf{Z}^{\text{(vac)}(\sigma)}\cdot\boldsymbol{\mathcal{H}}^{(\sigma)}_+\right)^{-1},
    \end{equation}
    \begin{equation}
        \textbf{r}^{(\sigma)} = \left(\boldsymbol{\mathcal{E}}-\textbf{Z}^{\text{(vac)}(\sigma)}\cdot\boldsymbol{\mathcal{H}}^{(\sigma)}_+\right)\cdot\left(\boldsymbol{\mathcal{E}}+\textbf{Z}^{\text{(vac)}(\sigma)}\cdot\boldsymbol{\mathcal{H}}^{(\sigma)}_+\right)^{-1} = \frac{1}{2}\left(\boldsymbol{\mathcal{E}}-\textbf{Z}^{\text{(vac)}(\sigma)}\cdot\boldsymbol{\mathcal{H}}^{(\sigma)}_+\right)\cdot\textbf{t}^{(\sigma)}.
    \end{equation}
    \label{eq:sm_tr_vac}
\end{subequations}
If we now do the same for the case in which light is incident from the time-varying half-space, the corresponding transmission and reflection Floquet matrices are given by:
\begin{subequations}
    \begin{equation}
        \tilde{\textbf{t}}^{(\sigma)} = \left(\textbf{Z}^{\text{(vac)}(\sigma)}\cdot\boldsymbol{\mathcal{H}}^{(\sigma)}_+\right)\cdot \textbf{t}^{(\sigma)}\cdot\boldsymbol{\mathcal{E}},
    \end{equation}
    \begin{equation}
        \tilde{\textbf{r}}^{(\sigma)} = -\boldsymbol{\mathcal{E}}^{-1}\cdot\textbf{r}^{(\sigma)}\cdot\boldsymbol{\mathcal{E}}.
    \end{equation}
    \label{eq:sm_tr_med}
\end{subequations}
Now, if we define the propagation matrix inside the medium $P^\text{(med)}_{p,p'}(-z') = e^{-ik_{z,p}z'}\delta_{p,p'}$\footnote{The minus sign in $-z'$ comes from the fact that the optical path from the source point to the interface is $\Delta z = z_\text{int}-z'$, where $z_\text{int}$ is the $z$-coordinate of the interface. Since $z_\text{int}=0$, $\Delta z = -z'$.}, where $\text{Im}(k_{z,p})>0$, and in free-space $P^\text{(vac)}_{n,n'}(z) = e^{ik^\text{(vac)}_{z,n}z}\delta_{n,n'}$, the Green's dyadic $\textbf{G}(z,z')$ for $z'<0$ and $z>0$ is given by:
\begin{equation}
    \textbf{G}(z,z') = \begin{bmatrix}
        \textbf{G}_{ss}(z,z') && \boldsymbol{0} && \boldsymbol{0} \\
        \boldsymbol{0} && \textbf{G}_{kk}(z,z') && \textbf{G}_{kz}(z,z') \\
        \boldsymbol{0} && \textbf{G}_{zk}(z,z') && \textbf{G}_{zz}(z,z')
    \end{bmatrix}
    \label{eq:sm_green},
\end{equation}
where the different components read:
\begin{subequations}
    \begin{equation}
        \textbf{G}_{ss}(z,z') = \textbf{P}^\text{(vac)}(z)\cdot\tilde{\boldsymbol{t}}^{\text{(TE)}}\cdot\textbf{P}^\text{(med)}(-z')\cdot\textbf{g}^\text{(TE)},
    \end{equation}
    \begin{equation}
        \textbf{G}_{kk}(z,z') = \textbf{P}^\text{(vac)}(z)\cdot\tilde{\boldsymbol{t}}^{\text{(TM)}}\cdot\textbf{P}^\text{(med)}(-z')\cdot\textbf{g}^\text{(TM)}_k,
    \end{equation}
    \begin{equation}
        \textbf{G}_{kz}(z,z') = \textbf{P}^\text{(vac)}(z)\cdot\tilde{\boldsymbol{t}}^{\text{(TM)}}\cdot\textbf{P}^\text{(med)}(-z')\cdot\textbf{g}^{\text{(TM)}(r)}_z,
    \end{equation}
    \begin{equation}
        \textbf{G}_{zk}(z,z') = -\text{diag}\left(K_\parallel / \textbf{k}^\text{(vac)}_z\right)\cdot\textbf{P}^\text{(vac)}(z)\cdot\tilde{\boldsymbol{t}}^{\text{(TM)}}\cdot\textbf{P}^\text{(med)}(-z')\cdot\textbf{g}^\text{(TM)}_k,
    \end{equation}
    \begin{equation}
        \textbf{G}_{zz}(z,z') = -\text{diag}\left(K_\parallel / \textbf{k}^\text{(vac)}_z\right)\cdot\textbf{P}^\text{(vac)}(z)\cdot\tilde{\boldsymbol{t}}^{\text{(TM)}}\cdot\textbf{P}^\text{(med)}(-z')\cdot\textbf{g}^{\text{(TM)}(r)}_z.
    \end{equation}
\end{subequations}
In Eq.~\eqref{eq:sm_green}, $\boldsymbol{0}$ is a square matrix of size $N_\text{tot}\cross N_\text{tot}$ whose elements are all zeroes. Crucially, all the calculations in this section have been done in the $\{\hat{\textbf{s}},\hat{\textbf{k}},\hat{\textbf{z}}\}$ basis. To transform back to the original $\{\hat{\textbf{x}},\hat{\textbf{y}},\hat{\textbf{z}}\}$ Cartesian basis, we must use the following rotation matrix:
\begin{equation}
    \textbf{R} = \begin{bmatrix}
    \cos(\varphi)\textbf{I} && \sin(\varphi)\textbf{I} && \boldsymbol{0} \\
    -\sin(\varphi)\textbf{I} && \cos(\varphi)\textbf{I} && \boldsymbol{0}
    \\
    \boldsymbol{0} && \boldsymbol{0} && \textbf{I}
    \end{bmatrix},
    \label{eq:sm_rot_mat}
\end{equation}
which undoes the  $\{\hat{\textbf{x}},\hat{\textbf{y}},\hat{\textbf{z}}\}\longrightarrow \{\hat{\textbf{s}},\hat{\textbf{k}},\hat{\textbf{z}}\}$ transformation. In Eq.~\eqref{eq:sm_rot_mat}, $\varphi$ is the azimuthal angle, so that the in-plane momentum reads $\textbf{K}_\parallel = K_\parallel[\cos(\varphi), \sin(\varphi), 0]^\text{T}$, where T stands for the transpose. Then, to return to the original Cartesian basis, we do $\textbf{G}(z,z')\longrightarrow\textbf{R}\cdot\textbf{G}(z,z')\cdot\textbf{R}^\text{T}$.
\section{Dispersive vs non-dispersive temporal modulations}
Here, we compare the relative strength between dispersive and non-dispersive temporal modulations. We will consider both a polar material with a time-periodic Lorentz frequency and a plasmonic medium with a time-periodic plasma frequency, as in the main text.
\par
Let us first consider the case of a polar insulator. For the dispersive time modulation, the Floquet susceptibility matrix reads:
\begin{equation}
    \chi_{n,n'}(\tilde{\omega}) =  \omega_\text{p}^2\big[D^{-1}(\tilde{\omega})\big]_{n,n'},
    \label{eq:sm_chi_lor_1}
\end{equation}
where $D^{-1}(\tilde{\omega})$ is the inverse of $D(\tilde{\omega})$, with elements:
\begin{equation}
    \begin{split}
        D_{n,n'}(\tilde{\omega}) &= \left(-\omega_n^2-i\gamma\omega_n+\omega^2_\ell\right)\delta_{n,n'}
        + \frac{\alpha}{2}\omega^2_\ell\left(\delta_{n+1,n'} + \delta_{n-1,n'}\right).
    \end{split}
    \label{eq:sm_chi_lor_2}
\end{equation}
Then, the Floquet permittivity matrix is given by:
\begin{equation}
    \begin{split}
        \varepsilon_{n,n'}(\tilde{\omega})= \varepsilon_\infty\delta_{n,n'} +  \omega_\text{p}^2\big[D^{-1}(\tilde{\omega})\big]_{n,n'}.
        \label{eq:sm_eps_lor}
    \end{split}
\end{equation}
Let us now assume $\alpha\ll1$ and make a two-band approximation. Then, the inverse of $D$ is approximately given by:
\begin{equation}
    D^{-1} \simeq \begin{bmatrix}
       \frac{1}{-\omega^2-i\gamma\omega + \omega_\ell^2} && -\frac{\alpha}{2}\frac{\omega_\ell^2}{\text{det}(D)} \\
       -\frac{\alpha}{2}\frac{\omega_\ell^2}{\text{det}(D)} &&  \frac{1}{-(\omega-\Omega)^2-i\gamma(\omega-\Omega) + \omega_\ell^2}
    \end{bmatrix},
\end{equation}
 where $\text{det}(D) = [-\omega^2-i\gamma\omega + \omega_\ell^2][-(\omega-\Omega)^2-i\gamma(\omega-\Omega) -\omega_\ell^2] - (\alpha/2)^2\omega_\ell^4$. In the Floquet sidebands regime, $\Omega\ll\omega$. Moreover, if $\alpha\ll1$, one can further approximate $\text{det}(D)\simeq(-\omega^2-i\gamma\omega + \omega_\ell^2)^2$. Finally, it is readily found that:
 \begin{equation}
     \varepsilon_{0,-1} \simeq -\frac{\alpha}{2}\frac{\omega^2_\ell}{\omega_\text{p}^2}\chi_\text{bg}^2(\omega).
 \end{equation}
 Meanwhile, for a dispersionless temporal modulation, $\varepsilon_{0,-1} = \alpha_\text{nd}/2$, which is exact. Then, the ratio between dispersive and non-dispersive $\varepsilon_{0,-1}$ is $R = -\big[(\omega_\ell/\omega_\text{p})^2\chi_\text{bg}^2(\omega)\big]\alpha / \alpha_\text{nd}$. Equating the absolute value of this ratio to one, we finally arrive to $\alpha_\text{nd} = (\omega_\ell/\omega_\text{p})^2\abs{\chi_\text{bg}(\omega)}^2\alpha$, as in the main text.
 \par
 Now, let us study the plasmonic medium. For a periodic temporal modulation of the plasma frequency, the Floquet permittivity matrix reads:
 \begin{equation}
    \begin{split}
        \varepsilon_{n,n'}(\tilde{\omega})& = \varepsilon_\infty-\frac{\omega_\text{p}^2}{\omega_n^2+i\gamma\omega_n}\delta_{n,n'}
        -\frac{\alpha}{2}\frac{\omega_\text{p}^2}{\omega_n^2+i\gamma\omega_n}\left(\delta_{n+1,n'} + \delta_{n-1,n'}\right).
    \end{split}
    \label{eq:sm_chi_plasma}
\end{equation}
Thus, the modulated part of the linear permittivity is $\varepsilon_\text{mod}(\omega) = -(\alpha/2)\chi_\text{bg}(\omega)$. On the other hand, for the non-dispersive temporal modulation, $\varepsilon_\text{mod} = \alpha_\text{nd} / 2$. Then, the ratio between dispersive and non-dispersive $\varepsilon_\text{mod}$ is $R = -\chi_\text{bg}(\omega)\alpha / \alpha_\text{nd}$. Finally, setting $\abs{R} = 1$ leads to $\alpha_\text{nd} = \abs{\chi_\text{bg}(\omega)}\alpha$, as in the main text.
\section{Thermal Radiation in Dispersive Time-Varying Media}
\subsection{Dispersionless temporal modulation}
Here, we analyze the spectral density of a periodically time-varying SiC half-space for the case of a dispersionless temporal modulation. As in the main text, the spectral density of the electromagnetic field is defined as $\mathcal{S}(\textbf{r},\omega) \equiv\sum_{i=x,y,z}\text{G}^{(1)}_{ii}(\textbf{r},\textbf{r};\omega)$. On the other hand, the Floquet permittivity matrix is given by:
\begin{equation}
    \begin{split}
        \varepsilon_{n,n'}(\tilde{\omega}) = \varepsilon_\text{bg}(\omega_n)\delta_{n,n'} + \frac{\alpha_\text{nd}}{2}\left(\delta_{n+1,n'} + \delta_{n-1,n'}\right),
    \end{split}
    \label{eq:sm_eps_nondisp}
\end{equation}
where, $\varepsilon_\text{bg}(\omega_n)=\varepsilon_\infty-\omega_\text{p}^2/\left(\omega_n^2 +i\gamma\omega_n-\omega_\ell^2\right)$ is the non-modulated background permittivity of the material, and $\alpha_\text{nd}$ is the strength of the non-dispersive temporal modulation. On the other hand, we take $\Omega = 0.05\omega_\text{ENZ}$ and assume the half-space to be initially (before the start of the time modulation) in a thermal state at a temperature $T = 300$K, similar to the main text. We also use the rescaling $\alpha_\text{nd} = (\omega_\ell/\omega_\text{p})^2\abs{\chi_\text{bg}(\omega_\text{SPh})}^2\alpha\simeq17.3\alpha$.
\begin{figure*}[t]
    \centering
    \hspace{0 mm}
    \includegraphics[scale = 0.50]{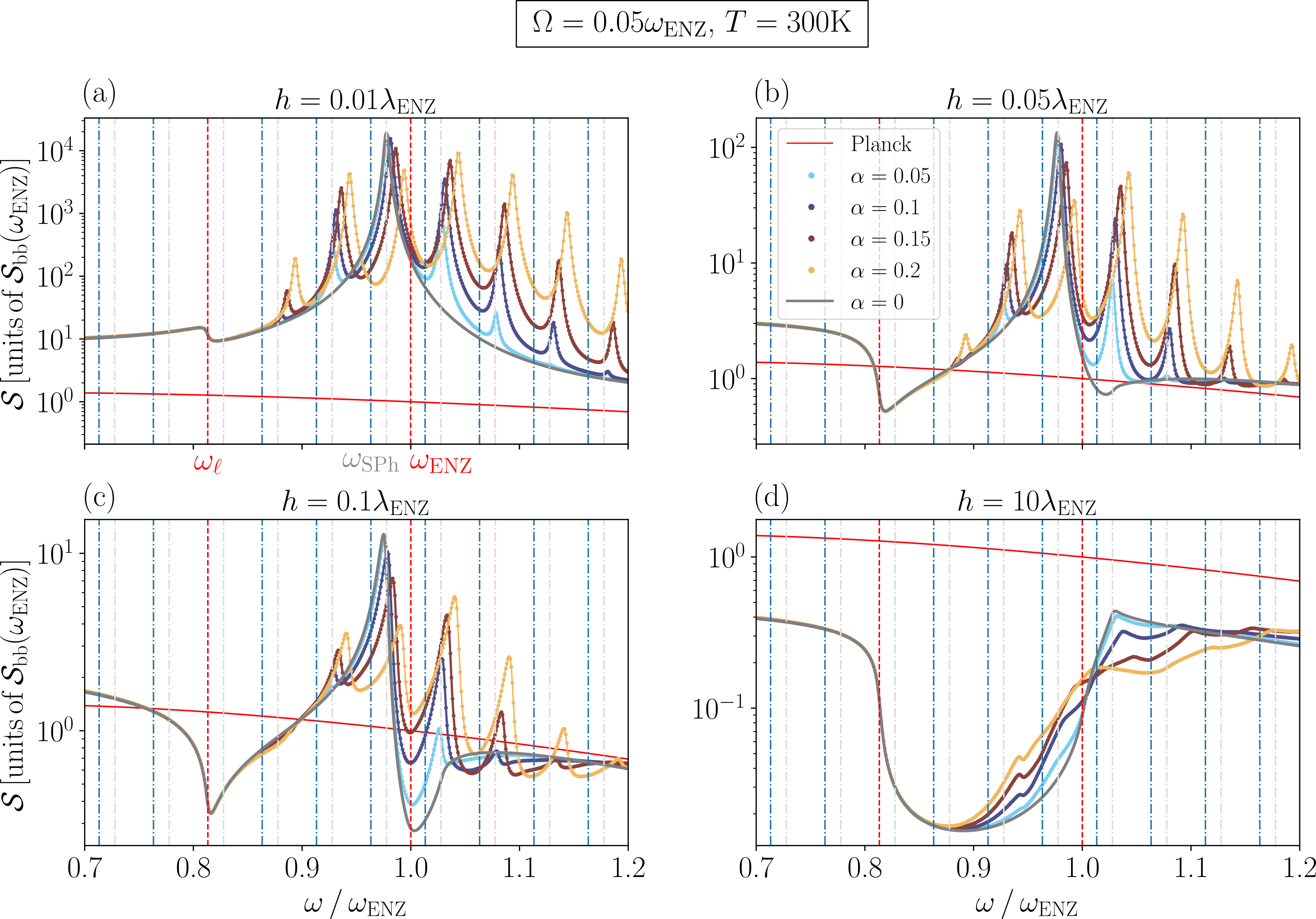} 
    \caption{\justifying Spectral density $\mathcal{S}$ [in units of the black-body spectral density evaluated at $\omega_\text{ENZ}$, $\mathcal{S}_\text{bb}(\omega_\text{ENZ})$] vs frequency for a SiC half-space with a non-dispersive periodic temporal modulation. We consider values different values of $\alpha$ and compare with Planck's law (red line) as well. On the other hand, we study different values of the height above the surface, $h =0.01\lambda_\text{ENZ}\text{(a), }0.05\lambda_\text{ENZ}\text{(b), }0.1\lambda_\text{ENZ}\text{(c), }10\lambda_\text{ENZ}\text{(d)}$. Finally, the modulation frequency is set to $\Omega=0.05\omega_\text{ENZ}$ and the polar insulator is initially (prior to the start of the temporal modulation) in a thermal state with temperature $T = 300$K.}
    \label{fig:sm_fig1} 
\end{figure*}
In Fig.~\ref{fig:sm_fig1} we plot the spectral density as a function of the frequency and for different values of the height above the surface, with $h=0.01\lambda_\text{ENZ}\text{(a), }0.05\lambda_\text{ENZ}\text{(b), }0.1\lambda_\text{ENZ}\text{(c), }\\10\lambda_\text{ENZ}\text{(d)}$. Furthermore, for each height we consider different values of the modulation strength $\alpha=0,0.05,0.1,0.2$ (where we recall that $\alpha_\text{nd} \simeq 17.3\alpha$). The vertical grey dashed-dotted lines are $\omega_\text{SPh}$ and its Floquet sidebands $\omega_\text{SPh} +l\Omega$, while the vertical red dashed lines are $\omega_\ell$ and $\omega_\text{ENZ}$, which limit the Reststrahlen band. Furthermore, the vertical blue dashed-dotted lines are the replicas of the Lorentz frequency, $\omega_\ell+l\Omega$, while the red curve is the black-body spectrum $\mathcal{S}_\text{bb}(\omega)$. 
\par
Let us first concentrate on panel (a), for which $h = 0.01\lambda_\text{ENZ}$, in the near-field of the structure. As was the case for the dispersive temporal modulation, the non-dispersive time modulation enhances the spectral density at the sidebands of the surface phonon frequency. This enhancement comes at the expense of the central peak at $\omega_\text{SPh}$, which weakens as $\alpha$ increases. However, for increasing $\alpha$, the sidepeaks progressively blueshift with respect to the lines at $\omega_\text{SPh}+l\Omega$. Moreover, the height of the positive sidebands (i.e., $l>0$) is larger than in the dispersive case, while the height of the negative ones (i.e., $l<0$) is comparatively smaller. This is rooted in the fact that for a dispersive time modulation, $\varepsilon_\text{mod}\sim1 / \omega_n^2$, whereas for a non-dispersive one, $\varepsilon_\text{mod}$ is frequency-independent. Consequently, a dispersionless temporal modulation is comparatively weaker at small frequencies and stronger at large ones. Finally, in the near-field the spectral density is super-Planckian as well for dispersionless $\varepsilon_\text{mod}$.
\par
Next, we consider $h = 0.05\lambda_\text{ENZ}\text{(b), }0.1\lambda_\text{ENZ}$(c), in the intermediate field of the structure. Similar to the case of dispersive $\varepsilon_\text{mod}$, as the distance to the surface increases, the coupling to the SPh and its sidebands weakens, and the spectral density becomes overall smaller. For $\alpha=0$, $\mathcal{S}$ is sub-Planckian for frequencies in the vicinity of $\omega_\ell$ and $\omega_\text{ENZ}$, at the lower and upper limits of the Reststrahlen band. On the other hand, for large enough $\alpha$, thermal emission becomes super-Planckian for frequencies near the top of the Reststrahlen band. However, the spectrum remains sub-Planckian near the bottom of the band even for strong temporal modulations, as opposed to the $\varepsilon_\text{mod}$ case, for which $\mathcal{S}$ was super-Planckian in the vicinity of $\omega_\ell$ for large enough $\alpha$. This is rooted in $\varepsilon_\text{mod}$ being frequency-independent for a non-dispersive temporal modulation. Additionally, the spectrum is featureless at $\omega_\ell+l\Omega$, in contrast to the dispersive case. Also, contrary to the findings of Refs.~\cite{vazquez2023incandescent,ren2026clarification}, there is no enhancement of $\mathcal{S}$ at the sidebands of the $\epsilon$-near-zero frequency. This is also consistent with the results for dispersive $\varepsilon_\text{mod}$ presented in the main text. Lastly, in panel (d) we study the far-field of the structure, with $h = 10\lambda_\text{ENZ}$. Similar to the case of dispersive $\varepsilon_\text{mod}$, the spectrum is not enhanced at $\omega_\text{ENZ}+l\Omega$, nor does it exhibit any super-Planckian features. 
\par
Therefore, in this section we have demonstrated that accounting for dispersion in the time-modulated permittivity is necessary to provide an accurate description of thermal emission in dispersive time-varying media.
\begin{figure*}[t]
    \centering
    \hspace{0 mm}
    \includegraphics[scale = 0.50] {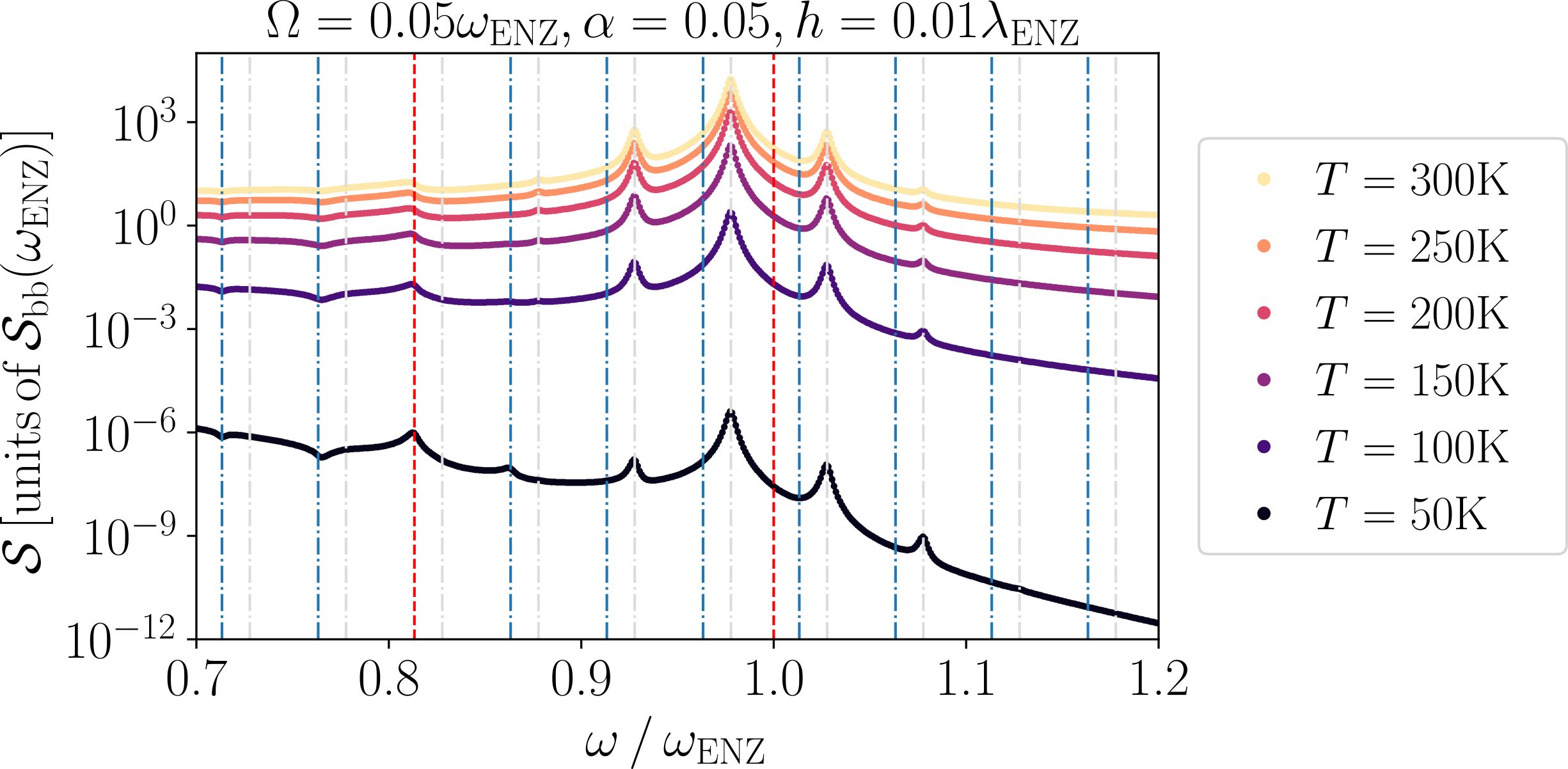} 
    \caption{\justifying Spectral density $\mathcal{S}$ [in units of the black-body spectral density evaluated at $\omega_\text{ENZ}$, $\mathcal{S}_\text{bb}(\omega_\text{ENZ})$ for a temperature of $T =300$K] vs frequency for a SiC half-space with a dispersive temporal modulation. We consider different values of the temperature, where $T=50,100,150,200,250,300$K. The modulation frequency and strength are set to $\Omega=0.05\omega_\text{ENZ}\text{, }\alpha=0.05$, and the height above the surface is $h = 0.01\lambda_\text{ENZ}$.}
    \label{fig:sm_fig2} 
\end{figure*}
\subsection{Dispersive temporal modulation: dependence with the temperature}
In Fig.~\ref{fig:sm_fig2} we study the dependence with the temperature of the spectral density $\mathcal{S}$ of a SiC half-space that is periodically modulated in time; in particular, we consider a dispersive temporal modulation, with the Lorentz frequency of the polar material varying periodically in time, as in the main text. The modulation frequency and strength are set to $\Omega=0.05\omega_\text{ENZ}$, $\alpha=0.05$, and the height above the surface is $h = 0.01\lambda_\text{ENZ}$. As in Fig.~\ref{fig:sm_fig1}, the vertical dashed red lines are $\omega_\ell$ and $\omega_\text{ENZ}$, whereas the grey dashed-dotted lines are the surface phonon frequency and its sidebands. Moreover, the blue dashed-dotted lines are the sidebands of the Lorentz frequency.
\par
We consider different values of the temperature, where $T=50,100,150,200,250,300$K. Importantly, we do not plot the $T=0$K case because in our theory the spectral density is identically zero in that limit. Moreover, for all different values of $T$ we normalize $\mathcal{S}$ to $\mathcal{S}_\text{bb}(\omega_\text{ENZ})$ evaluated at a temperature of $T=300$K [the same normalization as in Fig.~\ref{fig:sm_fig1}]. First, it is seen that the spectral density exhibits sidebands of the SPh at all the temperatures considered. Thus, the basic features of the physics of the time modulation are independent of the temperature of the system. On the other hand, the figure shows that, as the temperature decreases and the SiC half-space subsequently cools down, the spectral density weakens. In particular, $\mathcal{S}$ decreases monotonically with decreasing $T$. Hence, $\mathcal{S}$ indeed vanishes in the $T\longrightarrow0$ limit, in sharp contrast with the findings of Ref.~\cite{vazquez2023incandescent}. This is consistent with our theory, which shows that having $\text{G}^{(1)}>0$ in the $T\longrightarrow0$ limit necessitates a fast modulation, which must be large enough to couple positive and negative frequencies.
\bibliographystyle{unsrt}
\bibliography{supmat}